\documentclass[twocolumn]{aastex631}

\usepackage{xspace}
\usepackage{amsmath} 
\usepackage{amssymb} 
\usepackage{savesym}
\savesymbol{tablenum}
\usepackage{siunitx}
\restoresymbol{SIX}{tablenum}
\newcommand\lya{Ly$\alpha$\xspace}{}
\newcommand\Lya{Ly$\alpha$\xspace}{}
\newcommand{\angstrom}{\mbox{\normalfont\AA}}

\newcommand{\intensityunit}{\mbox{\normalfont{erg\, s}$^{-1}$\, cm$^{-2}$\, arcsec$^{-2}$\, \angstrom$^{-1}$}}

\newcommand{\lowzbbi}{$(6.7 \pm 3.1)$\xspace}
\newcommand{\midzbbi}{$(11.7 \pm 1.4)$\xspace}
\newcommand{\highzbbi}{$(8.3 \pm 1.5)$\xspace}
\newcommand{\bbiunit}{$\times 10^{-22}\,\intensityunit$\xspace}
\newcommand{\NPClowz}{170}
\newcommand{\NPCmediumz}{130}
\newcommand{\NPChighz}{100}
\newcommand{\Nifusfall}{$132,051$\xspace}
\newcommand{\Nifusspring}{$241,969$\xspace}
\newcommand{\Nifusnep}{$34,239$\xspace}

\begin{document}

\title{\Lya Intensity Mapping in HETDEX: Galaxy-\Lya Intensity Cross-Power Spectrum}

\author[0000-0002-6907-8370]{Maja Lujan Niemeyer}
\affiliation{Max-Planck-Institut f\"{u}r Astrophysik, Karl-Schwarzschild-Str. 1, 85741 Garching, Germany}
\affiliation{Ludwig-Maximilians-Universität München, Schellingstr. 4, 80799 München, Germany}
\email{maja@mpa-garching.mpg.de}

\author[0000-0002-0136-2404]{Eiichiro Komatsu}
\affiliation{Max-Planck-Institut f\"{u}r Astrophysik, Karl-Schwarzschild-Str. 1, 85741 Garching, Germany}
\affiliation{Ludwig-Maximilians-Universität München, Schellingstr. 4, 80799 München, Germany}
\affiliation{Kavli Institute for the Physics and Mathematics of the Universe (WPI), The University of Tokyo Institutes for Advanced Study (UTIAS), The University of Tokyo, Chiba 277-8583, Japan}

\author[0000-0002-0961-4653]{José Luis Bernal}
\affiliation{Instituto de F{\'i}sica de Cantabria (IFCA), CSIC-Univ. de Cantabria, Avda. de los Castros s/n, E-39005 Santander, Spain}

\author[0000-0002-0885-8090]{Chris Byrohl}
\affiliation{Kavli Institute for the Physics and Mathematics of the Universe (WPI), The University of Tokyo Institutes for Advanced Study (UTIAS), The University of Tokyo, Chiba 277-8583, Japan}
\affiliation{Universität Heidelberg, Institut für Theoretische Astrophysik, ZAH, 
Albert-Ueberle-Str. 2, 69120 Heidelberg,
Germany}

\author[0000-0002-1328-0211]{Robin Ciardullo}
\affiliation{Department of Astronomy \& Astrophysics, The Pennsylvania State University, University Park, PA 16802, USA}
\affiliation{Institute for Gravitation and the Cosmos, The Pennsylvania State University, University Park, PA 16802}

\author[0000-0002-0212-4563]{Olivia Curtis}
\affiliation{Department of Astronomy \& Astrophysics, The Pennsylvania State University, University Park, PA 16802, USA}
\affiliation{Institute for Gravitation and the Cosmos, The Pennsylvania State University, University Park, PA 16802}

\author[0000-0003-2575-0652]{Daniel J. Farrow}
\affiliation{E. A. Milne Centre for Astrophysics
University of Hull, Cottingham Road, Hull, HU6 7RX, UK}
\affiliation{Centre of Excellence for Data Science,
Artificial Intelligence \& Modelling (DAIM),
University of Hull, Cottingham Road, Hull, HU6 7RX, UK}

\author[0000-0001-8519-1130]{Steven L. Finkelstein}
\affiliation{Department of Astronomy, The University of Texas at Austin, 2515 Speedway Boulevard, Austin, TX 78712, USA}
\affiliation{Cosmic Frontier Center, The University of Texas at Austin, Austin, TX, USA}

\author[0000-0002-8433-8185]{Karl Gebhardt}
\affiliation{Department of Astronomy, The University of Texas at Austin, 2515 Speedway Boulevard, Austin, TX 78712, USA}

\author[0000-0001-6842-2371]{Caryl Gronwall}
\affiliation{Department of Astronomy \& Astrophysics, The Pennsylvania State University, University Park, PA 16802, USA}
\affiliation{Institute for Gravitation and the Cosmos, The Pennsylvania State University, University Park, PA 16802}

\author[0000-0001-6717-7685]{Gary J. Hill} 
\affiliation{Department of Astronomy, The University of Texas at Austin, 2515 Speedway Boulevard, Austin, TX 78712, USA} 
\affiliation{McDonald Observatory, The University of Texas at Austin, 2515 Speedway Boulevard, Austin, TX 78712, USA}

\author[0000-0001-7039-9078]{Matt J. Jarvis}
\affiliation{Astrophysics, Department of Physics, University of Oxford, Keble Road, Oxford, OX1 3RH, UK}
\affiliation{Department of Physics and Astronomy, University of the Western Cape, Robert Sobukwe Road, 7535 Bellville, Cape Town, South Africa}

\author[0000-0002-8434-979X]{Donghui Jeong}
\affiliation{Department of Astronomy \& Astrophysics, The Pennsylvania State University, University Park, PA 16802, USA}
\affiliation{Institute for Gravitation and the Cosmos, The Pennsylvania State University, University Park, PA 16802}
\affiliation{School of Physics, Korea Institute for Advanced Study, Seoul 02455, Korea}

\author[0000-0002-2307-0146]{Erin Mentuch Cooper}
\affiliation{Department of Astronomy, The University of Texas at Austin, 2515 Speedway Boulevard, Austin, TX 78712, USA}

\author[0000-0002-1350-019X]{Deeshani Mitra}
\affiliation{Institute for Multi-messenger Astrophysics and Cosmology, Department of Physics, Missouri University of Science and Technology, 1315 N. Pine St., Rolla MO 65409, USA}

\author[0000-0003-3823-8279]{Shiro Mukae}
\affiliation{Department of Astronomy, The University of Texas at Austin, 2515 Speedway Boulevard, Austin, TX 78712, USA}
\affiliation{MIRAI Technology Institute, Shiseido Co., Ltd., 1-2-11, Takashima, Nishi-ku, Yokohama, Kanagawa, 222-0011, Japan}

\author[0000-0002-8984-0465]{Julian B.~Mu\~noz}
\affiliation{Department of Astronomy, The University of Texas at Austin, 2515 Speedway Boulevard, Austin, TX 78712, USA}

\author[0000-0002-1049-6658]{Masami Ouchi}
\affiliation{Kavli Institute for the Physics and Mathematics of the Universe (WPI), The University of Tokyo Institutes for Advanced Study (UTIAS), The University of Tokyo, Chiba 277-8583, Japan}
\affiliation{National Astronomical Observatory of Japan, 2-21-1 Osawa, Mitaka, Tokyo 181-8588, Japan}
\affiliation{Institute for Cosmic Ray Research, The University of Tokyo, 5-1-5 Kashiwanoha, Kashiwa, Chiba 277-8582, Japan}
\affiliation{Department of Astronomical Science, SOKENDAI (The Graduate University for Advanced Studies), Osawa 2-21-1, Mitaka, Tokyo, 181-8588, Japan}

\author[0000-0002-6186-5476]{Shun Saito}
\affiliation{Kavli Institute for the Physics and Mathematics of the Universe (WPI), The University of Tokyo Institutes for Advanced Study (UTIAS), The University of Tokyo, Chiba 277-8583, Japan}
\affiliation{Institute for Multi-messenger Astrophysics and Cosmology, Department of Physics,
Missouri University of Science and Technology, 1315 N. Pine St., Rolla MO 65409, USA}

\author[0000-0001-7240-7449]{Donald P. Schneider}
\affiliation{Department of Astronomy \& Astrophysics, The Pennsylvania State University, University Park, PA 16802, USA}
\affiliation{Institute for Gravitation and the Cosmos, The Pennsylvania State University, University Park, PA 16802}

\author[0000-0003-2977-423X]{Lutz Wisotzki}
\affiliation{Leibniz-Institut for Astrophysik Potsdam (AIP), An der Sternwarte 16, 14482 Potsdam, Germany}

\begin{abstract}

We present a measurement of the Lyman-$\alpha$ (\Lya) intensity mapping power spectrum from the Hobby-Eberly Telescope Dark Energy Experiment (HETDEX\null). We measure the cross-power spectrum of the \Lya intensity and \Lya-emitting galaxies (LAEs) in a redshift range of $1.9\le z\le 3.5$.
We calculate the intensity from HETDEX spectra that do not contain any detected LAEs above a signal-to-noise ratio of $5.5$. To produce a power spectrum model and its covariance matrix, we simulate the data using lognormal mocks for the LAE catalog and \Lya intensity in redshift space. The simulations include the HETDEX sensitivity, selection function, and mask. The measurements yield the product of the LAE bias, the intensity bias, the mean intensity of undetected sources, and the ratio of the actual and fiducial redshift-space distortion parameters, $b_\mathrm{g} b_I \langle I \rangle \bar{F}_{\rm RSD} / \bar{F}^{\rm fid}_{\rm RSD}=$\lowzbbi,  \midzbbi, and \highzbbi~\bbiunit in three redshift bins centered at $\bar z=2.1$, 2.6, and 3.2, respectively. The results are reasonably consistent with cosmological hydrodynamical simulations that include \Lya radiative transfer. They are, however, significantly smaller than previous results from cross-correlations of quasars with \Lya intensity. These results demonstrate the statistical power of HETDEX for \Lya intensity mapping and pave the way for a more comprehensive analysis.
They will also be useful for constraining models of \Lya emission from galaxies used in modern cosmological simulations of galaxy formation and evolution.
\end{abstract}


\section{Introduction} 
\label{sec:intro}

Line intensity mapping (LIM) is a novel tool for studying cosmology and the astrophysics of galaxies and intergalactic gas \citep[see][for reviews]{kovetz/etal:2017,bernal/kovetz:2022}. 
Intensity maps of one or more emission lines in large volumes can be used as biased tracers of the underlying matter distribution.
Instead of detecting galaxies as peaks in the intensity above the noise level, LIM can use noisy data to measure summary statistics of the intensity such as $N$-point correlation functions. 
It thus incorporates photons from all galaxies and diffuse gas within the survey volume that would otherwise remain undetected.

Several LIM surveys are currently in operation \citep{dore/etal:2014,santos/etal:2016,deboer/etal:2017,concerto/etal:2020,cleary/etal:2022,karkare/etal:2022},
or in preparation \citep{vieira/etal:2020,sun/etal:2021,switzer/etal:2021,ccat_prime/etal:2023,renard/etal:2024,renard/etal:2025}. 
The emission lines targeted by these surveys span from the ultraviolet (Lyman-$\alpha$, or short, \Lya) to the radio ($21$ cm) and are emitted by atomic or molecular gas. LIM has produced detections of cross- and auto-power spectra for the 21 cm and CO lines \citep[e.g.,][]{chang/etal:2010,keating/etal:2016,keating/etal:2020,cunnington/etal:2023a,paul/etal:2023}.
Other surveys have provided upper limits on the auto-power spectrum \citep[e.g.,][]{cleary/etal:2022,keenan/etal:2022}
or employ stacking \citep[e.g.,][]{dunne/etal:2024,dunne/etal:2025,chen/etal:2025}.

The \Lya emission line of atomic hydrogen enables observations of high-redshift galaxies, called \Lya emitters (LAEs),
and of diffuse gas in the circumgalactic and intergalactic media \citep[CGM and IGM; e.g.,][]{ouchi/etal:2020}.
Stacking of LAEs \citep[e.g.,]{wisotzki/etal:2018,kakuma/etal:2021,kikuchihara/etal:2022,lujanniemeyer/etal:2022a,kikuta/etal:2023,trainor/etal:2025} and other galaxies \citep{steidel/etal:2011,kusakabe/etal:2022} has revealed the ubiquity of \Lya emission out to projected distances of $\simeq 1\,\mathrm{Mpc}$ from these galaxies.
However, few \Lya LIM measurements have been performed on larger scales.
All distance units refer to comoving distances unless otherwise specified.

\citet{croft/etal:2016,croft/etal:2018} and \citet{lin/etal:2022} present a measurement of the cross-correlation of quasars (QSOs) from the Baryon Oscillation Spectroscopic Survey (BOSS) and the Extended Baryon Oscillation Spectroscopic Survey (eBOSS) with \Lya intensity from a projected radius of $r_\perp = 0.5\, h^{-1} \mathrm{Mpc}$ to $150\, h^{-1} \mathrm{Mpc}$ \citep[see][for a measurement from $r_\perp = 0.1\, h^{-1} \mathrm{Mpc}$]{lin/etal:2022}. The \Lya intensity is measured by subtracting the best-fit galaxy spectrum from spectra of luminous red galaxies (LRGs), assuming that the residual consists of background \Lya intensity and noise. By measuring the QSO-\Lya intensity cross-correlation, they constrain the average cosmic \Lya luminosity density. 
However, these results are inconsistent with the upper limit on the cosmic \Lya luminosity density inferred from the \Lya forest, as reported in \citet{croft/etal:2018}.
While \citet{croft/etal:2018} argue that the intensity is dominated by \Lya emission from QSOs,
\citet{lin/etal:2022} conclude that the bulk of the \lya luminosity originates from star-forming galaxies.

The Hobby-Eberly Telescope Dark Energy Experiment \citep[HETDEX;][]{gebhardt/etal:2021} provides an ideal data set for \Lya LIM\null. HETDEX is a galaxy survey aimed at detecting $\sim 10^6$ LAEs at $1.88 < z < 3.52$ in a comoving volume of $10.9\,\mathrm{Gpc}^3$. HETDEX observes its survey area with the Visible Integral Field Replicable Unit Spectrograph \citep[VIRUS;][]{hill/etal:2021} on the Hobby-Eberly Telescope \citep[HET;][]{ramsey/etal:1998, hill/etal:2021} without target preselection. Therefore, most fiber spectra do not contain sources that are bright enough to be detected, with detected LAEs comprising only $\sim 0.01\%$ of the signal (Mentuch Cooper et al. in preparation).
This makes LIM a unique scientific target for HETDEX.

\citet{lujanniemeyer/etal:2022a,lujanniemeyer/etal:2022b} stacked the \Lya emission in spectra around HETDEX LAEs and independently detected \Lya-faint galaxies selected by their [\ion{O}{3}] emission, and detected \Lya emission out to $\simeq 100\,\mathrm{kpc}$ (proper) in both samples.
This stacking signal is closely related to the small-scale angular cross-correlation of galaxies with \Lya intensity. 
\citet{khanlari/etal:2025} increased the sample size and stacked HETDEX spectra as a function of line-of-sight (LOS) distance and angular separation, finding \Lya absorption halos (see Section \ref{subsec:lya_absorption} for a discussion).  
These studies confirm the suitability of HETDEX data for LIM\null. The simulation study presented in \citet{lujanniemeyer/bernal/komatsu:2023} \citep[see also][]{fonseca/etal:2017} predicts a high-significance detection of the LAE-\Lya intensity cross-power spectrum at wavenumbers of $k \simeq0.08 - 1\, h\,\mathrm{Mpc}^{-1}$ for an ideal HETDEX survey without systematics, while masking the spectra of detected LAEs. 

In this paper, we present \Lya LIM results from  HETDEX observations. Using the HETDEX LAE catalog and HETDEX spectra, we detect the LAE-\Lya intensity cross-power spectrum with high statistical significance. By masking LAEs that are detected above a certain signal-to-noise ratio (SNR) threshold, we ensure that the intensity contains only undetected sources. We generate mocks with the \textsc{Simple} code \citep{lujanniemeyer/bernal/komatsu:2023} to calculate the transfer function and estimate the power spectrum covariance matrix. By shuffling the on-sky positions of the HETDEX intensity, we calculate the uncertainty due to HETDEX noise. We constrain the product of the LAE bias, the intensity bias, the mean intensity of undetected sources, and the ratio of the actual and fiducial redshift-space distortion factors,
$b_\mathrm{g} b_I \langle I \rangle \bar{F}_{\rm RSD} / \bar{F}^{\rm fid}_{\rm RSD}$. Finally, we compare our findings to the expected mean intensity from integrating extrapolated LAE luminosity functions, and to a cosmological \Lya radiative transfer (RT) simulation.

Our measurement is complementary to those of \citet{croft/etal:2016,croft/etal:2018} and \citet{lin/etal:2022} because we cross-correlate the \Lya intensity with different sources, with an emphasis on larger scales, and with higher sensitivity.
Because this work does not exclusively use QSOs, but LAEs, our results should not be strongly affected by \Lya emission around QSOs.

This work represents a LIM detection in HETDEX data. We expect to improve data processing and increase the sample size of LAEs and spectra in the future.

This paper is structured as follows. Section \ref{sec:hetdex_data} describes the LAE catalog and spectra obtained by HETDEX and the data processing. It also describes the separation of the observations into smaller regions and the creation of galaxy and intensity maps. Section \ref{sec:mocks} presents lognormal mocks from the \textsc{Simple} code. 
Section \ref{sec:power_spectrum_measurement} explains how we estimate the LAE-\Lya intensity cross-power spectrum. 
Section \ref{sec:amplitude} describes how we fit the power spectrum model to the data.
We show the power spectra and constraints on $b_\mathrm{g} b_I \langle I \rangle \bar{F}_{\rm RSD} / \bar{F}^{\rm fid}_{\rm RSD}$ in Section \ref{sec:results}.
We compare our constraints with the QSO-\Lya intensity and \Lya forest-\Lya intensity cross-correlations in Section \ref{sec:comparison_previous_constraints}. 
We explore origins of the \Lya intensity in Section \ref{sec:origins_of_lya_emission}.
Section \ref{sec:discussion} discusses potential improvements in data processing and modeling, and the possible effect of \Lya absorption on our measurement. 
We conclude in Section \ref{sec:summary}.

We use the following Fourier convention:
 \begin{equation}
 \begin{split}
     \tilde{f}(\mathbf{k}) &= \int \mathrm{d}^3\mathbf{x} f(\mathbf{x}) e^{i \mathbf{k} \cdot \mathbf{x}}\,, \\
     f(\mathbf{x}) &= \int \frac{\mathrm{d}^3 \mathbf{k}}{(2\pi)^3} \tilde{f}(\mathbf{k})e^{-i\mathbf{k} \cdot \mathbf{x}}\,,
\end{split}
\end{equation}
where the tilde denotes quantities in Fourier space. We adopt a flat $\Lambda$-cold-dark-matter cosmology with $H_0 = 67.66\,\mathrm{km\,s^{-1}\,Mpc^{-1}}$, 
$\Omega_{\mathrm{b},0}h^2 = 0.022$, $\Omega_{\mathrm{m},0}h^2=0.142$, $\sum m_\nu = 0.06\,\mathrm{eV}$, 
$\ln\left(10^{10} A_s\right)=3.047$, and $n_s = 0.9665$ \citep{planck/etal:2020}. We refer to the specific intensity $I_\lambda$, the intensity per unit observed wavelength, as the intensity $I$ for simplicity.

\section{HETDEX Data}
\label{sec:hetdex_data}

Our analysis uses the data of the internal HETDEX data release HDR5 (Mentuch Cooper et al. in preparation). 
These data are collected from 2017-01-01 to 2024-07-31 and comprise the full main HETDEX survey.
The spectra cover $90\,\mathrm{deg}^2$ of sky, $87.7\mathrm{deg}^2$ of which meet the science quality criteria.
These data include observations within
an equatorial ``Fall'' field, whose footprint covers $150\,\mathrm{deg}^2$, a higher-declination ``Spring'' field, which covers $390\,\mathrm{deg}^2$, and several smaller fields, which were used for HETDEX science verification \citep{gebhardt/etal:2021}. We also use observations of the Texas Euclid Survey for \Lya (TESLA), which covers a $10\,\mathrm{deg}^2$ region of
the North Ecliptic Pole \citep[NEP;][]{chavezortiz/etal:2023}. The TESLA data were observed in the same manner, processed in the same way, and have the same exposure times as the HETDEX spectra.

The HETDEX spectra were obtained with VIRUS on the $10\,\mathrm{m}$ HET \citep{ramsey/etal:1998,hill/etal:2021}.
VIRUS is composed of up to 78 integral field unit fiber arrays (IFUs), each of which spans $\ang{;;51}\times \ang{;;51}$ on the sky and contains $448$ fibers that are $\ang{;;1.5}$ in diameter. The fibers from each IFU feed into a spectrograph with two spectral channels. The spectrographs have a resolution of $R\simeq 800$, which corresponds to a Full-Width-at-Half-Maximum (FWHM) of $5.6\,\angstrom$ at $\lambda = 4500\,\angstrom$, and cover the wavelengths between $3470\,\angstrom$ and $5540\,\angstrom$\null. Each spectral channel is read out by a CCD with two amplifiers; thus, the fibers from each IFU are split into four amplifiers. Each HETDEX observation consists of three six-minute exposures in a triangular dithering pattern that fills the gaps between the fibers on each IFU. 

The IFUs are spread out over the HET's $\ang{;18;}$ field of view on a $\ang{;;100}$-grid, so that the gap between each IFU is roughly equal to the size of the IFU\null. As a result, the gaps between the IFUs lead to a fill factor of $\sim 1/4.6$ for each observation.  The observations in the NEP field are dithered to fill the gaps between the IFUs, while these gaps remain in the Spring and Fall fields. 

\begin{figure*}
    \includegraphics[width=\textwidth]{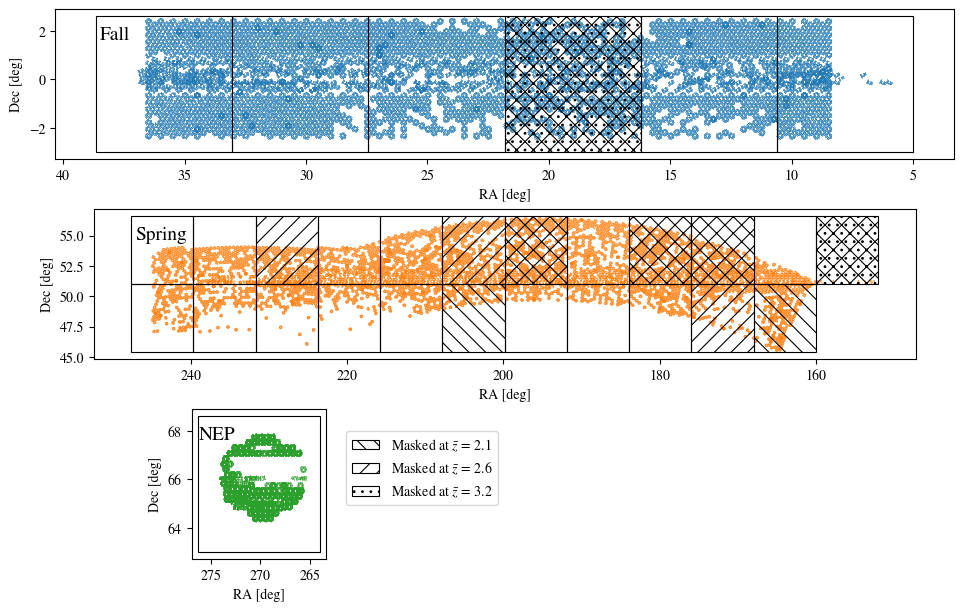}
    \caption{\label{fig:subboxes}IFU coordinates in the Fall (top), Spring (middle), and NEP (bottom) fields. The fields are divided into $\ang{5.6;;}$ wide regions for the power spectrum, shown as black squares (see Section \ref{sec:map_creation}). Masked maps are shown as hatched regions (see Section \ref{subsec:covariance_and_weights}). The backward-facing (forward-facing) diagonal hatchings represent boxes masked in the low-$z$ (medium-$z$) bin, and the dotted hatchings indicate those masked in the high-$z$ bin.}
\end{figure*}

For this work, we only include observations that are not masked by the HETDEX flag for low-quality observations.
As HETDEX requires a minimum effective throughput of 0.08 at $4540\,\angstrom$ for an LAE to be included in the catalog, we only select spectra from  observations that meet this requirement. Additionally, we exclude HETDEX observations before 2018 because these early data contain more artifacts and larger sky emission residuals. This approach leaves $1,949$ observations in the Fall field, $3,797$ in the Spring field, and $484$ in the NEP field. Figure \ref{fig:subboxes} shows the IFU coordinates used in the Fall, Spring, and NEP fields.

\subsection{LAE Catalog}
\label{subsec:lae_catalog}

\begin{figure*}
\includegraphics[width=\textwidth]{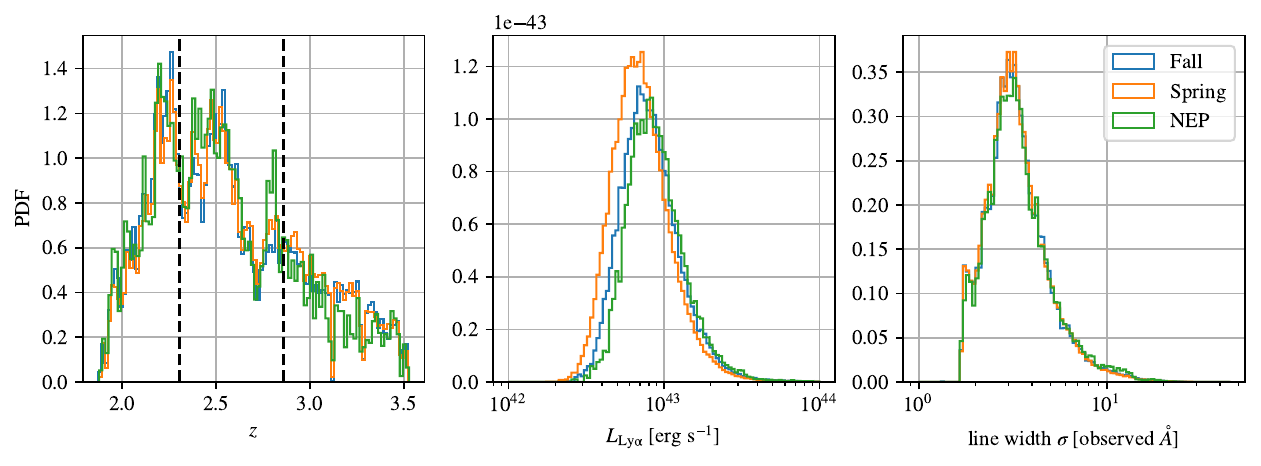}
\caption{\label{fig:lae_z_lum_lw_hists}The normalized distributions of redshifts (left panel), luminosities (middle panel), and the best-fit Gaussian $\sigma$ line widths in observed wavelength units (right panel) of HETDEX LAEs in the Fall (blue), Spring (orange), and NEP (green) fields. The black dashed lines in the left panel show the boundaries of the three redshift bins used to calculate the power spectra.}
\end{figure*}

We select LAEs from catalog 5.0.1 of the internal HETDEX data release HDR5 (Mentuch Cooper et al. in preparation). A public release of an early subset of this sample is found in \citet{mentuchcooper/etal:2023}, including a description of source detection and classification. For each line detection, the catalog provides the central wavelength; the sky coordinates of the detection; the line width of the best-fit Gaussian to the emission line ($\sigma$); the line flux; the SNR of the detection; an aperture correction factor, which quantifies the fraction of a source's light that falls onto an IFU (as opposed to off an edge);  and the source type, e.g., LAE or [\ion{O}{2}]-emitting galaxy, and the confidence of this classification.

Because the LAE detections will be masked in the \Lya intensity map, there will be no shot noise in the cross-power spectrum. We restrict our LAE sample to high-confidence objects (`flag\_best' detections with $\mathrm{SNR}>5.5$, see Mentuch Cooper et al. in preparation), preferring a smaller, higher-quality set of LAEs to a larger set with more contamination. 
The `flag\_best' criterion of the HETDEX detection catalog summarizes quality criteria based on known bad data; masks of large galaxies, meteors, and satellites; goodness-of-fit of the line detection; detection confidence based on Machine Learning and citizen science \citep{house/etal:2023,house/etal:2024}; and other quality criteria (details can be found in \citealt{mentuchcooper/etal:2023} and Mentuch Cooper et al. in preparation).
We also select only those sources with an aperture correction factor $>0.4$, and require the updated SNR, which incorporates systematics, to be $\mathrm{SNR}_\mathrm{rres}>4.5$.
The noise model used for $\mathrm{SNR}_\mathrm{rres}$ is the third model described in Section 6.20 of \citet{gebhardt/etal:2021}.

We further impose thresholds on the LAE detection confidence parameters derived from machine learning models. We require $p_\mathrm{conf} > 0.5$ from a Random Forest classifier based on emission-line measurements (Mentuch Cooper et al., in preparation) and a score $>0.4$ from a convolutional neural network (CNN) trained on HETDEX spectra (Mukae et al., in preparation). These classifiers are developed to distinguish LAEs from false-positive detections due to systematic errors in the spectra and sky residuals.
For the $\mathrm{SNR}> 5.5$ sample, the adopted CNN threshold yields a precision of $\sim 96\%$ when compared with visual classifications as described in \citet{house/etal:2023}, effectively classifying LAEs while minimizing false detections.
Using these criteria, $90.4\%$ of detections in the COSMOS science verification field are confirmed in a different HETDEX observation of the same field or a separate galaxy catalog. Thus, the fraction of false-positive detections using these criteria is $<9.6\%$.

We then remove detections within $\pm 10\,\angstrom$ of the [\ion{O}{3}] emission line at $5007\,\angstrom$ to mitigate contamination from high-Galactic latitude planetary nebulae. We also exclude LAE candidates within $\pm 10\,\angstrom$ of the sky emission lines at $5200\,\angstrom$ and $5457\,\angstrom$. 

Figure \ref{fig:lae_z_lum_lw_hists} shows the redshift, luminosity, and line-width distributions of the LAE samples in the Fall, Spring, and NEP fields. 
The dip at $z\simeq 2.7$ in the redshift distributions is due to a mask that is applied at the center of $50\%$ of the detectors as well as an
increase in night sky emission \citep{mentuchcooper/etal:2023}.
The dip at $z\simeq 2.3$ is also due to bright sky emission.
There remain $37,773$ LAEs in the Fall field, $95,774$ in the Spring field, and $9,372$ in the NEP field. After removing detections in IFUs that we mask (see Section \ref{subsec:spectra}), $36,303$ LAEs remain in the Fall field, $92,192$ in the Spring field, and $8,932$ in the NEP field.
This number is much lower than the total number of detected LAEs in HETDEX because of our strict selection criteria.

We mask these LAEs in the intensity map for the power spectrum measurement, as described in Section \ref{subsubsec:masking_laes}.

\subsection{Intensity Spectra}
\label{subsec:spectra}

In this section, we describe the additional data reduction steps we apply to the HETDEX spectra. The intensity mapping results can be sensitive to reduction artifacts and our goal is to reduce those as much as possible.

\subsubsection{Removing Sky Emission}

\citet{gebhardt/etal:2021} describe the HETDEX data processing pipeline. A crucial part of the processing for low-surface brightness measurements is sky subtraction. ``Sky'' emission, including airglow, zodiacal light, and foreground emission from the Milky Way, dominates the observed spectra and must be removed. 

HETDEX employs two methods for sky subtraction. The first method uses a `local' sky estimator, measured using the 112 fibers present on each CCD amplifier; these fibers cover a $\sim \ang{;;13}\times \ang{;;51}$ field. This sky subtraction technique is optimized for LAE detection. The second method computes a `full-frame' sky, derived from the photons detected on all $\sim 35$k fibers of the $\ang{;18;}$ field of view of the VIRUS instrument within one exposure \citep[see][for details]{lujanniemeyer/etal:2022a}. As demonstrated in \citet{lujanniemeyer/bernal/komatsu:2023}, the galaxy-intensity cross-power spectrum is suppressed on scales larger than the sky subtraction scale. To keep information on a scale as large as possible, we use the `full-field' sky subtraction in this work.

Our IFU data can be visualized in a two-dimensional figure, where the $x$-axis is the wavelength, and the $y$-axis is the fiber number. When averaging the full-frame sky-subtracted spectra of each amplifier over all observations taken within a month, one would expect a mean of zero. However, we find systematic two-dimensional patterns that are unique to each amplifier. To remove these residuals, we first divide each fiber spectrum by the full-frame sky spectrum of the same fiber.
We then calculate the biweight location (BL) of these relative residuals across observations within a month. The BL and biweight scale ($\sigma_{\mathrm{bw}}$) are estimators of the central location and scatter of a distribution that are robust to outliers \citep{beers/flynn/gebhardt:1990}. In each observation, we subtract this BL relative residual multiplied by the respective sky spectrum for each fiber.

\subsubsection{Masking Bad Detectors, Bright Voxels, Continuum, and [O II] Detections}
\label{sec:masking_individual_fibers}

Processed HETDEX spectra are output in $2\,\angstrom$ bins between $3470\,\angstrom$ and $5540\,\angstrom$. When all 78 IFUs are operational, $3\times 34,944$ spectra are obtained from each observation.  However, some of the CCD amplifiers and a few of the individual fibers produce bad data in certain observations.  The bad amplifiers, fibers, and pixels (including those affected by cosmic rays) are flagged by the HETDEX data processing software as described by criteria set in Table\,2 of \citet{mentuchcooper/etal:2023}, and are masked in our analysis.
We also mask spectra containing large galaxies and meteors using the fiber mask provided by the HETDEX Application Programming Interface \citep[API; see Section 2 of][]{mentuchcooper/etal:2023}

For each observation, we also mask outlier pixels individually. First, we calculate the BL and $\sigma_\mathrm{bw}$ of all the spectra, and then mask all pixels with fluxes that deviate by $>3\sigma_\mathrm{bw}$ from the BL of the fluxes.

Since HETDEX is an untargeted survey, the spectra of continuum objects, such as foreground stars and galaxies, are also in the HETDEX database, and these objects can contaminate the \Lya intensity map. We therefore remove all fiber spectra containing continuum emission.  To do this, we estimate the continuum within each fiber by calculating the BL of its spectrum within the wavelength range of $4100-5100\,\angstrom$ and mask out those fibers with a $3\sigma_\mathrm{bw}$ deviation from the BL of the continuum values within each observation.

Because light can scatter within the detector, we also mask out fibers adjacent to each masked fiber with continuum emission on the CCD\null. Spectra adjacent in the array obtained from the HETDEX Application Programming Interface (API) are adjacent on the CCD, unless they are the first or last fiber on the CCD\null. We mask all adjacent fibers in the array, which sometimes includes the first or last fiber on an adjacent amplifier.

We set the spectra of the fibers within $\ang{;;10}$ of all [\ion{O}{2}]-emitting galaxies to zero. The detection criteria match those used for LAEs, as described in Section \ref{subsec:lae_catalog}, except that the source type indicates [\ion{O}{2}]-emission.  For these objects, we do not apply cuts based on the CNN score, $p_\mathrm{conf}$, or $\mathrm{SNR}_\mathrm{rres}$.

\subsubsection{Separating Intensity Contributions from LAEs and Undetected Sources}
\label{subsubsec:masking_laes}

We separate the spectra into contributions from detected LAEs and undetected sources as described in this subsection.
In this step, we consider all LAEs selected in Section \ref{subsec:lae_catalog} as `detected.'
We select voxels within $\ang{;;10}$ of an LAE's positional centroid and within $2.5$ times the line width of its wavelength centroid. For the `undetected' spectra, we set these voxels to zero. For the `detected' spectra, we instead set all other voxels to zero. Adding the `detected' and `undetected' spectra therefore yields the total spectra. This process translates directly to the power spectra: adding the cross-power spectrum of LAEs with the `detected' intensity to that of LAEs with the `undetected' intensity has to be equal to the cross-power spectrum of LAEs with the `total' intensity. We use this property as a check for the power spectrum pipeline. Henceforth, we apply the same data processing to the `detected,' 'total,' and `undetected' spectra, unless specified otherwise.

\subsubsection{Lowering Angular Resolution}
\label{sec:lowering_angular_resolution}

The angular resolution of the HET is much better than that of typical LIM experiments: the FWHM of the VIRUS point-spread function (PSF) is between $1\farcs 2$ and $\simeq \ang{;;4}$. For computational simplicity, we condense $ 3 \times 448$ spectra taken by each IFU during a single HETDEX observation into one spectrum.
We have \Nifusfall IFU spectra in the Fall field, \Nifusspring in the Spring field, and \Nifusnep in the NEP field.
Each spectrum represents the average flux within $\ang{;;51}\times\ang{;;51}\times 2\,\angstrom$ voxels, which corresponds to $1.5\,\mathrm{Mpc}\times 1.5 \,\mathrm{Mpc} \times 1.9\,\mathrm{Mpc}$ at $z=2.5$. We will henceforth refer to these as `IFU spectral elements.'

We calculate the mean of the `total,' `detected,' and `undetected' fiber spectra within each IFU\null.
We also calculate the standard deviation of the `total' fiber spectra within each IFU and save the mean coordinates of all the IFU's fibers.

Finally, we calculate the inverse variance of the `total' fiber spectra of each pixel. Some voxels have much higher values than the typical distribution because few unmasked elements of the fiber spectra contribute to it.  We therefore set all pixels with inverse variances of $>3600 / u_I^2$ to zero, where $u_I = 10^{-17} \,\mathrm{erg}\,\mathrm{s}^{-1}\,\mathrm{cm}^{-2}\,\mathrm{arcsec}^{-2}\,\angstrom^{-1}$. 
This removes everything with a deviation $\gtrsim 5.7 \sigma_{\rm bw}$ from the BL of the inverse variance distribution.

\subsubsection{Calculating the Intensity and Removing Extinction}
\label{subsubsec:calculating_intensity_removing_extinction}
The spectra are output as specific fluxes, $f_\lambda$, per $2\,\angstrom$ bin in each fiber. The intensity is obtained from $f_\lambda$ as $I(z) = f_\lambda(z) / A_\mathrm{fiber}$, where the fiber area is $A_\mathrm{fiber} = \pi \left(0\farcs 75\right)^2$.

To obtain the intensity as would be seen outside the Milky Way, we remove the reddening caused by Galactic extinction. We collect the color-excess values, $E(B-V)$, as measured by \citet{schlegel/finkbeiner/davis:1998} at the coordinates of each IFU, using the \emph{dustmaps} API \citep{green:2018}. Following \citet{schlafly/finkbeiner:2011}, we assume an $R_V = 3.1$ reddening model of \citet{fitzpatrick:1999} and set $A_V = 2.742 E(B-V)$. We calculate and apply the extinction correction using the open-source Python module extinction\footnote{\url{https://github.com/sncosmo/extinction/}}.

\subsubsection{Removing More Systematics}
\label{subsubsec:removing_systematics}
We combine all IFU spectra within each of the three fields for the following steps.

First, we again mask those voxels that are much fainter or brighter than the rest of the distribution, i.e., those with values 
above $0.0125 u_I$ or below $-0.0125 u_I$. 
We determined this threshold empirically; this cut removes everything with a $\gtrsim 7\sigma_{\rm bw}$ deviation from the BL of the pixel distribution, which is $\simeq 0$.

We expect that, after the subtraction of foreground sources and sky emission, the spectra of each IFU will be dominated by sky noise. However, some spectra exhibit systematic patterns, indicating imperfect sky subtraction or other detector problems. Since these spectra typically have a higher standard deviation along the wavelength direction, we mask out those spectra with a standard deviation along the wavelength direction $>0.0035\,u_I$.
This maximum value was determined empirically; the cut removes spectra whose standard deviation along the wavelength direction deviates $\gtrsim 3.3\sigma_{\rm bw}$ from the BL of the distribution. 

The residuals from sky subtraction are most prominent around the emission lines in the sky foreground, including light from street lamps. We therefore mask all pixels of the IFU spectra at  $4350 - 4370\,\angstrom$, $5190-5210\,\angstrom$, and $5447-5467\,\angstrom$\null. We also mask $\pm 10\,\angstrom$ around the rest-frame [\ion{O}{3}] emission line at $5007\,\angstrom$ to mitigate contamination from planetary nebulae and diffuse gas within the Milky Way.

\subsubsection{Principal Component Analysis}
\label{subsubsec:pca}

Despite the extensive masking described above, the spectra still contain sky emission residuals.
We use a principal component analysis (PCA) to remove these sky emission residuals from the spectra.

We perform the PCA on the Fall, Spring, and NEP fields separately.  We first subtract the mean spectrum of each field, $\langle{I}(\lambda)\rangle=N^{-1}_\mathrm{IFU}\sum_{i=1}^{N_\mathrm{IFU}}I_i(\lambda)$, and normalize
\begin{equation}
    X_i(\lambda) = \frac{I_i(\lambda) - \langle{I}(\lambda)\rangle}{\sigma_I(\lambda)},
\end{equation}
where the subscript $i$ refers to the IFU and $\sigma_I(\lambda)$ is the standard deviation of the intensity within each wavelength slice.
$N_\mathrm{IFU}$ is the number of IFU spectra used for the analysis, which is \Nifusfall in the Fall field, \Nifusspring in the Spring field, and \Nifusnep in the NEP field.  
Of these IFU spectra, $5,882$ (Fall), $13,851$ (Spring), and $1,803$ (NEP) are fully masked in the steps described in Sections \ref{sec:masking_individual_fibers}, \ref{sec:lowering_angular_resolution}, and \ref{subsubsec:removing_systematics}. In the remaining fibers, $\simeq 7\%$ of the voxels are masked.
We then set all masked intensity values to zero.

We define a vector of length $N_\lambda$ as
\begin{equation}
    \mathbf{X}_i= \sum_{j=1}^{N_\lambda} X_i(\lambda_j) \hat{\mathbf{e}}^\lambda_j=\left(X_i(\lambda_1),X_i(\lambda_2),\dots,X_i(\lambda_{N_\lambda})\right)^\top,
\end{equation}
where $N_\lambda=1036$ is the number of wavelength bins per spectrum. The $j$th value of the vector $\mathbf{X}_i$ is the normalized intensity fluctuation in the $j$th wavelength bin, $X_i(\lambda_j)$. 
The value of the basis vector, $\hat{\mathbf{e}}^\lambda_j$, is equal to $1$ in the $j$th wavelength bin and zero otherwise.

We define an $N_\mathrm{IFU}$-by-$N_\lambda$ matrix, $\mathbf{X}$, which contains the vectors $\mathbf{X}_i$ as its rows. PCA then calculates an $N_\lambda$-by-$N_\lambda$ covariance matrix, $\mathbf{X}^\top\mathbf{X}$. The eigenvectors of the covariance matrix are the weight vectors. These form an orthonormal basis $\hat{\mathbf{e}}^\mathrm{PCA}_j$, such that 
\begin{equation}
    \mathbf{X}_i = \sum_{j=1}^{N_\lambda} y_{ij} \hat{\mathbf{e}}^\mathrm{PCA}_j.
\end{equation}
PCA arranges the eigenvectors $\hat{\mathbf{e}}^\mathrm{PCA}_j$ in order of decreasing variance of $y_{ij}$, with the variance being highest for $j=1$. The variance of $y_{ij}$ is proportional to the $j$th eigenvalue.

\begin{figure*}
    \includegraphics[width=\textwidth]{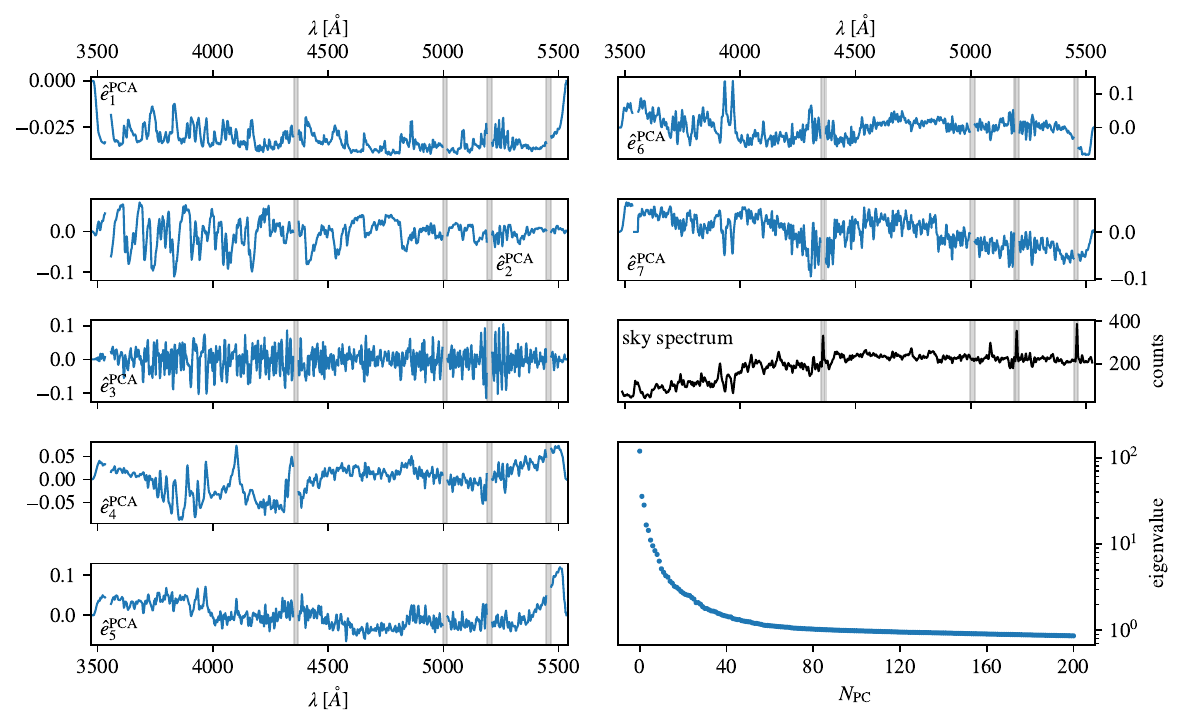}
    \caption{First seven weight vectors, $\hat{\mathbf{e}}^\mathrm{PCA}_j$ with $j=1-7$, of the Fall field are displayed as functions of wavelength. All $\hat{\mathbf{e}}^\mathrm{PCA}_j$ are normalized vectors. For comparison, we also show an example `full-field' sky spectrum in counts (third panel on the right). Masked wavelength regions are shown in gray. The bottom right panel shows the first $200$ eigenvalues, which are proportional to the variance along the corresponding weight vectors.}
    \label{fig:pca}
\end{figure*}

Figure \ref{fig:pca} shows the first seven PCA weight vectors, $\hat{\mathbf{e}}^\mathrm{PCA}_j$ with $j=1-7$, of the `total' spectra in the fall field; an example `full-field' estimated sky spectrum; and the eigenvalues corresponding to the first $200$ weight vectors. The first weight vectors contain features of the sky spectrum, such as the K and H lines at $3934\,\angstrom$ and $3968\,\angstrom$.

The high-variance weight vectors are dominated by systematics such as sky emission residuals. We therefore set the first $N_\mathrm{PC} \in \left\{10, 20, ..., 200\right\}$ of the principal components (PCs) to zero and work with
\begin{equation}
    \mathbf{X}^\mathrm{PCA}_i = \sum_{j=N_\mathrm{PC}+1}^{N_\lambda} y_{ij} \hat{\mathbf{e}}^\mathrm{PCA}_j.
\end{equation}
To obtain clean spectra, we perform an inverse coordinate transformation,
\begin{equation}
    \mathbf{X}^\mathrm{PCA}_i = 
    \sum_{k=1}^{N_\lambda} \sum_{j=N_\mathrm{PC}+1}^{N_\lambda} y_{ij}
    \left(\hat{\mathbf{e}}^{\lambda \top}_k
    \hat{\mathbf{e}}^\mathrm{PCA}_j \right)
    \hat{\mathbf{e}}^{\lambda}_k,
\end{equation}
so that 
\begin{equation}
    X^\mathrm{PCA}_i (\lambda_k) = \sum_{j=N_\mathrm{PC}+1}^{N_\lambda} y_{ik}
    \left(\hat{\mathbf{e}}^{\lambda \top}_k
    \hat{\mathbf{e}}^\mathrm{PCA}_j \right),
\end{equation}
and undo the normalization:
\begin{equation}
    I_i^\mathrm{PCA}(\lambda) = {X}^\mathrm{PCA}_i(\lambda) \sigma_I(\lambda).
\end{equation}
We then re-apply the mask.

We calculate the weight vectors, $\hat{\mathbf{e}}^\mathrm{PCA}_j$, from the covariance matrix $\mathbf{X}^\top\mathbf{X}$ of the normalized `total' spectra. We then use these vectors to clean the `total,' `detected,' and `undetected' spectra. This ensures that the sum of the `detected' and `undetected' spectra equals the `total' spectra. Since most spectral elements do not contain a detected LAE, the `total' and `undetected' spectra are similar.

The PC removal also removes cosmological signal. We model this loss in the mocks in Sections \ref{sec:insert_data_into_mock} and \ref{sec:loss_of_power_from_pca}.

\subsubsection{Comparison with Noise}
\label{subsubsec:noise_comparison}

\begin{figure}
    \includegraphics[width=0.5\textwidth]{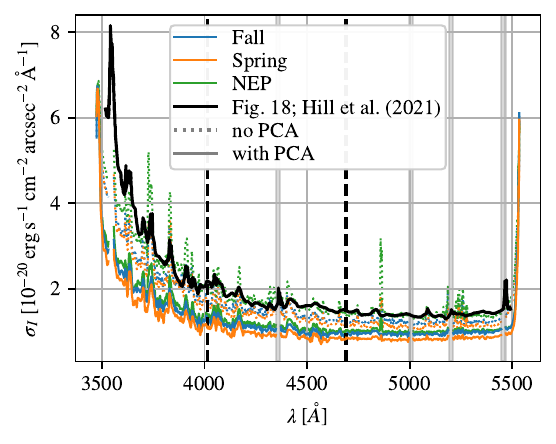}
    \caption{Standard deviation of the IFU spectra in each wavelength slice in the Fall (blue), Spring (orange), and NEP (green) fields compared to the pure-noise expectation inferred from Figure 18 of \citet{hill/etal:2021} (black). 
    The dashed colored lines show the standard deviation of the IFU spectra without PCA ($N_\mathrm{PC} = 0$). 
    The solid colored lines are the standard deviation of the IFU spectra after PCA cleaning, where $N_\mathrm{PC} = \NPClowz$ in the low-$z$ bin, $N_\mathrm{PC} = \NPCmediumz$ in the medium-$z$ bin, and $N_\mathrm{PC} = \NPChighz$ in the high-$z$ bin.
    The vertical black dashed lines indicate the boundaries of the three redshift bins used for power spectrum measurements. The gray shaded areas show masked wavelength values.}
    \label{fig:sigma_noise_vs_hilletal}
\end{figure}

Figure \ref{fig:sigma_noise_vs_hilletal} compares the measured and expected standard deviations of the IFU spectra
in the absence of astrophysical sources, as a function of wavelength. We translate the measured $5\sigma$ sky and read noise per fiber per spectral resolution element (FWHM$=5.6 \angstrom$) of Figure 18 in \citet{hill/etal:2021} to the intensity noise per fiber by dividing it by $(5 \times \pi (0\farcs 75)^2 \times 5.6 \angstrom)$. 
This assumes that the noise of the pixels within a resolution element is fully correlated, as expected if the noise is dominated by sky photons. The noise is indeed dominated by sky photons at $\lambda \gtrsim 4000\,\angstrom$ \citep{hill/etal:2021}. 
At shorter wavelengths, it is read-noise dominated, which is not correlated within a spectral resolution element.
The expected intensity noise per IFU spectrum is obtained by dividing this value by $\sqrt{3\times 448}$ for $3$ dithers and $448$ fibers per IFU per dither. 
This measured standard deviation of the IFU spectra without PCA cleaning is similar to the expectation inferred from \citet{hill/etal:2021} in the NEP field, but slightly smaller at most wavelengths in the Fall and Spring fields. This discrepancy could be due to masking outlier pixels and high-variance spectra.
After PCA cleaning, the standard deviation is smaller at most wavelengths. 
This is a natural consequence of our removal of PCs with the highest variance.

\subsection{Creation of Intensity Maps}
\label{sec:map_creation}

Due to the loss of power from sky subtraction \citep[see][]{lujanniemeyer/bernal/komatsu:2023}, we measure the power spectrum at wavenumbers $k>0.08 \, h\mathrm{Mpc}^{-1}$. To accommodate the required high angular resolution, we split the three fields into smaller regions with a side length of $\ang{5.6;;}$ for the maps. 

We divide the redshift range into three redshift bins from $1.85$ to $2.3$ ($\bar{z} = 2.07$), from $2.3$ to $2.86$ ($\bar{z} = 2.58$), and from $2.86$ to $3.56$ ($\bar{z} = 3.21$). This binning yields $23 \times 3 = 69$ maps for the Spring field and $6 \times 3 = 18$ maps for the Fall field. The NEP field fits within one such map in each of the three redshift bins. Figure \ref{fig:subboxes} shows the map division in the three fields.

Each map is contained within a cubic box with a side length of $432\, h^{-1}\,\mathrm{Mpc}$ for the low-$z$ and medium-$z$ bins and $463\, h^{-1}\,\mathrm{Mpc}$ for the high-$z$ bin. The fundamental frequencies are $k_\mathrm{F} = 0.015 \, h\,\mathrm{Mpc}^{-1}$ and $0.014 \, h\,\mathrm{Mpc}^{-1}$, respectively. We create cubic voxels that are $2\, h^{-1}\mathrm{Mpc}$ long, corresponding to a Nyquist frequency of $k_\mathrm{Ny} = 1.57\, h\,\mathrm{Mpc}^{-1}$.

To create the intensity map, we collect all the IFU spectral elements contained within each region and redshift bin. Then, we transform the sky and redshift coordinates into Cartesian coordinates and collect the IFU spectral elements in each voxel. 

For each voxel, we calculate the mean intensity $I(\mathbf{x})$ of the IFU spectral elements within.
We then calculate the intensity fluctuation, $\delta I(\mathbf{x})$, in each voxel,
\begin{equation}
    \delta I(\mathbf{x}) = I(\mathbf{x}) - \langle{I}\rangle(x_0),
\end{equation} 
where $\mathbf{x}=(x_0, x_1, x_2)$, $x_0$ is the position along the LOS axis of the map, $x_1$ and $x_2$ are the coordinates perpendicular to the LOS, and $\langle{I}\rangle(x_0) $ is the mean intensity in each slice along the LOS direction of the map. We calculate $\langle{I}\rangle(x_0)$ as
\begin{equation}
\langle{I}\rangle(x_0) = \frac{1}{N_\mathrm{vox}^2}\sum_{i=1}^{N_\mathrm{vox}} \sum_{j=1}^{N_\mathrm{vox}} I\left(x_0, x_1^i, x_2^j\right),
\end{equation}
where $x_1^i$ and $x_2^i$ are the coordinates of the $i$th voxel, and 
$N_\mathrm{vox} = L_\mathrm{box} / L_{\rm vox}$ is the number of voxels in each direction of $\mathbf{x}$. Here, $L_\mathrm{box}$ is the length of the box and $L_\mathrm{vox}$ is the length of a voxel.

\subsection{Mean Expected Number of LAEs per Voxel}
\label{subsec:N_bar}

We define the galaxy overdensity as 
\begin{equation}
    \delta_\mathrm{g}(\mathbf{x}) = \frac{N_\mathrm{LAE}(\mathbf{x}) - \bar{N}(\mathbf{x})}{\bar{N}(\mathbf{x})}.
\end{equation}
The number of LAEs detected per voxel is $N_\mathrm{LAE}(\mathbf{x})$ and the expected number of LAEs detected per voxel at location $\mathbf{x}$ in the absence of clustering is $\bar{N}(\mathbf{x})$. This parameter can be calculated as
\begin{equation}
\label{eq:Nbar_integral}
    \bar{N}(\mathbf{x}) = \delta V(\mathbf{x})\int_{L_\mathrm{min}}^{L_\mathrm{max}} \mathrm{d}L\, \frac{\mathrm{d}n}{\mathrm{d}L} C\left(\frac{L}{4\pi D_L^2(z)},f_{50}(\mathbf{x})\right),
\end{equation}
where $\delta V(\mathbf{x})$ is the volume covered by spectra within the voxel at position $\mathbf{x}$, $\frac{\mathrm{d}n}{\mathrm{d}L}$ the luminosity function of LAEs, $D_L$ the luminosity distance, and $f_{50}$ the flux limit at which $50\%$ of LAEs will be detected. The completeness factor, $C$, is a function of LAE line flux and is characterized by the $50\%$ completeness, $f_{50}$. Examples of completeness curves as a function of flux are shown in Figure 28 of \citet{gebhardt/etal:2021}.

We use the completeness model v4 of the HETDEX API\footnote{\url{https://github.com/HETDEX/hetdex_api}} \citep[][Farrow et al. in preparation]{gebhardt/etal:2021}. This completeness function depends on the ratio of the LAE flux to the flux at $50\%$ completeness; $C(f, f_{50}) = C_\mathrm{v4}(f/f_{50})$, which is a nonparametric model to map $f/f_{50}$ to detection completeness based on inserting simulated LAEs into the data.
The value of $f_{50}$ is obtained by multiplying the flux noise, $\sigma_f(\mathbf{\Theta},\lambda)$, at the sky coordinate $\mathbf{\Theta}$ and wavelength $\lambda$ by the minimum SNR for an emission-line detection (here $5.5$, as described in Section~\ref{subsec:lae_catalog}), a polynomial dependence on the wavelength of the detection $a_\lambda(\lambda)$, and a line-width dependent adjustment (Farrow et al. in preparation).
The value of $\sigma_f(\mathbf{\Theta},\lambda)$ is calculated as a PSF-weighted sum of the spectral errors within an aperture $\ang{;;3.5}$ in diameter and $14\,\angstrom$ deep.

We collect the values of $f_{50}$ in each IFU spectral element. We extract the flux noise $\sigma_f(\mathbf{\Theta},\lambda)$ on a $\ang{;;2}\times \ang{;;2}\times 2\,\angstrom$ grid for each IFU using the HETDEX API\null.
The redshift bins correspond to the $2\,\angstrom$ grid of wavelengths between $3470\,\angstrom$ and $5540\,\angstrom$ of the spectra. Because the API also provides a mask on the grid, we obtain the volume covered per IFU spectral element, $\delta V_i(z)$, for line detections. We transform these values to $f_{50}(\mathbf{\Theta},\lambda) = 5.5 \sigma_f(\mathbf{\Theta},\lambda) a_\lambda(\lambda)$, not accounting for the line-width adjustment. For each IFU and wavelength, we calculate a histogram of $f_{50}$ values in $5001$ logarithmically spaced bins between $10^{-18}$ and $10^{-13}\,\mathrm{erg}\,\mathrm{s}^{-1}\,\mathrm{cm}^{-2}$.

We calculate the number density, $\bar{n}(f_{50},z)$, for each bin center of the $f_{50}$ histogram and for each redshift bin corresponding to the $2\,\angstrom$ wavelength grid. We use the best-fit Schechter function component \citep{schechter:1976} of the $z=2.2$ \lya luminosity function measured by \citet{umeda/etal:2025}; with $L^\ast = 10^{42.8}\,\mathrm{erg}\,\mathrm{s}^{-1}$, $\phi^\ast = 10^{-3.16}\,\mathrm{Mpc}^{-3}$ and $\alpha = -1.53$ for $z\leq2.3$, corresponding to the low-$z$ bin for the power spectrum measurement. In the medium-$z$ and high-$z$ bins ($z>2.3$), we adopt the $z=3.3$ luminosity function of \citet{umeda/etal:2025} with $L^\ast = 10^{42.29}\,\mathrm{erg}\,\mathrm{s}^{-1}$, $\phi^\ast = 10^{-2.13}\,\mathrm{Mpc}^{-3}$ and $\alpha = -1.19$.

We evaluate equation \eqref{eq:Nbar_integral} by translating the $f_{50}$ histogram, $N^{(i)}_{f_{50}}(z)$, in the $i$th IFU into the number of expected LAEs, $\bar{N}_i(z)$, in the absence of clustering:
\begin{equation}
\label{eq:Nbar}
    \bar{N}_i(z) = \sum_{f_{50}} \delta V_i(z) \bar{n}(f_{50},z) N^{(i)}_{f_{50}}(z).
\end{equation}
We save $\bar{N}_i(z)$, the covered volume $\delta V_i(z)$, and the mean value of $f_{50}$,
\begin{equation}
\bar{f}^{(i)}_{50}(z) = \frac{\sum_{f_{50}} f_{50} N^{(i)}_{f_{50}}(z) }{ \sum_{f_{50}} N^{(i)}_{f_{50}}(z)}.
\end{equation}

The value of $\bar{N}_i(\lambda)$
predominantly depends on the observing conditions and the sky spectrum. 
It therefore varies from IFU to IFU,
although it is similar for different IFUs within the same observation.

Equation \eqref{eq:Nbar} overestimates the number of detected LAEs by a factor of $\simeq 2$. 
This could be due to an overestimated luminosity function, especially at the faint end, or an imperfect model for $f_{50}$. 
To avoid a systematic power spectrum signal resulting from this discrepancy, we ensure that the redshift distribution of $\bar{N}(z)$, averaged over the IFUs, matches that of the detected LAEs in each field. However, we still need $\bar{N}_i(z)$ to model the IFU-to-IFU variations. We therefore multiply $\bar{N}_i(z)$ by the ratio, $f_{N}(z) = N_\mathrm{LAE}(z) / \bar{N}(z)$. We calculate this ratio separately for the Spring and Fall fields. We use $f_{N}(z)$ from the Fall field for the NEP field due to the low number of LAEs.
The values of $f_{N}(z)$ are mostly around $0.5$, with individual spikes up to $\simeq 3$.
After this correction, $84\%$ ($95\%$) of the values of $\bar{n}_i(z) = \bar{N}_i(z) /\delta V_i(z)$ are below $4 $, $6$, and $4$ ($7$, $10$, and $6$) $\times 10^{-4}\mathrm{Mpc}^{-3} h^3$ in the Fall, Spring, and NEP fields, respectively.

To create a map of the LAE overdensity, we first transform the number of LAEs, $N_\mathrm{LAE}$, and the expected $\bar{N}$ into the same format as the spectra. If the mean total spectrum at an IFU spectral element is masked, then all other values, including $N_\mathrm{LAE}$ and $\bar{N}$, are also masked at  this IFU spectral element.
We then calculate the total number of detected LAEs $N_\mathrm{LAE}(\mathbf{x})$, the expected number of detected LAEs in the absence of clustering $\bar{N}(\mathbf{x})$, in each map voxel to create a map of $\delta_\mathrm{g}$.
We also save the mean $f_{50}$ in each voxel and use it for the mocks.

\section{Mock Data}
\label{sec:mocks}

We generate mock data to compare the measured LIM power spectra with a theoretical expectation.  These mocks allow us to quantify the effect of the data cleaning, and to calculate the covariance matrix of the power spectra.

\subsection{General Setup}
\label{sec:generalsetup}

To create a model for the power spectra and estimate their covariance matrices, we generate mock data using the \textsc{Simple} code\footnote{\url{https://github.com/mlujnie/simple}} \citep{lujanniemeyer/bernal/komatsu:2023}. This package generates galaxy and intensity maps with nonlinear redshift-space distortions (RSD) due to peculiar velocities \citep{agrawal/etal:2017}. 

We use the same box size, voxel size, and redshifts as in the HETDEX maps (see Section \ref{sec:map_creation}). The redshifts used to calculate the intensity of the mock data vary along one axis. We generate $100$ mocks for each redshift bin and each region shown in Figure \ref{fig:subboxes}, henceforth called boxes, i.e., $6900$ for the Spring field, $1800$ for the Fall field, and $300$ for the NEP field. We input the nonlinear matter power spectrum calculated at the three mean redshifts using the public CLASS code\footnote{\url{http://class-code.net}} \citep{blas/lesgourges/tram:2011} with the default Halofit model parameters \citep{smith/etal:2003,takahashi/etal:2012,bird/etal:2012}.
We compare the mock power spectra from the nonlinear and linear matter power spectra in Appendix \ref{sec:effect_nonlinear_pk}.

We set the LAE bias to $b_\mathrm{mock} = 2$, which is consistent with the measurements of $b_{\rm LAE} = 1.7^{+0.3}_{-0.4}$ at $z=3.1$ \citep{gawiser/etal:2007}, $1.8 \pm 0.3$ at $z=2.1$ \citep{guaita/etal:2010}, and $1.72_{‑0.27}^{+0.26}$ and $2.01_{‑0.29}^{+0.26}$ at $z=2.4$ and $3.1$, respectively \citep{herrera/etal:2025}.
In this setup, detected and undetected LAEs in the mock have the same bias parameter; thus, the LAE bias is $b_\mathrm{g}=2$ and the intensity bias is $b_I=2$.
Our simulations use the same luminosity functions of \citet{umeda/etal:2025} as we did for the $\bar{N}$ calculation described in Section~\ref{subsec:N_bar}, and we 
set the  minimum \Lya luminosity to $4 \times 10^{40}\,\mathrm{erg}\,\mathrm{s}^{-1}$.
This minimum luminosity mainly determines the number of simulated LAEs and the mean intensity.
However, a smaller value barely changes the mean intensity.
We apply a $\sigma_\lambda = 2.38\,\angstrom$ LOS Gaussian smoothing algorithm to the intensity maps in order to imitate the VIRUS spectral resolution. Since the VIRUS PSF is much smaller than the voxel size, we do not smooth in the angular direction.

To determine which simulated LAEs are detected, we randomly draw from the simulated LAEs. The probability that a simulated LAE with flux $f$ at position $\mathbf{x}$ will be detected is given by the completeness function, $C(f,f_{50}(\mathbf{x}))$, described in Section \ref{subsec:N_bar}, where $f_{50}(\mathbf{x})$ is the mean of the values of $f_{50}$ within the voxel.

\subsection{Inserting the Observed Data into the Mock Data}
\label{sec:insert_data_into_mock}

To create the same mask for the data and mocks, we convert the mock galaxy and intensity maps to the same format as the IFU spectra. For each IFU spectral element, we select the corresponding map and voxel in the mock. We save the mock `total,' 'detected,' and `undetected' intensities and the number of detected galaxies in the voxel. 

The volume of a voxel, $V_\mathrm{vox}$, is larger than the volume covered by one IFU spectral element, $V_{\mathrm{IFU}\times\Delta\lambda}$. Therefore, the number of galaxies that can be detected in the spectra is less than the number calculated for the voxels by a factor of $f_\mathrm{cov} = V_{\mathrm{IFU}\times\Delta\lambda}/V_\mathrm{vox}$. Ideally, we could account for this effect through binomial sampling with a probability of $p=f_\mathrm{cov}$. However, the redshift distribution of the detected mock galaxies differs from the actual LAE distribution similarly to $\bar{N}$ (see Section \ref{subsec:N_bar}). We therefore sample from the detected mock LAEs using a binomial distribution with a probability of $p=f_\mathrm{cov}f_{N}(z)$. We use the redshift of the voxel for $z$.

We imitate the sky subtraction process by subtracting the mean mock intensity spectrum of each observation from the mock intensity. We calculate the mean mock ``sky'' spectrum for the mock `total,' `detected,' and `undetected' intensity maps separately, and subtract it from the same mock intensity map.
We refer to the mock intensity spectra as $I^\mathrm{m}(\lambda)$ and the actual data intensity spectra as $I^\mathrm{d}(\lambda)$. We add the mock intensity to the data intensity ($I^\mathrm{m+d}_\mathrm{undet}(\lambda) = I^\mathrm{m}_\mathrm{undet}(\lambda) + I^\mathrm{d}_\mathrm{undet}(\lambda)$, etc.), where the data have been processed until, and including, the step described in Section \ref{subsubsec:calculating_intensity_removing_extinction}.
The intensity noise is thus naturally included in the mock spectra, $I^\mathrm{m+d}_\mathrm{undet}(\lambda)$.
Then, we perform the steps detailed in Section \ref{subsubsec:removing_systematics} (masking outlier pixels and high-variance spectra) and Section \ref{subsubsec:pca} (PCA) on $I^\mathrm{m}(\lambda)$ and $I^\mathrm{m+d}(\lambda)$. Finally, we calculate the PCA weight vectors of the $I^\mathrm{m+d}_\mathrm{tot}(\lambda)$ spectra and apply the PCA cleaning to the $I^\mathrm{m}_\mathrm{undet}(\lambda)$ and $I^\mathrm{m+d}_\mathrm{undet}(\lambda)$ spectra.

\section{Power Spectrum Measurement}
\label{sec:power_spectrum_measurement}

\subsection{Power Spectrum Estimator}

We calculate the cross-power spectrum of the detected LAE overdensity, ${\delta}_\mathrm{g}$, and the \Lya intensity fluctuation, $\delta I$, in each box.
We use a fast Fourier transform (FFT) to calculate $\widetilde{\delta I}(\mathbf{k})$ and $\widetilde{\delta}_\mathrm{g}(\mathbf{k})$
and compute the power spectrum,
\begin{equation}
    \hat{P}^{(i)}(\mathbf{k}) = V_\mathrm{box} \widetilde{\delta}^\ast_\mathrm{g}(\mathbf{k}) \widetilde{\delta I}(\mathbf{k}),
\end{equation}
where $V_\mathrm{box}$ is the volume of the box, the asterisk denotes the complex conjugate,
and the superscript $i$ identifies each box, as shown in Figure \ref{fig:subboxes}.

We calculate the power spectrum monopole, $\hat{P}^{(i)}_0(k)$,
by averaging $\hat{P}^{(i)}(\mathbf{k})$ over angles,
\begin{equation}
\label{eq:pk_angular_averaging}
\hat{P}_0^{(i)}(k) = \frac1{4\pi}\int d^2\hat{\mathbf{k}}~\hat{P}^{(i)}(\mathbf{k}),
\end{equation}
in $23$ linearly spaced $k$ bins from
$k_\mathrm{min} = 0.08\,h \mathrm{Mpc}^{-1}$ to $k_\mathrm{max} = 1 \,h \mathrm{Mpc}^{-1}$
with $\Delta k = 0.04\,h \mathrm{Mpc}^{-1}$.
We calculate $\hat{P}^{(i)}_0(k)$ for the cross-power spectrum of LAEs with the intensity of undetected sources for $N_\mathrm{PC} \in \{0, 10, ..., 200 \}$. We also calculate $\hat{P}^{(i)}_0(k)$ of detected mock galaxies and the mock intensity of undetected sources, $\delta I^{\rm m+d}_\mathrm{undet}$ and $\delta I^{\rm m}_\mathrm{undet}$. 

We do not use weights for the intensity because we find that the generated inverse-variance weights do not improve statistical significance of the power spectrum and complicate the analysis.

$\bar{N}$ naturally includes duplicate detections in repeated observations. The effective mask, $m(\mathbf{x})$, for both the galaxy and intensity maps is therefore equal to $1$ in the voxels that contain IFU spectral elements and equal to zero otherwise ($m \in \{0,1\}$). We adjust the power spectra of the boxes by the fraction of observed voxels, $f_V = \langle m \rangle =  \langle m^2\rangle$, and calculate the weighted mean in each redshift bin:
\begin{equation}
    \hat{P}_0(k) = \frac{ \displaystyle\sum_{i=1}^{N_\mathrm{boxes} } \hat{P}^{(i)}_0(k) f_{V,i}^{-1} w_i(k)} {\displaystyle\sum_{i=1}^{N_\mathrm{boxes}} w_i(k) },
    \label{eq:meanPk}
\end{equation}
where $w_i(k)$ are the weights defined in the next section.

\subsection{Covariance Matrix, Weights, and Masking}
\label{subsec:covariance_and_weights}

We calculate the covariance matrix of $\hat P_0(k)$ as the sum of two matrices: the first from shuffling the data ($C_\mathrm{shuffle}$), and the second from the mocks ($C_\mathrm{mock}$).  $C_\mathrm{shuffle}$ quantifies the statistical covariance arising from noise in the data, while $C_\mathrm{mock}$ quantifies cosmic variance. The total covariance matrix is given by $C = C_\mathrm{shuffle} + C_\mathrm{mock}$.

For $C_\mathrm{shuffle}$ and the weights, $w_i(k)$, we
randomly shuffle intensities of the IFU spectral elements in each of the three fields while keeping the redshift fixed, then calculate the power spectra, $\hat{P}^{s,(i)}_0(k)$. We repeat this process $100$ times, and compute weights as the inverse variance of the power spectrum monopole of each box as a function of wavenumber,
\begin{equation}
    w_i(k) = \frac{1}{\mathrm{var}\left(\hat{P}^{s,(i)}_0(k) f_{V,i}^{-1} \right)}.
\end{equation}
For each shuffling instance, we calculate the weighted mean, $\hat{P}_0^\mathrm{s}(k)$,
of the shuffled box power spectra, $\hat{P}^{s,(i)}_0(k)$. Finally, we construct the covariance matrix of the weighted mean power spectrum of each field in each redshift bin.

For $C_\mathrm{mock}$, we calculate the covariance matrix using the mock-only intensity, $I^\mathrm{m}_\mathrm{undet}(z)$, of $100$ mocks per box. This covariance quantifies cosmic variance of the intensity and the LAE catalog. 

The power spectra of some boxes exhibit a much higher variance along the wavenumbers than the rest, indicating the presence of residual systematic errors. To remain cautious, we exclude these boxes from the analysis. We also mask one box in the Spring field due to its small number of observations. The masked boxes are marked by the hatched regions in Figure \ref{fig:subboxes}.

\subsection{Loss of Power from PCA}
\label{sec:loss_of_power_from_pca}

\begin{figure*}
    \includegraphics[width=\textwidth]{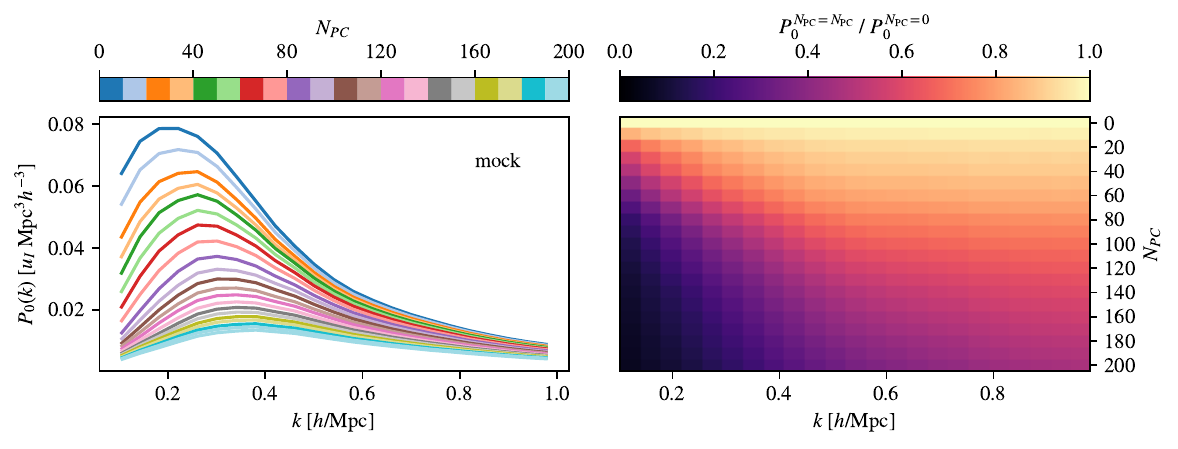}
    \caption{Left panel: suppression of the power spectrum monopole due to PCA cleaning. We show the mean power spectrum monopole obtained from the fiducial mocks in the Spring field in the medium-$z$ bin. The different colors correspond to different values of $N_\mathrm{PC}$, as indicated in the color bar. Here, $u_I = 10^{-18}\,\mathrm{erg\, s^{-1}\, cm^{-2}\, arcsec^{-2}\,}\angstrom^{-1}$. Right panel: the ratio of the power spectrum monopole with and without $N_\mathrm{PC}$ PCs removed as a function of $k$ and $N_\mathrm{PC}$.}
    \label{fig:mock_pk_Npc_v19}
\end{figure*}

Figure \ref{fig:mock_pk_Npc_v19} displays the effect of removing PCs on the cross-power spectrum of galaxies and intensity from the mocks. The left panel shows the cross-power spectrum of mock galaxies with $I^\mathrm{m}_\mathrm{undet}(z)$ after sky subtraction and removal of $N_\mathrm{PC} \in \{0, 10, ..., 200\}$ PCs in the Spring field in the medium-$z$ bin. The suppression of power at $k \lesssim 0.4 \, h\mathrm{Mpc}^{-1}$ is caused by sky subtraction. The right panel shows the ratio of the sky-subtracted mock power spectrum monopole with and without PCA. 

Removing PCs suppresses power, but does not completely eliminate it. Mocks are useful for capturing the effect of PCA on the signal power spectrum. We will use this information when we compare theoretical predictions with observed power spectra.

\section{Fitting the Power Spectrum Model}
\label{sec:amplitude}

\subsection{Model}
We use mock power spectra to calculate theoretical predictions and compare them with observed power spectra. The mock power spectrum can be written as 
\begin{equation}
\label{eq:model_pk}
\begin{split}
P_\mathrm{mock}(\mathbf{k}) &=  b_\mathrm{g} b_I \langle I \rangle 
\mathcal{T}(\mathbf{k})\\
    &\times
                                \int \,\frac{d^3\mathbf{k}'}{(2\pi)^3} P_\mathrm{m}(\mathbf{k}')
                                F_{\rm RSD}(\mathbf{k}')
                                \tilde{D}(\mathbf{k}')\\
                                &\times
                                \widetilde{W}^\ast_I(\mathbf{k} - \mathbf{k}') \widetilde{W}_g(\mathbf{k} - \mathbf{k}'),
\end{split}                            
\end{equation}
where $b_\mathrm{g}$ and $b_I$ are the linear bias parameters of
LAEs and intensity, respectively,
$\langle I\rangle$ is the mean intensity of the mock, and $P_\mathrm{m}(\mathbf{k})$ is the matter power spectrum computed for the fiducial cosmological model. 

The factor $F_{\rm RSD}(\mathbf{k})$ accounts for the effects of nonlinear RSD and a potential \Lya RT \citep[][]{lujanniemeyer:2025}.
In the mock, we account for nonlinearity in the Jacobian of the coordinate transformation from real to redshift space nonperturbatively \citep{agrawal/etal:2017}. 
In a linear model, it would be $F_{\rm RSD}^{\rm lin} = (1 + f\mu^2 b_\mathrm{g}^{-1}) (1 + f\mu^2 b_I^{-1})$ \citep{kaiser:1987}, where $\mu = \hat{\mathbf{k}} \cdot \hat{\mathbf{k}}_\parallel$ is the cosine of the angle between the unit wavevector $\hat{\mathbf{k}}$ and the LOS $\hat{\mathbf{k}}_\parallel$.

In Appendix \ref{app:lya_rt_rsd} we show a linear model for \Lya RT in $F_{\rm RSD}$. We do not, however, account for this effect in this paper because we do not detect the quadrupole power spectrum and the RT effect is degenerate with the bias parameters.

The factor $\tilde{D}(\mathbf{k})$ accounts for the LOS damping due to limited spectral resolution (see Section \ref{sec:generalsetup}). The transfer function, $\mathcal{T}(\mathbf{k})$, accounts for the loss of power due to data processing, especially sky subtraction and PCA (see the right panel of Figure~\ref{fig:mock_pk_Npc_v19} for the loss of power due to PCA).
Note that we include the transfer function in our forward model and do not attempt to reconstruct a PCA-free power spectrum from the data. 
The window functions, $\widetilde{W}^\ast_I(\mathbf{k})$ and $\widetilde{W}_g(\mathbf{k})$, account for the observing mask and weights. Since the mask and weights are the same for the mocks and the data, we assume that the window functions are fully modeled by the mocks. 

In our model for the monopole power spectrum, the only free parameter is the overall factor, $b_\mathrm{g} b_I \langle I \rangle \bar{F}_{\rm RSD}$,
where $\bar{F}_{\rm RSD}$ is the effective nonlinear RSD factor from the integral given in Equation \eqref{eq:model_pk} and the angular averaging given in Equation \eqref{eq:pk_angular_averaging}.

In the absence of a mask, a linear model gives $\bar{F}_{\rm RSD}^{\rm lin} = 1 + \frac{f}{3}\left(b_\mathrm{g}^{-1} + b_I^{-1}\right) + \frac{f^2}{5} b_\mathrm{g}^{-1}b_I^{-1} \simeq 1.3$. Therefore, a slight mismatch in the values of $b_\mathrm{g}$ and $b_I$ between the mock and the data only gives a minor correction to this factor.

In practice, we define the fitting parameter, $A = b_\mathrm{g} b_I \langle I \rangle \bar{F}_{\rm RSD} / (b_\mathrm{g}^\mathrm{fid} b_I^\mathrm{fid} \langle I^\mathrm{fid} \rangle \bar{F}_{\rm RSD}^{\rm fid})$, where $b^\mathrm{fid}_\mathrm{g}$, $b^\mathrm{fid}_I$, and $\langle I^\mathrm{fid} \rangle$ are the fiducial LAE bias, the intensity bias, and the mean intensity of undetected sources assumed in the mocks. $\bar{F}^{\rm fid}_{\rm RSD}$ is the effective nonlinear RSD factor in the mocks.
If the mocks were in perfect agreement with the data, we would find that $A = 1$.

Due to nonlinearity, $\bar{F}_{\rm RSD}$ depends on the wavenumber $k$. However, the ratio, $\bar{F}_{\rm RSD}/\bar{F}_{\rm RSD}^{\rm fid}$, depends on $k$ only weakly if the mock and the data are in reasonable agreement, which is the case (see Section \ref{sec:results}). Therefore, we neglect the scale dependence of $\bar{F}_{\rm RSD}/\bar{F}_{\rm RSD}^{\rm fid}$ and treat it as a constant factor. This is a good approximation given our limited ability in determining the shape of the power spectrum due to the large statistical uncertainty.

\subsection{Parameter Inference}
To determine $A$,
we fit the fiducial mock power spectrum multiplied by an amplitude $A$ to the data. 
We define $\chi^2 = \mathbf{Y}^\top \mathbf{C}^{-1} \mathbf{Y}$,
where $\mathbf{Y} = \mathbf{Y}_\mathrm{data} - A \mathbf{Y}_\mathrm{fid}$ is the difference between the measured power spectrum and the model power spectrum,
\begin{equation}
\begin{split}
\mathbf{Y}^\top_\mathrm{data} &= (P^\mathrm{data}_0(k_1), P^\mathrm{data}_0(k_2), \dots, P^\mathrm{data}_0(k_{23})), \\
\mathbf{Y}^\top_\mathrm{fid} &= (P^\mathrm{fid}_0(k_1), P^\mathrm{fid}_0(k_2), \dots, P^\mathrm{fid}_0(k_{23})),
\end{split}
\end{equation}
and $P^\mathrm{fid}_0(k)$ is the monopole of the mock power spectrum $P_\mathrm{mock}(\mathbf{k})$ (Equation \eqref{eq:model_pk}), calculated in the same $23$ $k$ bins as the data. 

We also combine the power spectra of the fields and fit a single amplitude $A$ to all of the fields within each redshift bin simultaneously. Specifically, we use 
\begin{equation}
\begin{split}
\mathbf{Y}^\top_\mathrm{data,all} &= ( P^\mathrm{data,Fall}_0(k_1), \dots, P^\mathrm{data,Fall}_0(k_{23}),\\
 & P^\mathrm{data,Spring}_0(k_1), \dots, P^\mathrm{data,Spring}_0(k_{23}),\\
 & P^\mathrm{data,NEP}_0(k_1), \dots, P^\mathrm{data,NEP}_0(k_{23}))
\end{split}
\end{equation}
and the equivalent vector for the model.
For the covariance matrix, we combine the individual covariance matrices:
\begin{equation}
C_{\rm all} = \begin{pmatrix}
C_{\rm Fall} & 0 & 0\\
0 & C_{\rm Spring} & 0 \\
0 & 0 & C_{\rm NEP}
\end{pmatrix}.
\end{equation}
The best-fit amplitude that minimizes $\chi^2$ for the individual fields is given by
\begin{equation}
\label{eq:Abest}
    A_\mathrm{best} = \frac{\mathbf{Y}^\top_\mathrm{fid} C^{-1} \mathbf{Y}_\mathrm{data}}{\mathbf{Y}^\top_\mathrm{fid} C^{-1} \mathbf{Y}_\mathrm{fid}}.
\end{equation}
The best-fit amplitude that minimizes $\chi^2$ for the combined fit is given by
\begin{equation}
\label{eq:Abestall}
    A^\mathrm{all}_\mathrm{best} = \frac{\mathbf{Y}^\top_\mathrm{fid,all} C^{-1}_\mathrm{all} \mathbf{Y}_\mathrm{data,all}}{\mathbf{Y}^\top_\mathrm{fid,all} C^{-1}_\mathrm{all} \mathbf{Y}_\mathrm{fid,all}}.
\end{equation}

\subsection{Error Estimation and Goodness-of-fit}

The uncertainty of the best-fit amplitude can be estimated as
\begin{equation}
    \sigma_{\rm ls} = \frac{1}{\sqrt{\mathbf{Y}^\top_\mathrm{fid} C^{-1} \mathbf{Y}_\mathrm{fid}}}.
\end{equation}
We calculate $A_\mathrm{best}$ and $\sigma_{\rm ls}$ for each field, as well as $A_\mathrm{best}^\mathrm{all}$ and $\sigma_{\rm ls}^\mathrm{all}$, for each $N_\mathrm{PC}$ and redshift bin.

To ensure that we do not underestimate the uncertainty of the measured power spectra, we empirically estimate the uncertainty, referred to as $\sigma_\mathrm{emp}$. This method assumes that the best-fit model is correct, but that the data contain an unaccounted-for systematic error that increases the point-to-point variance. We generate fake power spectra by randomly sampling from a Gaussian probability density function (PDF) at each wavenumber bin. The mean of the Gaussian is given by the power spectrum of the best-fit model, $\mathbf{Y}_\mathrm{model} = A_\mathrm{best}\mathbf{Y}_\mathrm{fid}$; the standard deviation of the point in the $j$th $k$ bin is given by $\sqrt{N_\mathrm{RMS}^{-1} \sum_{i=j-4}^{j+4} (Y_{\mathrm{data},i} - Y_{\mathrm{model},i})^2}$, where we sum over the $N_\mathrm{RMS}=9$ nearest wavenumber bins. We fit the model to the data using equations \eqref{eq:Abest} and \eqref{eq:Abestall} and repeat this process $1000$ times. The standard deviation $\sigma_\mathrm{emp}$ of the $1000$ best-fit amplitudes is typically larger than $\sigma_\mathrm{ls}$ with the mean ratio $\langle \sigma_\mathrm{emp}/\sigma_\mathrm{ls}\rangle \simeq 1.6$.
To remain conservative, we select the larger of the two estimated uncertainties, 
\begin{equation}
    \sigma_A = \mathrm{max}\{\sigma_\mathrm{emp},\sigma_\mathrm{ls}\}.
\end{equation}

To determine the goodness-of-fit, we calculate the reduced $\chi^2$, 
$\chi^2_\nu=\chi^2_\mathrm{min}/\nu$, where $\chi^2_\mathrm{min}$ is the $\chi^2$ minimum and $\nu= N_k - N_\mathrm{fit}$ is the number of degrees of freedom, assuming that the covariance matrix is diagonal (see Appendix \ref{sec:correlation_matrices} for the correlation matrices). Each individual field has $N_k = 23$ wavenumber bins per power spectrum. Since we are fitting a single parameter, $N_\mathrm{fit} = 1$, there are 
$\nu=22$ degrees of freedom for the fit to individual fields. The combined fit has
$\nu = 3 \times 23 - 1 = 68$ degrees of freedom per redshift bin. 

We also calculate the probability-to-exceed (PTE, also called p-value), from the integral of a $\chi^2$ PDF, $\mathcal{P}_\nu(\chi^2)$, 
from $\chi^2_\mathrm{min}$ of the best fit to infinity:
\begin{equation}
    \mathrm{PTE}(\chi^2_\mathrm{min}) = \int_{\chi^2_\mathrm{min}}^\infty 
    \mathcal{P}_\nu(\hat{\chi}^2)d\hat{\chi}^2.
\end{equation}
If a fit yields $\mathrm{PTE}>0.05$, the probability of obtaining a higher value of $\chi^2_\mathrm{min}$ than this fit is $>5\%$ (assuming the model is correct), indicating consistency with the model.

\section{Results}
\label{sec:results}

\begin{figure*}
    \includegraphics[width=\textwidth]{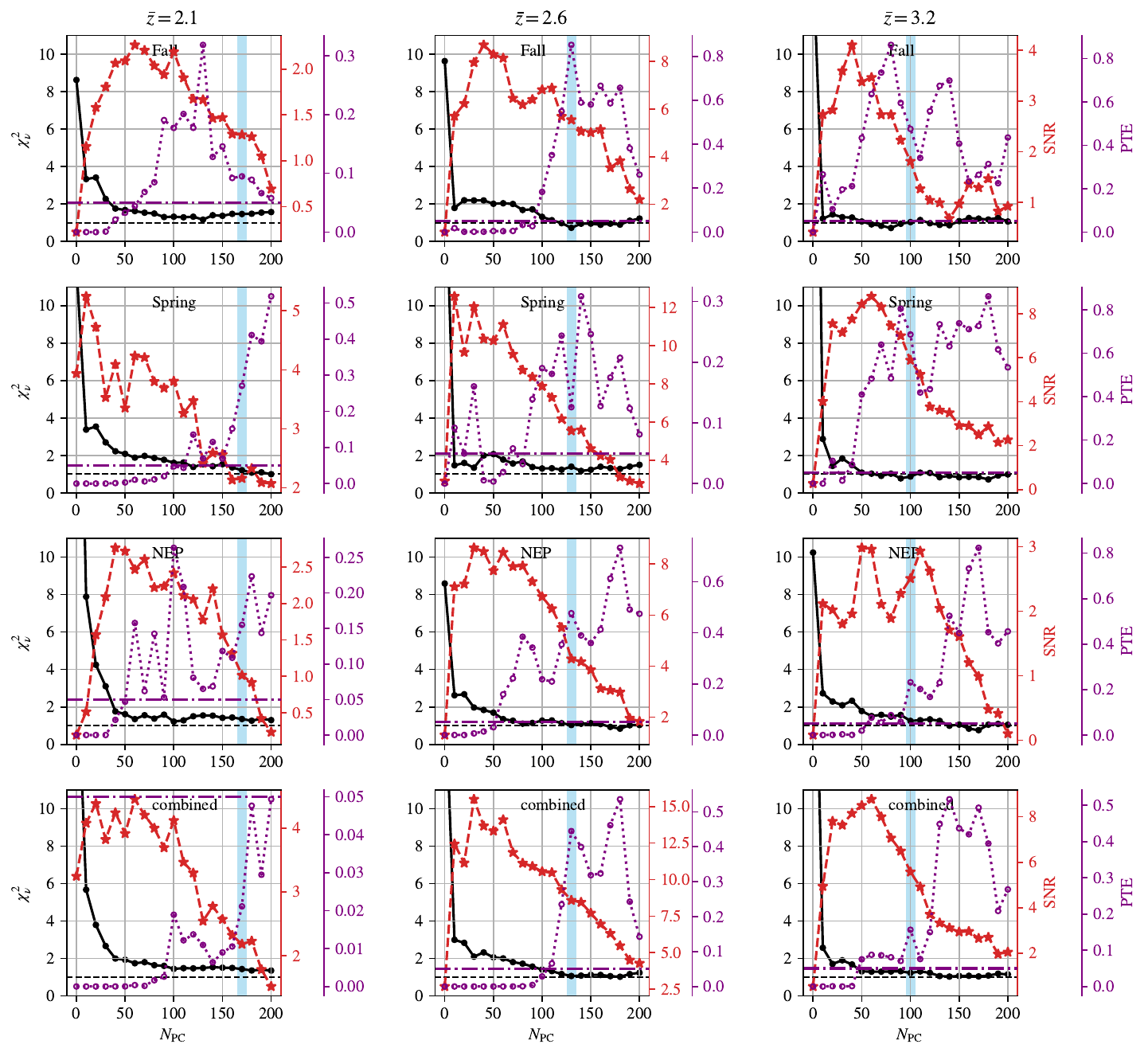}
    \caption{Reduced $\chi^2_\nu$ (black dots), SNR (red stars), and PTE (purple empty circles) of the fit as a function of $N_\mathrm{PC}$ in the Fall (first row), Spring (second row), and NEP (third row) fields, and that combining the three fields (fourth row). The columns show the three redshift bins. The horizontal black dashed lines indicate $\chi^2_\nu = 1$. The horizontal purple dotted-dashed lines indicate $\mathrm{PTE} = 0.05$.
    }
    \label{fig:red_chi2_sn}
\end{figure*}

Choosing the optimal $N_\mathrm{PC}$ is a crucial decision in the analysis.  We want $N_\mathrm{PC}$ to be high enough that most systematics are removed. We use goodness-of-fit as a guide: good agreement between the model and the data implies a low value of $\chi_\nu^2 \simeq 1$ and a high value of $\mathrm{PTE}> 0.05$. However, removal of PCs also reduces the signal. We therefore want $N_\mathrm{PC}$ to be small enough to retain a high SNR\null. Finally, we want the amplitudes inferred in the three fields to be consistent with each other at the chosen $N_\mathrm{PC}$.

Figure \ref{fig:red_chi2_sn} shows $\chi_\nu^2$, the PTE, and the $SNR(A)= A_\mathrm{best}/\sigma_A$ as a function of $N_\mathrm{PC}$ for the three fields and the combined fit, for the three redshift bins. $\chi_\nu^2$ decreases with increasing $N_\mathrm{PC}$, indicating a better agreement between the measured and model power spectra. In the low-$z$ bin of the combined fit, $\chi^2_\nu$ approaches $\simeq 1.5$ at $N_\mathrm{PC} \geq 100$; in the medium- and high-$z$ bins of the combined fit, $\chi^2_\nu$ is close to $1$ at $N_\mathrm{PC} \geq 120$. The PTE generally also improves (increases) with increasing $N_\mathrm{PC}$. The SNR, however, decreases with $N_\mathrm{PC}$ above a threshold value of $N_\mathrm{PC} \simeq 50$.

The goodness-of-fit is acceptable in all cases except for the fit of the combined power spectrum in the low-$z$ bin, which yields a low $\mathrm{PTE}<0.05$ at all $N_{\rm PC}$. This result could indicate the presence of systematic errors that are not quantified by the covariance matrix or an underestimation of the statistical covariance matrix of the power spectrum. Otherwise, the fit is excellent.

\begin{figure*}
    \includegraphics[width=\textwidth]{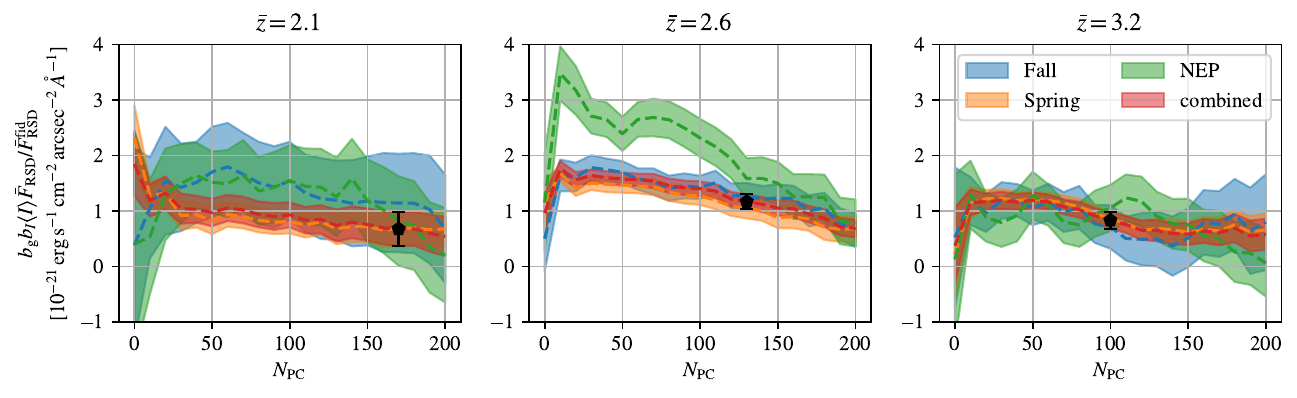}
    \caption{Constraints on $b_\mathrm{g} b_I \langle I \rangle \bar{F}_{\rm RSD} / \bar{F}^{\rm fid}_{\rm RSD}$ inferred from the best-fit amplitude as a function of $N_\mathrm{PC}$ in the Fall (blue), Spring (orange), and NEP (green) fields, along with the combination of all three fields (red). The dashed lines show the mean, while the shaded areas show the $1\sigma$ uncertainties. The black polygons indicate the constraints from the combined fit, as determined from the best combination of $\chi^2_\nu$, SNR, and PTE values.}
    \label{fig:bestfit_bgbImeanI}
\end{figure*}

We also want to choose $N_\mathrm{PC}$ so that the inferred amplitudes in the three fields agree. Figure \ref{fig:bestfit_bgbImeanI} displays the best-fit $b_\mathrm{g} b_I \langle I \rangle \bar{F}_{\rm RSD} / \bar{F}^{\rm fid}_{\rm RSD} = A_\mathrm{best} b_\mathrm{g}^\mathrm{fid} b_I^\mathrm{fid} \langle I^\mathrm{fid} \rangle$ and its uncertainty for the three fields and the combined fit as a function of $N_\mathrm{PC}$.
The fiducial bias values used for the mocks are $b_I^{\rm fid} = b_{\rm g}^{\rm fid} = 2$ (Section~\ref{sec:generalsetup}). The fiducial mean intensity of undetected sources obtained from the mocks is $\langle I^\mathrm{fid}\rangle = 1.5$, $2.2$, and $1.4 \times 10^{-22}\,\intensityunit$ in the redshift bins centered around $\bar{z} = 2.1$, $2.6$, and $3.2$, respectively.

In principle, we could choose different values of $N_\mathrm{PC}$ for each field and redshift bin because the PCA is performed independently in each field. However, for simplicity, we choose the same $N_\mathrm{PC}$ for all fields in each redshift bin. Based on the small $\chi^2_\nu$ and high SNR shown in Figure \ref{fig:red_chi2_sn}, and the requirement that the best-fit amplitudes be consistent across fields, we choose $N_\mathrm{PC} = \NPClowz$, $\NPCmediumz$, and $\NPChighz$ in the low-, medium-, and high-$z$ bins, respectively. The best-fit values of $b_\mathrm{g} b_I \langle I \rangle \bar{F}_{\rm RSD} / \bar{F}^{\rm fid}_{\rm RSD}$ at these $N_\mathrm{PC}$ are $b_\mathrm{g} b_I \langle I \rangle \bar{F}_{\rm RSD} / \bar{F}^{\rm fid}_{\rm RSD} =$\lowzbbi, \midzbbi, and \highzbbi~\bbiunit in the low-, medium-, and high-$z$ bins, respectively. Our main result is that we significantly detect the LAE-\Lya intensity cross-power spectrum.

\begin{figure*}
    \includegraphics[width=\textwidth]{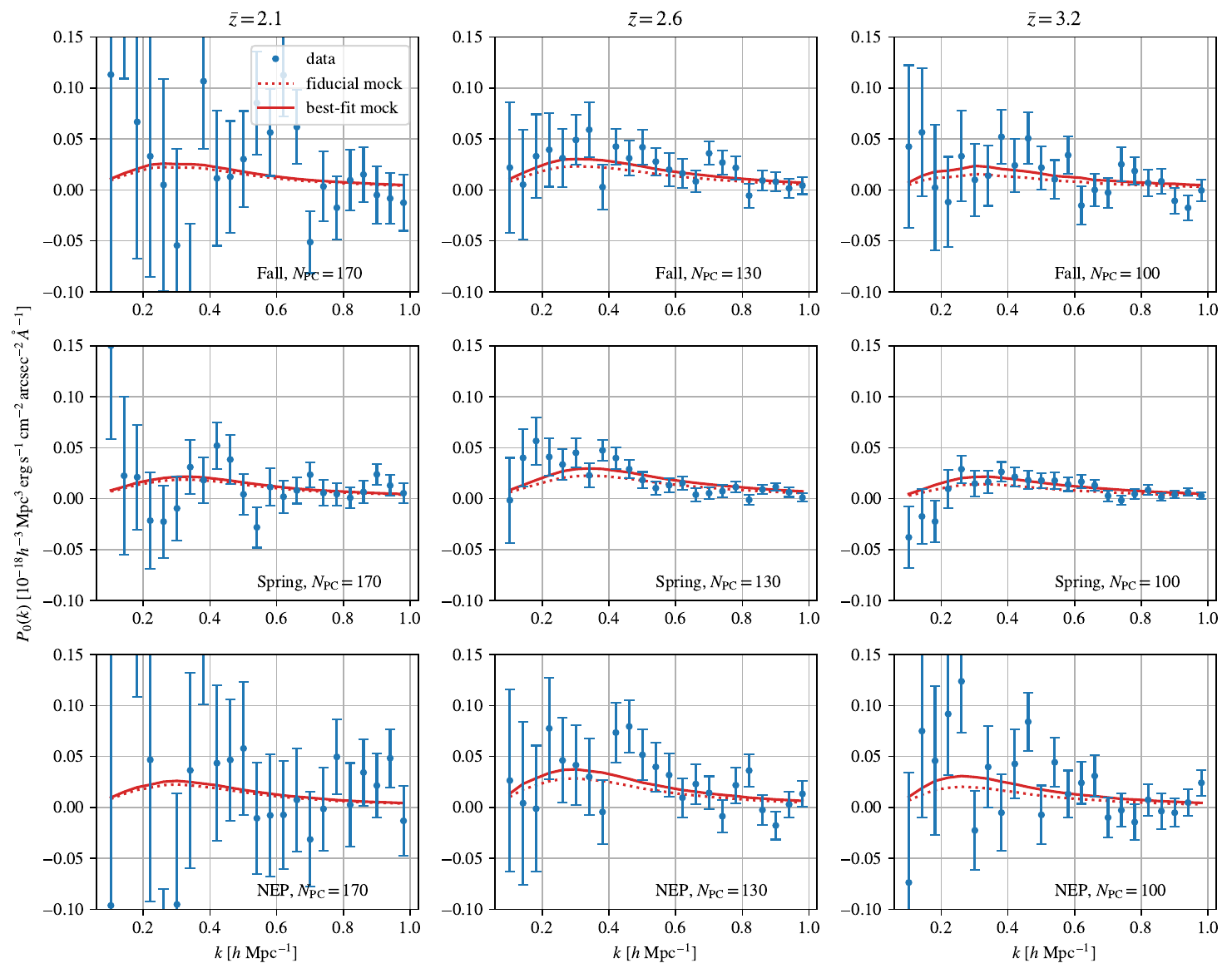}
    \caption{Power spectrum monopoles in the Fall (top row), Spring (middle row), and NEP (bottom row) fields and in the three redshift bins (different columns). Each panel shows the mean power spectrum monopole (Eq.~\eqref{eq:meanPk}) in each field and redshift bin. We have removed $N_\mathrm{PC} = \NPClowz$, $\NPCmediumz$, and $\NPChighz$ PCs in the low-, medium-, and high-$z$ bins, respectively.
    The dotted red lines indicate the fiducial mock power spectra using the same $N_\mathrm{PC}$. The solid red lines are the mock power spectra multiplied by the best-fit amplitude.}
    \label{fig:pk_mock_data_allfields}
\end{figure*}

Finally, Figure \ref{fig:pk_mock_data_allfields} shows the quality of the fits to the power spectrum monopoles from each field and redshift bin using the selected $N_\mathrm{PC}$. We compare the measured power spectra to the fiducial mock power spectrum and to the mock power spectrum multiplied by the best-fit amplitude. The error bars of the power spectra are given by the square root of the diagonal elements of the covariance matrix (see Section \ref{subsec:covariance_and_weights}). The mocks accurately describe the data.

\begin{figure}
    \centering
    \includegraphics[width=0.5\textwidth]{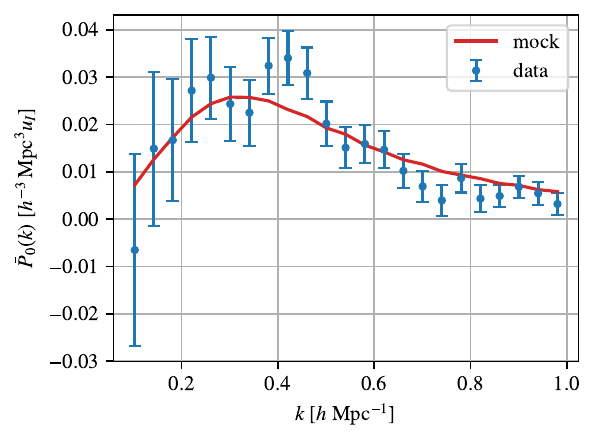}
    \caption{Weighted mean power spectrum monopole of the three fields and three redshift bins of the data (blue) and the best-fit mocks (red). Here, $u_I=10^{-18}\,\intensityunit$.}
    \label{fig:pk_combined}
\end{figure}

Figure \ref{fig:pk_combined} shows the weighted mean power spectrum monopole of the three fields and the three redshift bins shown in Figure \ref{fig:pk_mock_data_allfields}. We use inverse variance weights,
\begin{equation}
\label{eq:weighted_mean_pk_over_fields}
    \bar{P}_0(k_i) = \frac{\displaystyle\sum_{j=1}^{9} \hat{P}_0^{(j)}(k) \sigma_{\hat{P}_0^{(j)}}^{-2}(k_i)}{\displaystyle\sum_{j=1}^{9}\sigma_{\hat{P}_0^{(j)}}^{-2}(k_i)},
\end{equation}
where $\sigma_{\hat{P}_0^{(j)}}(k_i)$ is the square root of the diagonal element $i$ of the covariance matrix of the power spectrum monopole $\hat{P}_0^{(j)}$, and the index $j$ denotes a field and redshift bin.
The propagated uncertainty of the weighted mean power spectrum monopole, $\bar{P}_0(k)$, is given by 
\begin{equation}
    \sigma_{\bar{P}_0}(k_i) = \frac{1}{\sqrt{\displaystyle\sum_{j=1}^{9}\sigma_{\hat{P}_0^{(j)}}^{-2}(k_i)}}.
\end{equation}
We calculate the weighted mean of the best-fit mock power spectra using the same inverse weights of the data.
We use the same $N_{\rm PC}$ as in Figure \ref{fig:pk_mock_data_allfields} for the data and mocks.
As a test, we fit an amplitude times the weighted mean of the best-fit mock power spectrum monopole to the weighted mean power spectrum monopole. We find a best-fit amplitude of $0.97 \pm 0.07$, which is consistent with the expectation of $1$.
With $\nu=22$ degrees of freedom, we find $\chi^2 = 21$,  $\chi^2_\nu = 0.94$, and $\mathrm{PTE}=0.54$.

Although all quadrupole power spectra are consistent with zero, we do not use them to extract physical constraints, such as the \Lya RT effect \citep{lujanniemeyer:2025}, as the shape and amplitude of the quadrupole power spectra are strongly affected by sky subtraction and PCA, and are therefore not yet stable. We leave detailed investigations of the quadrupole power spectra for future study.

\section{Comparison to Previous Constraints}
\label{sec:comparison_previous_constraints}

\begin{figure}
    \includegraphics[width=0.5\textwidth]{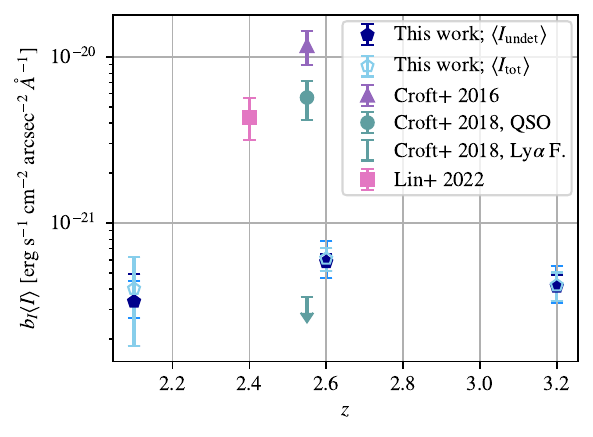}
    \caption{Constraints on $b_I \langle I \rangle$ obtained from this work (dark blue polygons). We assume $b^\mathrm{fid}_\mathrm{g} = 2$ and $\bar{F}_{\rm RSD} = \bar{F}^{\rm fid}_{\rm RSD}$. 
    The medium blue error bars span the the constraints when assuming $b^\mathrm{fid}_\mathrm{g} = 2\pm 0.5$.
    The light blue polygons show $b_I \langle I_\mathrm{tot} \rangle \simeq b_I^\mathrm{fid} (A_\mathrm{best} \langle I_\mathrm{undet}^\mathrm{fid} \rangle + \langle I_\mathrm{det}^\mathrm{fid} \rangle)$ (see text for more information). 
    The purple triangle, olive circle, and pink square are the constraints from the QSO-\Lya intensity cross-correlations of \citet{croft/etal:2016,croft/etal:2018} and \citet{lin/etal:2022}, respectively. The upper limit from the \Lya forest-\Lya intensity cross-correlation \citep{croft/etal:2018} is displayed as an olive arrow.}
    \label{fig:mean_bI_compare}
\end{figure}

To compare our results to previous work, we convert our measurements to $b_I \langle I \rangle$ assuming a fiducial LAE bias of $b_\mathrm{g} = 2$ (Section~\ref{sec:generalsetup}) and $\bar{F}_{\rm RSD} = \bar{F}^{\rm fid}_{\rm RSD}$.

Figure \ref{fig:mean_bI_compare} compares our combined best-fit values of $b_I \langle I \rangle$, i.e., the product of the linear intensity bias and the intensity of only undetected sources, with those obtained from cross-correlations between QSOs and \Lya intensity 
\citep{croft/etal:2016,croft/etal:2018,lin/etal:2022}, and the $95\%$ confidence upper limit on $b_I \langle I \rangle$ inferred from \Lya forest-\Lya intensity cross-correlation \citep{croft/etal:2018}. These studies are all performed at redshifts from $z=2$ to $3.5$. 

Since these analyses constrain the product of the intensity bias and total \Lya intensity, we also show $b_I^\mathrm{fid} (A_\mathrm{best} \langle I_\mathrm{undet}^\mathrm{fid} \rangle + \langle I_\mathrm{det}^\mathrm{fid}  \rangle)$, where $A_\mathrm{best}$ is the best-fit amplitude, $b_I^\mathrm{fid}=2$ is the fiducial intensity bias, and $ \langle I_\mathrm{undet}^\mathrm{fid} \rangle$ and $ \langle I_\mathrm{det}^\mathrm{fid} \rangle$ are the mean intensities of only undetected and only detected sources from the fiducial mocks, respectively. The luminosity function of the mocks does not include QSOs (see Section \ref{sec:mocks}).
These values are only slightly larger than $b_I \langle I_\mathrm{undet} \rangle$. 

Both $b_I \langle I_\mathrm{undet} \rangle$ and $b_I^\mathrm{fid} (A_\mathrm{best} \langle I_\mathrm{undet}^\mathrm{fid} \rangle + \langle I_\mathrm{det}^\mathrm{fid} \rangle)$ are significantly smaller than the values inferred from QSO-\Lya intensity correlations. The low- and high-$z$ best-fit values are consistent with the upper limit inferred from the \Lya forest-\Lya intensity cross-correlation, while the medium-$z$ bin is $2.6\sigma$ higher.

This discrepancy could be due to the QSO-\Lya cross-correlations being dominated by small-scale physics that are affected by QSOs, such as an elevated radiation field surrounding them \citep[e.g.,][]{miller/etal:2021,dong/etal:2023}. Since our galaxy sample is dominated by star-forming galaxies, our measurement is less affected by QSO-related issues and may provide a more accurate representation of the \Lya radiation field in the IGM.

\section{Origins of the \texorpdfstring{\Lya}{Lyα} Emission}
\label{sec:origins_of_lya_emission}

In this work, we masked the spectra within $10''$ of the detected LAEs, which, in our redshift range, corresponds to a proper distance of $\simeq 75-85\,\mathrm{kpc}$. Beyond this distance, the photons contributing to the radial profiles of the simulated LAEs originate from outside the stacked LAEs' dark-matter halos \citep{byrohl/etal:2021,byrohl/nelson:2023}. Therefore, the intensity should be dominated by photons unrelated to the detected LAEs. The intensity includes photons emitted within undetected galaxies, from detected LAEs that scatter many tens of kpc from the origin, and from the CGM and IGM\null. While we cannot directly constrain the origin and emission mechanism from our measurement, we can compare it to theoretical expectations.

\begin{figure}
\includegraphics[width=0.5\textwidth]{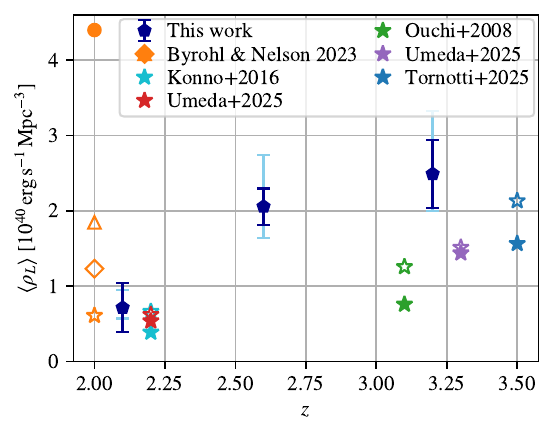}
\caption{
    Constraints on the mean comoving luminosity density of undetected sources from this work, $\langle \rho_L \rangle$, assuming $b^\mathrm{fid}_\mathrm{g} = b^\mathrm{fid}_I = 2$ and $\bar{F}_{\rm RSD} = \bar{F}^{\rm fid}_{\rm RSD}$ (dark-blue polygons). The dark-blue error bars are the statistical uncertainty of the fit, whereas the light-blue error bars indicate the best-fit values with $b^\mathrm{fid}_I = 2.0 \pm 0.5$ and $b^\mathrm{fid}_\mathrm{g}=2$. The filled stars show the luminosity density obtained by integrating the luminosity functions \citep{konno/etal:2016,umeda/etal:2025,ouchi/etal:2008,tornotti/etal:2025} from $L_\mathrm{min} = 4\times 10^{40}\,\mathrm{erg\, s^{-1}}$ to a typical HETDEX detection limit of $L_\mathrm{max} = 4\times 10^{42}\,\mathrm{erg\, s^{-1}}$.
    The empty stars of the same colors show the luminosity density obtained with $L_\mathrm{min} = 10^{36}\,\mathrm{erg\, s^{-1}}$, below which the integral barely changes.
    The orange symbols are predictions from cosmological simulations with \Lya RT \citep{byrohl/nelson:2023}: the filled circle shows the total mean luminosity density; the empty orange star represents photons from galaxies with luminosity $L<10^{41.75}\,\mathrm{erg\, s^{-1}}$; and the empty diamond and triangle denote photons that last scattered in the IGM and CGM, respectively.
    }
    \label{fig:bI_theory_comparison}
\end{figure}

We translate our constraint on $b_\mathrm{g} b_I \langle I \rangle \bar{F}_{\rm RSD} / \bar{F}^{\rm fid}_{\rm RSD}$ to a constraint on the comoving luminosity density 
\begin{equation}
\label{eq:rho_L_from_I}
    \rho_L = \frac{4 \pi}{c}{ \langle I \rangle  H(z) (1 + z)^2 \lambda_0},
\end{equation}
where $\lambda_0 = 1215.67\,\angstrom$ is the \Lya rest-frame wavelength. 
We assume $b^\mathrm{fid}_\mathrm{g} = b^\mathrm{fid}_I = 2$, as we do not know the bias of the detected LAEs or of the intensity. 
We will discuss a possible range of the intensity bias from simulation work in Section \ref{subsec:RTsim}. 
In addition, we assume that $\bar{F}_{\rm RSD} = \bar{F}^{\rm fid}_{\rm RSD}$. 
We compare our constraint on $\rho_L$ to theoretical expectations in Figure \ref{fig:bI_theory_comparison}.

\subsection{Expectations from Luminosity Functions}

The mock is based on our fiducial model, which only includes the intensity from faint, undetected galaxies, as expected from extrapolating the LAE luminosity function measured by \citet{umeda/etal:2025}. The mean comoving luminosity density from undetected galaxies can be calculated by integrating the luminosity function up to the HETDEX detection limit:
\begin{equation}
    \langle \rho_L^\mathrm{undet} \rangle = \int_{L_\mathrm{min}}^{L_\mathrm{det}} \mathrm{d}L\frac{\mathrm{d}n}{\mathrm{d}L} L ,
    \label{eq:lumdensity}
\end{equation}
where $L_\mathrm{min}$ is given by the lowest \Lya luminosity of galaxies and $L_\mathrm{det}$ is the minimum luminosity detectable by HETDEX\null. The corresponding mean specific intensity at redshift $z$ is then given by
\begin{equation}
\label{eq:I_from_rhoL}
    \langle I_{\mathrm{undet}}^\mathrm{gal} \rangle = \frac{c\langle \rho_L^\mathrm{undet} \rangle}{4 \pi H(z) (1 + z)^2 \lambda_0}.
\end{equation}

We integrate Equation~\eqref{eq:lumdensity} from $L_\mathrm{min} = 4\times 10^{40}\,\mathrm{erg}\,\mathrm{s}^{-1}$ to a typical HETDEX detection limit, $L_\mathrm{max} = 4 \times 10^{42}\,\mathrm{erg}\,\mathrm{s}^{-1}$ (see Figure \ref{fig:lae_z_lum_lw_hists}). 
To illustrate the uncertainty due to the unknown minimum luminosity, we also show the result for $L_\mathrm{min} = 10^{36}\,\mathrm{erg}\,\mathrm{s}^{-1}$.
Using a smaller value for $L_\mathrm{min}$ barely changes the integral.
Figure \ref{fig:bI_theory_comparison} shows $\langle \rho_L^\mathrm{undet} \rangle$ obtained by integrating various luminosity functions from the literature \citep{ouchi/etal:2008,konno/etal:2016,umeda/etal:2025,tornotti/etal:2025}. The luminosity densities of the different luminosity functions are similar to each other and slightly lower than that of our measurement.
Note that the luminosity functions of \citet{ouchi/etal:2008,konno/etal:2016,umeda/etal:2025} are obtained from narrowband surveys, which have different LAE selection criteria from spectroscopic surveys like HETDEX or \citet{tornotti/etal:2025}.
The value of $\rho_L$ from integrating luminosity functions is also strongly dependent on the faint-end slope $\alpha$ of the luminosity function, which is degenerate with $L_\star$ and $\phi_\star$ \citep[see Figure 3 of][]{tornotti/etal:2025}.

\subsection{Star Formation}

The comoving star-formation rate density at $z = 2 - 3.5$ is $\simeq 0.1 \, M_\odot\, \mathrm{yr^{-1}\,Mpc^{-3}}$ \citep{rowan-robinson/etal:2016}. We use the conversion of the star-formation rate, $\dot{M}_\star$, to the intrinsic \Lya luminosity \citep{dijkstra:2019}
\begin{equation}
\label{eq:lya_from_sfr}
L_{\mathrm{Ly}\alpha} \simeq 10^{42}\,\mathrm{erg\, s^{-1}}\left(\frac{\dot{M}_\star}{M_\odot\,\mathrm{yr}^{-1}}\right).
\end{equation}
Note that this conversion factor strongly depends on the initial stellar mass function and the assumption of case-B recombination and can deviate by factors of $2-3$ with other assumptions \citep{raiter/etal:2010}.

Using Equation \eqref{eq:lya_from_sfr}, we obtain a comoving intrinsic luminosity density of $\rho_L \simeq 10^{41}\,\mathrm{erg\, s^{-1}\, Mpc^{-3}}$. The corresponding observed intensity at redshift $z= 2.1\, (2.6/3.2$) is $\langle I_\mathrm{SF} \rangle \simeq f_\mathrm{esc} \times 2.4 \, (1.4/0.8) \times 10^{-21}\,\mathrm{erg\, s^{-1}\, cm^{-2}\, arcsec^{-2}}\,\angstrom^{-1}$, where $f_\mathrm{esc}$ is the mean escape fraction of \Lya photons from star-forming regions. 

Therefore, star formation can power the mean intensity inferred from our measurements if $b_I \langle I \rangle f_{\rm esc} / \langle I_{SF}\rangle = b_I f_\mathrm{esc} \simeq 0.1$, $0.4$, and $0.5$ in the low-, medium-, and high-$z$ bins, respectively, assuming $b_\mathrm{g}=2$ and $\bar{F}_{\rm RSD} = \bar{F}^{\rm fid}_{\rm RSD}$. These numbers are reasonable: much higher ($b_I f_\mathrm{esc}\gg 1$) and lower ($b_I f_\mathrm{esc}\ll 0.1$) values would not support the interpretation of \Lya emission powered by star formation.

\subsection{Comparison to a RT Simulation}
\label{subsec:RTsim}

\citet{byrohl/nelson:2023} post-processed a cosmological hydrodynamic simulation in a cosmological volume at $z=2$ with \Lya RT\null. The authors calibrated the escape fraction of \Lya emission from galaxies to an observed luminosity function, painted \Lya emission from recombination and collisional excitation into cells without star formation, and simulated the scattering process of the photons. In the simulation, $48\%$ of \Lya photons are produced by collisional excitation, making it the dominant emission mechanism. Star formation produces $28\%$ and recombination outside the interstellar medium (ISM) contributes $23\%$ of the photons. The vast majority, $98\%$, of \Lya photons are produced within halos, with more than half produced within the CGM\null. Only $2\%$ are produced in the IGM.

Figure \ref{fig:bI_theory_comparison} compares our constraints on the luminosity density with the total luminosity density of the simulation of \citet{byrohl/nelson:2023} and its contributions from photons that last scattered in faint galaxies, in the CGM, and in the IGM.

The total mean luminosity density of the \citet{byrohl/nelson:2023} simulation is $4.4\times 10^{40}\,\mathrm{erg\, s^{-1}\, Mpc^{-3}}$ (filled circle). After scattering, $30\%$ of the luminosity density originates from galaxies within an $1\farcs 5$ aperture, $42\%$ of photons last scatter in the CGM (empty triangle), and $28\%$ of \Lya photons reach the observer from the IGM (empty diamond). 

Of the photons that reach us from galaxies, $6.1\times 10^{39}\,\mathrm{erg\, s^{-1}\, Mpc^{-3}}$ (empty star) are due to galaxies with a \Lya luminosity of $L < 10^{41.75}\,\mathrm{erg}\,\mathrm{s}^{-1}$, which is fainter than the typical HETDEX detection limit. Given this luminosity density, the photons that reach us from faint galaxies, the CGM, and the IGM would produce an intensity of $9\,(5/3)\times 10^{-22}\,\mathrm{erg\, s^{-1}\, cm^{-2}\, arcsec^{-2}}\,\angstrom^{-1}$ at redshift $z = 2.1 (2.6/3.2)$. Assuming that this value is the mean intensity of undetected sources, the intensity bias necessary for our measurement is $b_I \simeq 0.4\,(1.1/1.3)$. Note, however, that the approximation that $\bar{F}_{\rm RSD} = \bar{F}^{\rm fid}_{\rm RSD}$ breaks for values of $b_I$ that differ significantly from $b_I^{\rm fid}=2$.

We masked spectra containing continuum emission in our measurement, which may include galaxies with nonzero \Lya emission. When we include only emission that last scattered in the CGM or IGM, the necessary intensity bias to match the simulation results is $b_I \simeq 0.5$, 1.3, and 1.6 at $z = 2.1$,  2.6, and 3.2, respectively.

We can approximate the intensity bias $b_{I}$ from the relation between the luminosity density as a function of dark-matter halo mass. 
As shown in Figure 3 in \citet{byrohl/nelson:2023}, their simulation predicts a tight relationship between the \Lya luminosity and halo mass.  
From this relation, the intensity bias $b_{I}$ can be estimated as 
\begin{equation}
b_{I} = \frac{\int^{M_{\rm h,max}}_{M_{\rm h,min}}dM_{\rm h}\,Ldn/dM_{\rm h}\,b(M_{\rm h})}{\int ^{M_{\rm h,max}}_{M_{\rm h,min}}dM_{\rm h}\,Ldn/dM_{\rm h}}, 
\end{equation}
where $M_{\rm h}$ is the halo mass and we adopt a fitting formula from \citet{tinker/etal:2010} to compute the halo bias, $b(M_{\rm h})$. 
We obtain $b_{I}\simeq 1.4$ for $(\log_{10}(M_{\rm h,min}/M_{\odot}),\log_{10}(M_{\rm h,max}/M_{\odot}))=(9,12)$ and $b_{I}\simeq 1.6$ for $(\log_{10}(M_{\rm h,min}/M_{\odot}),\log_{10}(M_{\rm h,max}/M_{\odot}))=(9,\infty)$. 
This is a consequence of the intensity bias being weighted more toward small-mass halos given the relation in the simulation, which can be compared with $b_{I}\simeq 3$ assumed in \citet{croft/etal:2016}. 
Given the 10-20\% accuracy of the fitting formula in \citet{tinker/etal:2010}, it is reasonable to estimate that $1.5\lesssim b_{I}\lesssim 2$. 

\section{Discussion}
\label{sec:discussion}

\subsection{Possible Data Processing Improvements}

This work has demonstrated the feasibility of detecting the LAE-\Lya intensity cross-power spectrum in HETDEX data and provides a constraint on $b_\mathrm{g} b_I \langle I \rangle \bar{F}_{\rm RSD} / \bar{F}^{\rm fid}_{\rm RSD}$. We have used extensive masking and PCA to remove systematics from our results.  This resulted in significant signal loss. This subsection describes three ideas for improving the HETDEX LIM measurement in the future.

\begin{itemize}
\item Implementing a better full-frame sky subtraction technique would reduce systematics. This includes identifying low-level systematic contributions associated with individual fiber spectra or entire IFUs.  With these improvements, we would not have to mask vast regions of the observed HETDEX volume or apply the PCA.
\item  Using more robust weights for the intensity and galaxy catalogs could improve the statistical significance of the power spectrum detection.
\item The completeness model and luminosity function used in this work could be updated to better match the redshift distribution of LAEs.
\end{itemize}

\subsection{Possible Modeling Improvements}

Within the statistical uncertainty,
the measured LAE-\Lya intensity cross-power spectra are consistent with our mock based on a lognormal model \citep{agrawal/etal:2017,lujanniemeyer/bernal/komatsu:2023}.  This allows us to constrain $b_\mathrm{g} b_I \langle I_\mathrm{undet}\rangle \bar{F}_{\rm RSD} / \bar{F}^{\rm fid}_{\rm RSD}$. Fast lognormal mocks are therefore essential because they allow us to study many systematic effects in the data and enable us to faithfully forward-model the data.

However, future higher-fidelity measurements with lower statistical uncertainty will require more detailed models with better treatment of the relevant physics. Although cosmological simulations with \Lya RT are still computationally expensive, they are essential for a more accurate modeling of physics \citep[see, e.g.,][]{behrens/etal:2018,byrohl/etal:2019,gurung-lopez/etal:2021,byrohl/nelson:2023,khoraminezhad/etal:prep}.

While our current model includes RSD, the LAE luminosity function, and the HETDEX sensitivity and masking function, it is otherwise simple. For example, all galaxies have the same bias, luminosity function, and emission-line width, regardless of their environment. The model only includes \Lya emission sourced by galaxies expected from extrapolating the luminosity function of \citet{umeda/etal:2025} to faint luminosities.
Future models could include diffuse \Lya emission of the CGM and IGM, and the scattering of \Lya photons into or out of the LOS\null. These emissions source \Lya halos around galaxies, which have been found ubiquitously at $z>2$ \citep[e.g.,][]{steidel/etal:2011,wisotzki/etal:2018,lujanniemeyer/etal:2022a,lujanniemeyer/etal:2022b}. They could also model the \Lya absorption around LAEs, which we discuss further in Section \ref{subsec:lya_absorption}.

\subsection{Contamination from \texorpdfstring{[\ion{O}{2}]}{[OII]}-emitting Galaxies}

Despite thoroughly masking spectra with continuum emission and detected low-redshift sources, some interloper contamination may remain.  If present in both the galaxy catalog and the intensity map, the signal of these interlopers would contaminate the cross-power spectrum of the detected LAEs and the \Lya intensity measurement.

However, we believe that any contamination in the HETDEX LAE catalog is likely to be small. \citet{davis/etal:2023a} estimated that [\ion{O}{2}]-emitting galaxies make up only $\simeq 1.2\% \pm 0.1\%$ of the HETDEX LAE catalog, and contamination from other sources is only $\simeq 0.8\% \pm 0.1\%$.

\subsection{\texorpdfstring{\Lya}{Lyα} Absorption around LAEs}
\label{subsec:lya_absorption}

By stacking the spectra of tens of thousands of LAEs detected in HETDEX, \citet{davis/etal:2023b} and \citet{weiss/etal:2024} found broad, negative absorption troughs around the stacked \Lya emission lines. 

\citet{weiss/etal:2025} argued that these troughs are due to the extragalactic background light (EBL) that is absorbed in the CGM surrounding the LAEs. If HETDEX detects LAEs preferentially on the near side of overdensities -- for example, because less \Lya emission is absorbed on the way to Earth -- then the EBL is enhanced behind the LAEs and absorbed around the \Lya wavelength. The sky subtraction removes the mean EBL spectrum from all spectra, including those where the EBL is completely absorbed. Consequently, the absorption troughs become negative, explaining the data.

\citet{khanlari/etal:2025} stacked the spectra as a function of the angular distance from the LAEs and find \Lya absorption out to $\simeq 350\,\mathrm{kpc}$ (proper). The \Lya absorption at LAE positions and in their surroundings is stronger for sources with a smaller detection significance, and is not observed for LAEs with $\mathrm{SNR}>6$.

The expected effect of this absorption on our measurement is twofold. First, the absorption around LAEs can be modeled as part of the emission-line profile. This profile can be written as $\phi(\lambda) = \phi_\mathrm{em}(\lambda) + \phi_\mathrm{abs}(\lambda)$, where $\phi_\mathrm{em}(\lambda)$ is the shape of the emission line, and $\phi_\mathrm{abs}(\lambda)$ is the shape of the absorption feature (e.g., a Voigt profile). One can then model an intensity map from delta functions at the locations of point sources and convolve this map with $\phi(\lambda)$ \citep[see, e.g.,][]{lujanniemeyer/bernal/komatsu:2023}. This convolution dampens the power spectrum, $P(\mathbf{k})$, along the LOS\null. In this work, we have neglected $\phi_\mathrm{abs}(\lambda)$ and modeled $\phi(\lambda) \simeq \phi_\mathrm{em}(\lambda)$, including the LOS resolution of HETDEX\null. We have also not included continuum emission or the EBL in the model. The measured cross-power spectrum of detected LAEs with undetected \Lya intensity is consistent with the model without the absorption feature, suggesting that \Lya absorption found in \citet{davis/etal:2023b} and \citet{weiss/etal:2024} does not dominate the cross-correlation signal.

Second, we only considered sources with $\mathrm{SNR}>5.5$ as detected; thus, galaxies with lower detection significance and undetected galaxies contribute to the \Lya intensity map. If the intensity map were dominated by a continuum background that is absorbed in overdensities, the cross-correlation between the detected LAEs and the \Lya intensity map excluding detected LAEs would be negative. However, since we detect a positive signal, we conclude that \Lya emission dominates the intensity map.

More sophisticated modeling of and exploration of the absorption feature will be left for future work.

\newpage
\section{Summary and Conclusions}
\label{sec:summary}

In this paper, we presented a \Lya LIM power spectrum measurement in HETDEX\null. We reported the detection of the LAE-\Lya intensity cross-power spectrum of LAEs with SNR~$>5.5$ and the intensity of undetected sources in three redshift bins centered around $\bar{z}=2.1$, $2.6$, and $3.2$.

To accomplish this measurement, we thoroughly cleaned the spectral data to remove systematic contributions. We created self-consistent lognormal mocks for the LAEs and the \Lya intensity, including RSD, using the HETDEX window function, sensitivity, and noise \citep{lujanniemeyer/bernal/komatsu:2023,lujanniemeyer:2025}. We estimated the covariance matrix of the power spectra from the mocks and from the shuffled data.

By fitting the fiducial mock prediction times the overall amplitude to the measured power spectra, we constrained the product of the detected LAE bias, the intensity bias, the mean intensity of undetected sources, and the ratio of the real and fiducial RSD factors, $b_\mathrm{g} b_I \langle I \rangle \bar{F}_{\rm RSD} / \bar{F}_{\rm RSD}^{\rm fid}$.

Assuming a fiducial LAE bias $b_\mathrm{g}^\mathrm{fid} = 2$ and $\bar{F}_{\rm RSD} = \bar{F}_{\rm RSD}^{\rm fid}$, we inferred lower values of $b_I \langle I \rangle$ than those from the QSO-\Lya cross-correlations \citep{croft/etal:2016,croft/etal:2018,lin/etal:2022}. Our constraint, however, is slightly higher than the upper limit inferred from the \Lya forest-\Lya intensity cross-correlation measurement \citep{croft/etal:2018}.

Assuming fiducial LAE and intensity biases of $b_\mathrm{g}^\mathrm{fid} = 2$ and $b_I^\mathrm{fid} = 2$, and $\bar{F}_{\rm RSD} = \bar{F}_{\rm RSD}^{\rm fid}$, our constraints on the \Lya luminosity density are slightly larger than those obtained by integrating extrapolated LAE luminosity functions. Our constraints are on the same order of magnitude as the intensity of faint galaxies and of the CGM and IGM emission in the cosmological simulations with \Lya RT at $z=2$ \citep{byrohl/nelson:2023}. Our measured intensity is smaller than the total \Lya intensity expected from the star formation rate density at $z\simeq 2-3$, and is consistent with an escape fraction of \Lya photons from the ISM being $f_\mathrm{esc}<1$.
These results will be useful for constraining models of galaxy formation and evolution.

This work represents initial LIM results from the HETDEX data\null. We will improve the data processing and modeling of the signal in the future.

\section*{}

HETDEX is led by the University of Texas at Austin McDonald Observatory and Department of Astronomy with participation from the Ludwig-Maximilians-Universität München, Max-Planck-Institut für Extraterrestrische Physik (MPE), Leibniz-Institut für Astrophysik Potsdam (AIP), Texas A\&M University, Pennsylvania State University, Institut für Astrophysik Göttingen, The University of Oxford, Max-Planck-Institut für Astrophysik (MPA), The University of Tokyo and Missouri University of Science and Technology. In addition to Institutional support, HETDEX is funded by the National Science Foundation (grant AST-0926815), the State of Texas, the US Air Force (AFRL FA9451-04-2- 0355), and generous support from private individuals and foundations.

The observations were obtained with the Hobby-Eberly Telescope (HET), which is a joint project of the University of Texas at Austin, the Pennsylvania State University, Ludwig-Maximilians-Universität München, and Georg-August-Universität Göttingen. The HET is named in honor of its principal benefactors, William P. Hobby and Robert E. Eberly.

VIRUS is a joint project of the University of Texas at Austin,
Leibniz-Institut f\"ur Astrophysik Potsdam (AIP), Texas A\&M University
(TAMU), Max-Planck-Institut f\"ur Extraterrestrische Physik (MPE),
Ludwig-Maximilians-Universit\"at Muenchen, Pennsylvania State
University, Institut f\"ur Astrophysik G\"ottingen, University of Oxford,
and the Max-Planck-Institut f\"ur Astrophysik (MPA). In addition to
Institutional support, VIRUS was partially funded by the National
Science Foundation, the State of Texas, and generous support from
private individuals and foundations.

The authors acknowledge the Texas Advanced Computing Center (TACC) at The University of Texas at Austin for providing high-performance computing, visualization, and storage resources that have contributed to the research results reported within this paper. URL: \url{http://www.tacc.utexas.edu}.

This work was supported in part by the Deutsche Forschungsgemeinschaft (DFG, German Research Foundation) under Germany's Excellence Strategy---EXC-2094---390783311.

J.L.B. acknowledges funding from the Ramón y Cajal Grant RYC2021-033191-I, financed by MCIN/AEI/10.13039/501100011033 and by
the European Union ``NextGenerationEU''/PRTR, as well as the project UC-LIME (PID2022-140670NA-I00), financed by MCIN/AEI/10.13039/501100011033/FEDER, UE.

MJJ acknowledges the support of a UKRI Frontiers Research Grant [EP/X026639/1], which was selected by the European Research Council, and the STFC consolidated grants [ST/S000488/1] and [ST/W000903/1]. MJJ  also acknowledges support from the Oxford Hintze Centre for Astrophysical Surveys which is funded through generous support from the Hintze Family Charitable Foundation. 

The Institute for Gravitation and the Cosmos is supported by the Eberly College of Science and the Office of the Senior Vice President for Research at the Pennsylvania State University.

R.C. acknowledges support from the National Science Foundation under grant AST-2408358.
S.S. acknowledges support from the National Science Foundation under grants NSF-2219212 and NSF-2511145.


\vspace{5mm}

\software{astropy \citep{2013A&A...558A..33A,2018AJ....156..123A},  
            dustmaps \citep{green:2018},
            extinction \url{https://extinction.readthedocs.io/en/latest/}
             matplotlib \citep{matplotlib},
            numpy \citep{harris2020array},
            scipy \citep{2020SciPy-NMeth}
            }
            
\appendix

\section{Effect of Nonlinearity on the Power Spectrum}
\label{sec:effect_nonlinear_pk}

To see the effects of nonlinear corrections to the matter power spectrum, we generate a new set of mocks with a linear input matter power spectrum generated by CLASS, i.e., not using Halofit. Otherwise, we keep the same settings as the fiducial mocks described in Section \ref{sec:mocks}. 
In both cases, nonlinearity in RSD due to the Jacobian of the coordinate transformation from real space to redshift space is included \citep{agrawal/etal:2017}.
We also process the mock maps in the same way, including sky subtraction.

\begin{figure*}
    \includegraphics[width=\textwidth]{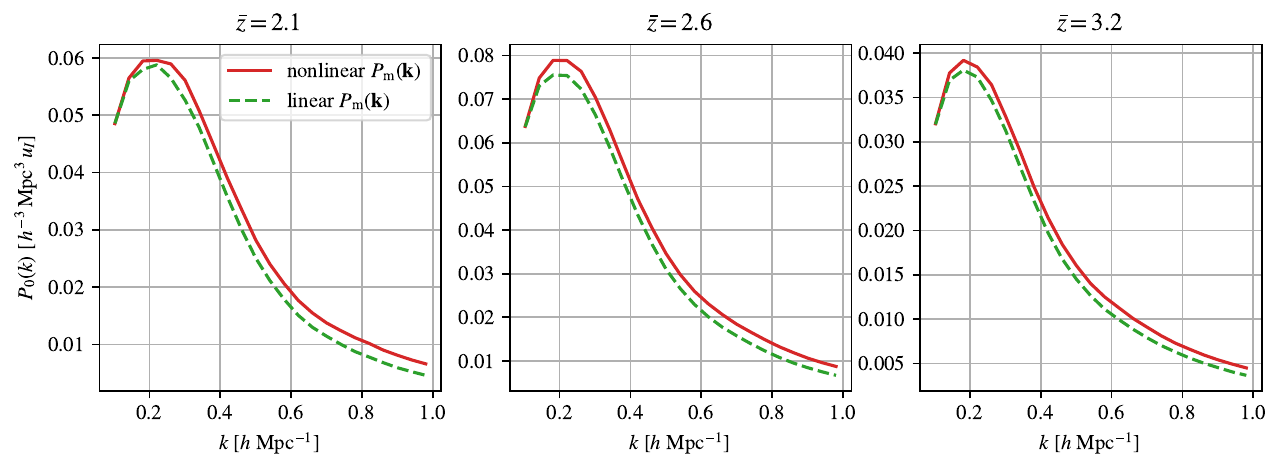}
    \caption{The red solid lines show the fiducial nonlinear mock cross-power spectra, while the green dashed lines show the linear mock cross-power spectra. These two power spectra are compared in the Spring field across three redshift bins. Here, $u_I = 10^{-18}\,\mathrm{erg\, s^{-1}\, cm^{-2}\, arcsec^{-2}\,}\angstrom^{-1}$.}
    \label{fig:mock_pk_linear}
\end{figure*}

Figure \ref{fig:mock_pk_linear} shows the LAE-\Lya intensity cross-power spectrum of the linear mock compared to that of the fiducial nonlinear mock in the Spring field. The power spectra are similar even at $k > 0.5\, h\, \mathrm{Mpc}^{-1}$ across all explored redshift bins from $z=1.88$ to $3.52$, as nonlinearity due to structure formation is mild at such high redshift \citep{jeong/komatsu:2006}.

\section{\texorpdfstring{$F_{\rm RSD}$}{FRSD} Including the \texorpdfstring{\Lya}{Lya} RT Effect}
\label{app:lya_rt_rsd}

When \Lya photons travel from a galaxy to an observer on Earth, they can be scattered out of the LOS by hydrogen atoms in the IGM, which can be modeled as an effective absorption. 
This reduces the observed \Lya luminosity of a galaxy, making it less likely to be detected, and the observed \Lya intensity in the intensity map.
This effect depends on the density and velocity gradient of the gas between the observed galaxy and us.
If a significant fraction of \Lya photons are subject to scattering in the IGM, this can affect the observed clustering statistics of LAEs and \Lya LIM \citep[e.g.,][]{zheng/etal:2011,behrens/niemeyer:2013,behrens/etal:2018}.
Specifically, it can change the anisotropy of the power spectrum in redshift space.
In a linear model for RSD and \Lya RT following \citet{wyithe/dijkstra:2011} and \citet{lujanniemeyer:2025}, the RSD factor reads
\begin{equation}
    F_{\rm RSD}(\mathbf{k}) = \left[b_I + b_\mathrm{ion} K_\lambda (k)C - c_\gamma C + (1-C) f\mu^2\right] 
    \left[b_\mathrm{g} + b_\mathrm{ion} K_\lambda(k) C^\mathrm{g_\alpha} -c_\gamma C^\mathrm{g_\alpha} + \left(1 - C^\mathrm{g_\alpha} \right) f \mu^2\right],
\end{equation}
where $b_I$ and $b_g$ are the intensity and LAE bias, respectively, $b_{\rm ion}$ is the bias of ionizing sources, and $c_\gamma \simeq 1.72$.
The smoothing kernel $K_\lambda (k) = \mathrm{arctan}(k\lambda_\mathrm{mfp})/\left(k\lambda_\mathrm{mfp}\right)$, with the mean free path of ionizing photons $\lambda_\mathrm{mfp}$, translates the locations of ionizing sources to a map of ionization rate fluctuations.
The constants $C$ and $C^\mathrm{g_\alpha}$ quantify the effect of the \Lya absorption in the IGM on the intensity fluctuation and the detected galaxy overdensity, respectively:
\begin{align}
    C &= \frac{F_\mathrm{abs} \tau_0 e^{-\tau_0}}{1 - F_\mathrm{abs} + F_\mathrm{abs} e^{-\tau_0}};\\
    C^\mathrm{g_\alpha} &= (\beta_\phi - 1) \frac{F_\mathrm{abs}\tau_0 e^{-\tau_0}}{1 - F_\mathrm{abs} + F_\mathrm{abs} e^{-\tau_0}}.
\end{align}
Here, $F_{\rm abs} \in [0,1]$ is the fraction of \Lya photons that can be absorbed in the IGM. The fraction $1 - F_{\rm abs}$ travels to the observer unobstructed, for example, because they have redshifted out of resonance before escaping the galaxy.
$\tau_0$ is the mean effective optical depth of the IGM; $\beta_\phi = -\alpha$ is $-1$ times the faint-end slope of the \Lya luminosity function.
Note that the linear model for \Lya RT is only valid for small overdensities and breaks down in the nonlinear environment around galaxies.

\section{Correlation Matrices}
\label{sec:correlation_matrices}

\begin{figure}
    \centering
    \includegraphics[width=\textwidth]{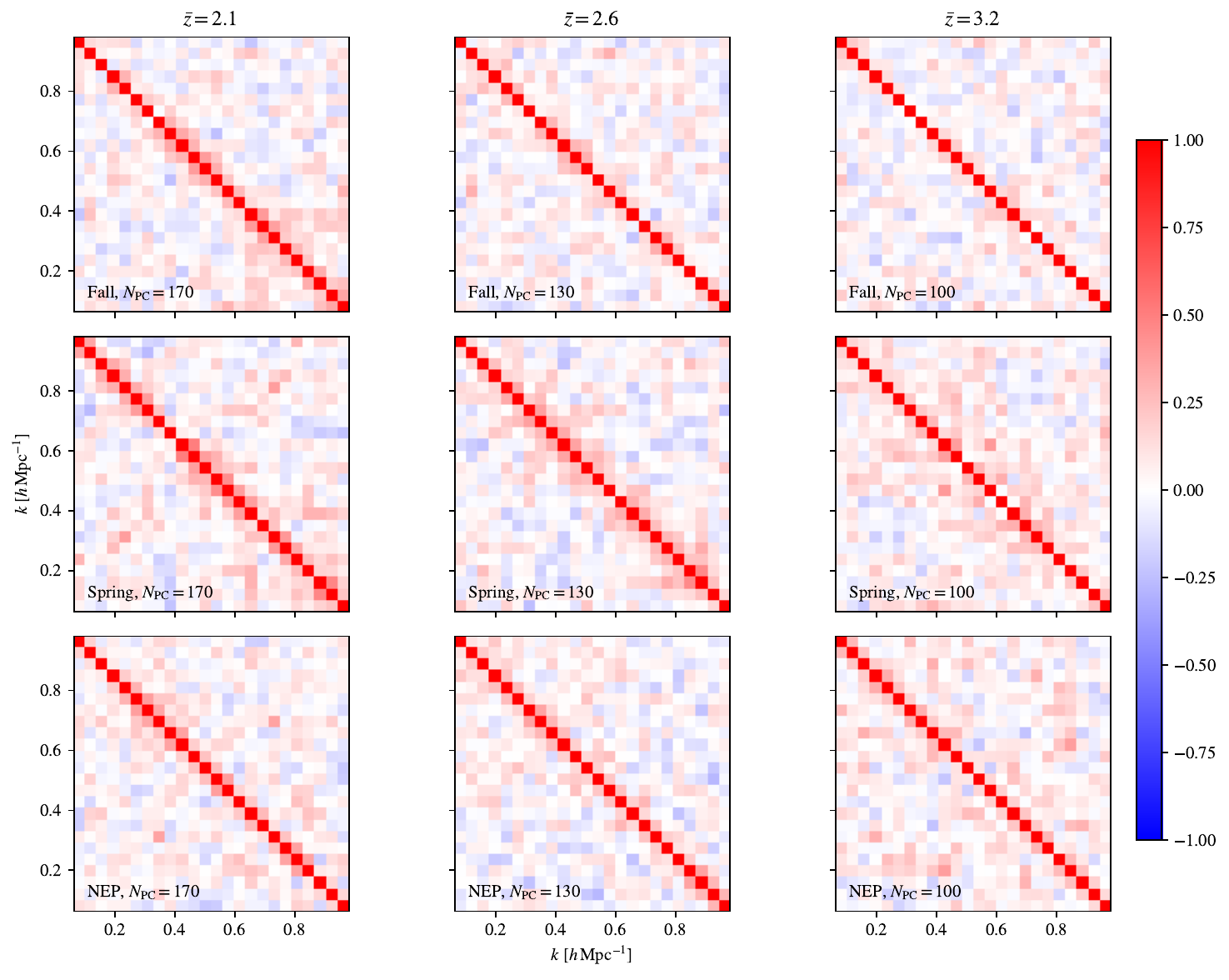}
    \caption{Correlation matrices of the power spectrum monopoles of Figure \ref{fig:pk_mock_data_allfields}, in the same order.}
    \label{fig:correlation_matrices}
\end{figure}

We calculate the correlation matrices $M_{ij}$ of the power spectrum monopoles shown in Figure \ref{fig:pk_mock_data_allfields} as $M_{ij} = C_{ij} / \sqrt{C_{ii} C_{jj}}$, where $C_{ij}$ is the total covariance matrix described in Section \ref{subsec:covariance_and_weights}.
Figure \ref{fig:correlation_matrices} shows the correlation matrices. They are mostly diagonal.

\bibliography{sample631}{}

@ARTICLE{chang/etal:2010,
       author = {{Chang}, Tzu-Ching and {Pen}, Ue-Li and {Bandura}, Kevin and {Peterson}, Jeffrey B.},
        title = "{An intensity map of hydrogen 21-cm emission at redshift z\raisebox{-0.5ex}\textasciitilde0.8}",
      journal = {\nat},
         year = 2010,
        month = jul,
       volume = {466},
       number = {7305},
        pages = {463-465},
          doi = {10.1038/nature09187},
       adsurl = {https://ui.adsabs.harvard.edu/abs/2010Natur.466..463C}
}

@ARTICLE{khoraminezhad/etal:prep,
       author = {{Khoraminezhad}, Hasti and {Saito}, Shun and {Gronke}, Max and {Byrohl}, Chris},
        title = "{Simulating realistic Lyman-$α$ emitting galaxies including the effect of radiative transfer}",
      journal = {arXiv e-prints},
         year = 2025,
        month = jul,
          eid = {arXiv:2507.16707},
        pages = {arXiv:2507.16707},
          doi = {10.48550/arXiv.2507.16707},
archivePrefix = {arXiv},
       eprint = {2507.16707},
 primaryClass = {astro-ph.GA},
       adsurl = {https://ui.adsabs.harvard.edu/abs/2025arXiv250716707K}
}

@ARTICLE{jeong/komatsu:2006,
       author = {{Jeong}, Donghui and {Komatsu}, Eiichiro},
        title = "{Perturbation Theory Reloaded: Analytical Calculation of Nonlinearity in Baryonic Oscillations in the Real-Space Matter Power Spectrum}",
      journal = {\apj},
         year = 2006,
        month = nov,
       volume = {651},
       number = {2},
        pages = {619-626},
          doi = {10.1086/507781},
archivePrefix = {arXiv},
       eprint = {astro-ph/0604075},
 primaryClass = {astro-ph},
       adsurl = {https://ui.adsabs.harvard.edu/abs/2006ApJ...651..619J}
}

@ARTICLE{takahashi/etal:2012,
       author = {{Takahashi}, Ryuichi and {Sato}, Masanori and {Nishimichi}, Takahiro and {Taruya}, Atsushi and {Oguri}, Masamune},
        title = "{Revising the Halofit Model for the Nonlinear Matter Power Spectrum}",
      journal = {\apj},
         year = 2012,
        month = dec,
       volume = {761},
       number = {2},
          eid = {152},
        pages = {152},
          doi = {10.1088/0004-637X/761/2/152},
archivePrefix = {arXiv},
       eprint = {1208.2701},
 primaryClass = {astro-ph.CO},
       adsurl = {https://ui.adsabs.harvard.edu/abs/2012ApJ...761..152T}
}

@ARTICLE{bird/etal:2012,
       author = {{Bird}, Simeon and {Viel}, Matteo and {Haehnelt}, Martin G.},
        title = "{Massive neutrinos and the non-linear matter power spectrum}",
      journal = {\mnras},
         year = 2012,
        month = mar,
       volume = {420},
       number = {3},
        pages = {2551-2561},
          doi = {10.1111/j.1365-2966.2011.20222.x},
archivePrefix = {arXiv},
       eprint = {1109.4416},
 primaryClass = {astro-ph.CO},
       adsurl = {https://ui.adsabs.harvard.edu/abs/2012MNRAS.420.2551B}
}

@ARTICLE{2018AJ....156..123A,
       author = {{Astropy Collaboration} and {Price-Whelan}, A.~M. and {Sip{\H{o}}cz}, B.~M. and {G{\"u}nther}, H.~M. and {Lim}, P.~L. and {Crawford}, S.~M. and {Conseil}, S. and {Shupe}, D.~L. and {Craig}, M.~W. and {Dencheva}, N. and {Ginsburg}, A. and {VanderPlas}, J.~T. and {Bradley}, L.~D. and {P{\'e}rez-Su{\'a}rez}, D. and {de Val-Borro}, M. and {Aldcroft}, T.~L. and {Cruz}, K.~L. and {Robitaille}, T.~P. and {Tollerud}, E.~J. and {Ardelean}, C. and {Babej}, T. and {Bach}, Y.~P. and {Bachetti}, M. and {Bakanov}, A.~V. and {Bamford}, S.~P. and {Barentsen}, G. and {Barmby}, P. and {Baumbach}, A. and {Berry}, K.~L. and {Biscani}, F. and {Boquien}, M. and {Bostroem}, K.~A. and {Bouma}, L.~G. and {Brammer}, G.~B. and {Bray}, E.~M. and {Breytenbach}, H. and {Buddelmeijer}, H. and {Burke}, D.~J. and {Calderone}, G. and {Cano Rodr{\'\i}guez}, J.~L. and {Cara}, M. and {Cardoso}, J.~V.~M. and {Cheedella}, S. and {Copin}, Y. and {Corrales}, L. and {Crichton}, D. and {D'Avella}, D. and {Deil}, C. and {Depagne}, {\'E}. and {Dietrich}, J.~P. and {Donath}, A. and {Droettboom}, M. and {Earl}, N. and {Erben}, T. and {Fabbro}, S. and {Ferreira}, L.~A. and {Finethy}, T. and {Fox}, R.~T. and {Garrison}, L.~H. and {Gibbons}, S.~L.~J. and {Goldstein}, D.~A. and {Gommers}, R. and {Greco}, J.~P. and {Greenfield}, P. and {Groener}, A.~M. and {Grollier}, F. and {Hagen}, A. and {Hirst}, P. and {Homeier}, D. and {Horton}, A.~J. and {Hosseinzadeh}, G. and {Hu}, L. and {Hunkeler}, J.~S. and {Ivezi{\'c}}, {\v{Z}}. and {Jain}, A. and {Jenness}, T. and {Kanarek}, G. and {Kendrew}, S. and {Kern}, N.~S. and {Kerzendorf}, W.~E. and {Khvalko}, A. and {King}, J. and {Kirkby}, D. and {Kulkarni}, A.~M. and {Kumar}, A. and {Lee}, A. and {Lenz}, D. and {Littlefair}, S.~P. and {Ma}, Z. and {Macleod}, D.~M. and {Mastropietro}, M. and {McCully}, C. and {Montagnac}, S. and {Morris}, B.~M. and {Mueller}, M. and {Mumford}, S.~J. and {Muna}, D. and {Murphy}, N.~A. and {Nelson}, S. and {Nguyen}, G.~H. and {Ninan}, J.~P. and {N{\"o}the}, M. and {Ogaz}, S. and {Oh}, S. and {Parejko}, J.~K. and {Parley}, N. and {Pascual}, S. and {Patil}, R. and {Patil}, A.~A. and {Plunkett}, A.~L. and {Prochaska}, J.~X. and {Rastogi}, T. and {Reddy Janga}, V. and {Sabater}, J. and {Sakurikar}, P. and {Seifert}, M. and {Sherbert}, L.~E. and {Sherwood-Taylor}, H. and {Shih}, A.~Y. and {Sick}, J. and {Silbiger}, M.~T. and {Singanamalla}, S. and {Singer}, L.~P. and {Sladen}, P.~H. and {Sooley}, K.~A. and {Sornarajah}, S. and {Streicher}, O. and {Teuben}, P. and {Thomas}, S.~W. and {Tremblay}, G.~R. and {Turner}, J.~E.~H. and {Terr{\'o}n}, V. and {van Kerkwijk}, M.~H. and {de la Vega}, A. and {Watkins}, L.~L. and {Weaver}, B.~A. and {Whitmore}, J.~B. and {Woillez}, J. and {Zabalza}, V. and {Astropy Contributors}},
        title = "{The Astropy Project: Building an Open-science Project and Status of the v2.0 Core Package}",
      journal = {\aj},
         year = 2018,
        month = sep,
       volume = {156},
       number = {3},
          eid = {123},
        pages = {123},
          doi = {10.3847/1538-3881/aabc4f},
archivePrefix = {arXiv},
       eprint = {1801.02634},
 primaryClass = {astro-ph.IM},
       adsurl = {https://ui.adsabs.harvard.edu/abs/2018AJ....156..123A}
}

@ARTICLE{2013A&A...558A..33A,
       author = {{Astropy Collaboration} and {Robitaille}, Thomas P. and
         {Tollerud}, Erik J. and {Greenfield}, Perry and {Droettboom}, Michael and
         {Bray}, Erik and {Aldcroft}, Tom and {Davis}, Matt and
         {Ginsburg}, Adam and {Price-Whelan}, Adrian M. and
         {Kerzendorf}, Wolfgang E. and {Conley}, Alexander and {Crighton}, Neil and
         {Barbary}, Kyle and {Muna}, Demitri and {Ferguson}, Henry and
         {Grollier}, Fr{\'e}d{\'e}ric and {Parikh}, Madhura M. and
         {Nair}, Prasanth H. and {Unther}, Hans M. and {Deil}, Christoph and
         {Woillez}, Julien and {Conseil}, Simon and {Kramer}, Roban and
         {Turner}, James E.~H. and {Singer}, Leo and {Fox}, Ryan and
         {Weaver}, Benjamin A. and {Zabalza}, Victor and {Edwards}, Zachary I. and
         {Azalee Bostroem}, K. and {Burke}, D.~J. and {Casey}, Andrew R. and
         {Crawford}, Steven M. and {Dencheva}, Nadia and {Ely}, Justin and
         {Jenness}, Tim and {Labrie}, Kathleen and {Lim}, Pey Lian and
         {Pierfederici}, Francesco and {Pontzen}, Andrew and {Ptak}, Andy and
         {Refsdal}, Brian and {Servillat}, Mathieu and {Streicher}, Ole},
        title = "{Astropy: A community Python package for astronomy}",
      journal = {\aap},
         year = "2013",
        month = "Oct",
       volume = {558},
          eid = {A33},
        pages = {A33},
          doi = {10.1051/0004-6361/201322068},
archivePrefix = {arXiv},
       eprint = {1307.6212},
 primaryClass = {astro-ph.IM},
       adsurl = {https://ui.adsabs.harvard.edu/abs/2013A&A...558A..33A}
}

@ARTICLE{lujanniemeyer:2025,
       author = {{Lujan Niemeyer}, Maja},
        title = "{Effect of Ly{\ensuremath{\alpha}} Radiative Transfer on Intensity Mapping Power Spectra}",
      journal = {\apj},
         year = 2025,
        month = feb,
       volume = {980},
       number = {2},
          eid = {250},
        pages = {250},
          doi = {10.3847/1538-4357/ada8a3},
archivePrefix = {arXiv},
       eprint = {2407.03060},
 primaryClass = {astro-ph.CO},
       adsurl = {https://ui.adsabs.harvard.edu/abs/2025ApJ...980..250L}
}

@ARTICLE{lujanniemeyer/bernal/komatsu:2023,
       author = {{Lujan Niemeyer}, Maja and {Bernal}, Jos{\'e} Luis and {Komatsu}, Eiichiro},
        title = "{SIMPLE: Simple Intensity Map Producer for Line Emission}",
      journal = {\apj},
         year = 2023,
        month = nov,
       volume = {958},
       number = {1},
          eid = {4},
        pages = {4},
          doi = {10.3847/1538-4357/acfef4},
archivePrefix = {arXiv},
       eprint = {2307.08475},
 primaryClass = {astro-ph.CO},
       adsurl = {https://ui.adsabs.harvard.edu/abs/2023ApJ...958....4L}
}

@ARTICLE{lujanniemeyer/etal:2022b,
       author = {{Lujan Niemeyer}, Maja and {Bowman}, William P. and {Ciardullo}, Robin and {Gronke}, Max and {Komatsu}, Eiichiro and {Fabricius}, Maximilian and {Farrow}, Daniel J. and {Finkelstein}, Steven L. and {Gebhardt}, Karl and {Gronwall}, Caryl and {Hill}, Gary J. and {Liu}, Chenxu and {Cooper}, Erin Mentuch and {Schneider}, Donald P. and {Tuttle}, Sarah and {Zeimann}, Gregory R.},
        title = "{Ly{\ensuremath{\alpha}} Halos around [O III]-selected Galaxies in HETDEX}",
      journal = {\apjl},
         year = 2022,
        month = aug,
       volume = {934},
       number = {2},
          eid = {L26},
        pages = {L26},
          doi = {10.3847/2041-8213/ac82e5},
archivePrefix = {arXiv},
       eprint = {2207.11098},
 primaryClass = {astro-ph.GA},
       adsurl = {https://ui.adsabs.harvard.edu/abs/2022ApJ...934L..26L}
}

@ARTICLE{lujanniemeyer/etal:2022a,
       author = {{Lujan Niemeyer}, Maja and {Komatsu}, Eiichiro and {Byrohl}, Chris and {Davis}, Dustin and {Fabricius}, Maximilian and {Gebhardt}, Karl and {Hill}, Gary J. and {Wisotzki}, Lutz and {Bowman}, William P. and {Ciardullo}, Robin and {Farrow}, Daniel J. and {Finkelstein}, Steven L. and {Gawiser}, Eric and {Gronwall}, Caryl and {Jeong}, Donghui and {Landriau}, Martin and {Liu}, Chenxu and {Cooper}, Erin Mentuch and {Ouchi}, Masami and {Schneider}, Donald P. and {Zeimann}, Gregory R.},
        title = "{Surface Brightness Profile of Lyman-{\ensuremath{\alpha}} Halos out to 320 kpc in HETDEX}",
      journal = {\apj},
         year = 2022,
        month = apr,
       volume = {929},
       number = {1},
          eid = {90},
        pages = {90},
          doi = {10.3847/1538-4357/ac5cb8},
archivePrefix = {arXiv},
       eprint = {2203.04826},
 primaryClass = {astro-ph.GA},
       adsurl = {https://ui.adsabs.harvard.edu/abs/2022ApJ...929...90L}
}

@ARTICLE{gebhardt/etal:2021,
       author = {{Gebhardt}, Karl and {Mentuch Cooper}, Erin and {Ciardullo}, Robin and {Acquaviva}, Viviana and {Bender}, Ralf and {Bowman}, William P. and {Castanheira}, Barbara G. and {Dalton}, Gavin and {Davis}, Dustin and {de Jong}, Roelof S. and {DePoy}, D.~L. and {Devarakonda}, Yaswant and {Dongsheng}, Sun and {Drory}, Niv and {Fabricius}, Maximilian and {Farrow}, Daniel J. and {Feldmeier}, John and {Finkelstein}, Steven L. and {Froning}, Cynthia S. and {Gawiser}, Eric and {Gronwall}, Caryl and {Herold}, Laura and {Hill}, Gary J. and {Hopp}, Ulrich and {House}, Lindsay R. and {Janowiecki}, Steven and {Jarvis}, Matthew and {Jeong}, Donghui and {Jogee}, Shardha and {Kakuma}, Ryota and {Kelz}, Andreas and {Kollatschny}, W. and {Komatsu}, Eiichiro and {Krumpe}, Mirko and {Landriau}, Martin and {Liu}, Chenxu and {Niemeyer}, Maja Lujan and {MacQueen}, Phillip and {Marshall}, Jennifer and {Mawatari}, Ken and {McLinden}, Emily M. and {Mukae}, Shiro and {Nagaraj}, Gautam and {Ono}, Yoshiaki and {Ouchi}, Masami and {Papovich}, Casey and {Sakai}, Nao and {Saito}, Shun and {Schneider}, Donald P. and {Schulze}, Andreas and {Shanmugasundararaj}, Khavvia and {Shetrone}, Matthew and {Sneden}, Chris and {Snigula}, Jan and {Steinmetz}, Matthias and {Thomas}, Benjamin P. and {Thomas}, Brianna and {Tuttle}, Sarah and {Urrutia}, Tanya and {Wisotzki}, Lutz and {Wold}, Isak and {Zeimann}, Gregory and {Zhang}, Yechi},
        title = "{The Hobby-Eberly Telescope Dark Energy Experiment (HETDEX) Survey Design, Reductions, and Detections}",
      journal = {\apj},
         year = 2021,
        month = dec,
       volume = {923},
       number = {2},
          eid = {217},
        pages = {217},
          doi = {10.3847/1538-4357/ac2e03},
archivePrefix = {arXiv},
       eprint = {2110.04298},
 primaryClass = {astro-ph.IM},
       adsurl = {https://ui.adsabs.harvard.edu/abs/2021ApJ...923..217G}
}

@ARTICLE{mentuchcooper/etal:2023,
       author = {{Mentuch Cooper}, Erin and {Gebhardt}, Karl and {Davis}, Dustin and {Farrow}, Daniel J. and {Liu}, Chenxu and {Zeimann}, Gregory and {Ciardullo}, Robin and {Feldmeier}, John J. and {Drory}, Niv and {Jeong}, Donghui and {Benda}, Barbara and {Bowman}, William P. and {Boylan-Kolchin}, Michael and {Ch{\'a}vez Ortiz}, {\'O}scar A. and {Debski}, Maya H. and {Dentler}, Mona and {Fabricius}, Maximilian and {Farooq}, Rameen and {Finkelstein}, Steven L. and {Gawiser}, Eric and {Gronwall}, Caryl and {Hill}, Gary J. and {Hopp}, Ulrich and {House}, Lindsay R. and {Janowiecki}, Steven and {Khoraminezhad}, Hasti and {Kollatschny}, Wolfram and {Komatsu}, Eiichiro and {Landriau}, Martin and {Niemeyer}, Maja Lujan and {Lee}, Hanshin and {MacQueen}, Phillip and {Mawatari}, Ken and {McKay}, Brianna and {Ouchi}, Masami and {Poppe}, Jennifer and {Saito}, Shun and {Schneider}, Donald P. and {Snigula}, Jan and {Thomas}, Benjamin P. and {Tuttle}, Sarah and {Urrutia}, Tanya and {Weiss}, Laurel and {Wisotzki}, Lutz and {Zhang}, Yechi and {HETDEX Collaboration}},
        title = "{HETDEX Public Source Catalog 1: 220 K Sources Including Over 50 K Ly{\ensuremath{\alpha}} Emitters from an Untargeted Wide-area Spectroscopic Survey}",
      journal = {\apj},
         year = 2023,
        month = feb,
       volume = {943},
       number = {2},
          eid = {177},
        pages = {177},
          doi = {10.3847/1538-4357/aca962},
archivePrefix = {arXiv},
       eprint = {2301.01826},
 primaryClass = {astro-ph.GA},
       adsurl = {https://ui.adsabs.harvard.edu/abs/2023ApJ...943..177M}
}

@ARTICLE{weiss/etal:2024,
       author = {{Weiss}, Laurel H. and {Davis}, Dustin and {Gebhardt}, Karl and {Gazagnes}, Simon and {Mirza Khanlari}, Mahan and {Mentuch Cooper}, Erin and {Chisholm}, John and {Berg}, Danielle and {Bowman}, William P. and {Byrohl}, Chris and {Ciardullo}, Robin and {Fabricius}, Maximilian and {Farrow}, Daniel and {Gronwall}, Caryl and {Hill}, Gary J. and {House}, Lindsay R. and {Jeong}, Donghui and {Khoraminezhad}, Hasti and {Kollatschny}, Wolfram and {Komatsu}, Eiichiro and {Lujan Niemeyer}, Maja and {Saito}, Shun and {Schneider}, Donald P. and {Zeimann}, Gregory R.},
        title = "{Absorption Troughs of Ly{\ensuremath{\alpha}} Emitters in HETDEX}",
      journal = {\apj},
         year = 2024,
        month = feb,
       volume = {962},
       number = {2},
          eid = {102},
        pages = {102},
          doi = {10.3847/1538-4357/ad1b51},
archivePrefix = {arXiv},
       eprint = {2401.02490},
 primaryClass = {astro-ph.GA},
       adsurl = {https://ui.adsabs.harvard.edu/abs/2024ApJ...962..102W}
}

@ARTICLE{davis/etal:2023b,
       author = {{Davis}, Dustin and {Gebhardt}, Karl and {Cooper}, Erin Mentuch and {Bowman}, William P. and {Garcia Castanheira}, Barbara and {Chisholm}, John and {Ciardullo}, Robin and {Fabricius}, Maximilian and {Farrow}, Daniel J. and {Finkelstein}, Steven L. and {Gronwall}, Caryl and {Gawiser}, Eric and {Hill}, Gary J. and {Hopp}, Ulrich and {House}, Lindsay R. and {Jeong}, Donghui and {Kollatschny}, Wolfram and {Komatsu}, Eiichiro and {Liu}, Chenxu and {Niemeyer}, Maja Lujan and {Saldana-Lopez}, Alberto and {Saito}, Shun and {Schneider}, Donald P. and {Snigula}, Jan and {Tuttle}, Sarah and {Weiss}, Laurel H. and {Wisotzki}, Lutz and {Zeimann}, Gregory},
        title = "{HETDEX Public Source Catalog 1-Stacking 50,000 Lyman Alpha Emitters}",
      journal = {\apj},
         year = 2023,
        month = sep,
       volume = {954},
       number = {2},
          eid = {209},
        pages = {209},
          doi = {10.3847/1538-4357/ace4c2},
archivePrefix = {arXiv},
       eprint = {2307.03096},
 primaryClass = {astro-ph.GA},
       adsurl = {https://ui.adsabs.harvard.edu/abs/2023ApJ...954..209D}
}

@ARTICLE{davis/etal:2023a,
       author = {{Davis}, Dustin and {Gebhardt}, Karl and {Cooper}, Erin Mentuch and {Ciardullo}, Robin and {Fabricius}, Maximilian and {Farrow}, Daniel J. and {Feldmeier}, John J. and {Finkelstein}, Steven L. and {Gawiser}, Eric and {Gronwall}, Caryl and {Hill}, Gary J. and {Hopp}, Ulrich and {House}, Lindsay R. and {Jeong}, Donghui and {Kollatschny}, Wolfram and {Komatsu}, Eiichiro and {Landriau}, Martin and {Liu}, Chenxu and {Saito}, Shun and {Tuttle}, Sarah and {Wold}, Isak G.~B. and {Zeimann}, Gregory R. and {Zhang}, Yechi},
        title = "{The HETDEX Survey Emission-line Exploration and Source Classification}",
      journal = {\apj},
         year = 2023,
        month = apr,
       volume = {946},
       number = {2},
          eid = {86},
        pages = {86},
          doi = {10.3847/1538-4357/acb0ca},
archivePrefix = {arXiv},
       eprint = {2301.01799},
 primaryClass = {astro-ph.GA},
       adsurl = {https://ui.adsabs.harvard.edu/abs/2023ApJ...946...86D}
}

@ARTICLE{hill/etal:2021,
       author = {{Hill}, Gary J. and {Lee}, Hanshin and {MacQueen}, Phillip J. and {Kelz}, Andreas and {Drory}, Niv and {Vattiat}, Brian L. and {Good}, John M. and {Ramsey}, Jason and {Kriel}, Herman and {Peterson}, Trent and {DePoy}, D.~L. and {Gebhardt}, Karl and {Marshall}, J.~L. and {Tuttle}, Sarah E. and {Bauer}, Svend M. and {Chonis}, Taylor S. and {Fabricius}, Maximilian H. and {Froning}, Cynthia and {H{\"a}user}, Marco and {Indahl}, Briana L. and {Jahn}, Thomas and {Landriau}, Martin and {Leck}, Ron and {Montesano}, Francesco and {Prochaska}, Travis and {Snigula}, Jan M. and {Zeimann}, Greg and {Bryant}, Randy and {Damm}, George and {Fowler}, J.~R. and {Janowiecki}, Steven and {Martin}, Jerry and {Mrozinski}, Emily and {Odewahn}, Stephen and {Rostopchin}, Sergey and {Shetrone}, Matthew and {Spencer}, Renny and {Mentuch Cooper}, Erin and {Armandroff}, Taft and {Bender}, Ralf and {Dalton}, Gavin and {Hopp}, Ulrich and {Komatsu}, Eiichiro and {Nicklas}, Harald and {Ramsey}, Lawrence W. and {Roth}, Martin M. and {Schneider}, Donald P. and {Sneden}, Chris and {Steinmetz}, Matthias},
        title = "{The HETDEX Instrumentation: Hobby-Eberly Telescope Wide-field Upgrade and VIRUS}",
      journal = {\aj},
         year = 2021,
        month = dec,
       volume = {162},
       number = {6},
          eid = {298},
        pages = {298},
          doi = {10.3847/1538-3881/ac2c02},
archivePrefix = {arXiv},
       eprint = {2110.03843},
 primaryClass = {astro-ph.IM},
       adsurl = {https://ui.adsabs.harvard.edu/abs/2021AJ....162..298H}
}

@ARTICLE{lin/etal:2022,
       author = {{Lin}, Xiaojing and {Zheng}, Zheng and {Cai}, Zheng},
        title = "{Probing the Diffuse Ly{\ensuremath{\alpha}} Emission on Cosmological Scales: Ly{\ensuremath{\alpha}} Emission Intensity Mapping Using the Complete SDSS-IV eBOSS}",
      journal = {\apjs},
         year = 2022,
        month = oct,
       volume = {262},
       number = {2},
          eid = {38},
        pages = {38},
          doi = {10.3847/1538-4365/ac82e8},
archivePrefix = {arXiv},
       eprint = {2207.10682},
 primaryClass = {astro-ph.GA},
       adsurl = {https://ui.adsabs.harvard.edu/abs/2022ApJS..262...38L}
}

@ARTICLE{croft/etal:2016,
       author = {{Croft}, Rupert A.~C. and {Miralda-Escud{\'e}}, Jordi and {Zheng}, Zheng and {Bolton}, Adam and {Dawson}, Kyle S. and {Peterson}, Jeffrey B. and {York}, Donald G. and {Eisenstein}, Daniel and {Brinkmann}, Jon and {Brownstein}, Joel and {Cen}, Renyue and {Delubac}, Timoth{\'e}e and {Font-Ribera}, Andreu and {Hamilton}, Jean-Christophe and {Lee}, Khee-Gan and {Myers}, Adam and {Palanque-Delabrouille}, Nathalie and {P{\^a}ris}, Isabelle and {Petitjean}, Patrick and {Pieri}, Matthew M. and {Ross}, Nicholas P. and {Rossi}, Graziano and {Schlegel}, David J. and {Schneider}, Donald P. and {Slosar}, An{\v{z}}e and {Vazquez}, Jos{\'e} and {Viel}, Matteo and {Weinberg}, David H. and {Y{\`e}che}, Christophe},
        title = "{Large-scale clustering of Lyman {\ensuremath{\alpha}} emission intensity from SDSS/BOSS}",
      journal = {\mnras},
         year = 2016,
        month = apr,
       volume = {457},
       number = {4},
        pages = {3541-3572},
          doi = {10.1093/mnras/stw204},
archivePrefix = {arXiv},
       eprint = {1504.04088},
 primaryClass = {astro-ph.CO},
       adsurl = {https://ui.adsabs.harvard.edu/abs/2016MNRAS.457.3541C}
}

@ARTICLE{croft/etal:2018,
       author = {{Croft}, Rupert A.~C. and {Miralda-Escud{\'e}}, Jordi and {Zheng}, Zheng and {Blomqvist}, Michael and {Pieri}, Matthew},
        title = "{Intensity mapping with SDSS/BOSS Lyman-{\ensuremath{\alpha}} emission, quasars, and their Lyman-{\ensuremath{\alpha}} forest}",
      journal = {\mnras},
         year = 2018,
        month = nov,
       volume = {481},
       number = {1},
        pages = {1320-1336},
          doi = {10.1093/mnras/sty2302},
archivePrefix = {arXiv},
       eprint = {1806.06050},
 primaryClass = {astro-ph.CO},
       adsurl = {https://ui.adsabs.harvard.edu/abs/2018MNRAS.481.1320C}
}

@INPROCEEDINGS{ramsey/etal:1998,
       author = {{Ramsey}, Lawrence W. and {Adams}, M.~T. and {Barnes}, Thomas G. and {Booth}, John A. and {Cornell}, Mark E. and {Fowler}, James R. and {Gaffney}, Niall I. and {Glaspey}, John W. and {Good}, John M. and {Hill}, Gary J. and {Kelton}, Philip W. and {Krabbendam}, Victor L. and {Long}, L. and {MacQueen}, Phillip J. and {Ray}, Frank B. and {Ricklefs}, Randall L. and {Sage}, J. and {Sebring}, Thomas A. and {Spiesman}, W.~J. and {Steiner}, M.},
        title = "{Early performance and present status of the Hobby-Eberly Telescope}",
    booktitle = {Advanced Technology Optical/IR Telescopes VI},
         year = 1998,
       editor = {{Stepp}, Larry M.},
       series = {Society of Photo-Optical Instrumentation Engineers (SPIE) Conference Series},
       volume = {3352},
        month = aug,
        pages = {34-42},
          doi = {10.1117/12.319287},
       adsurl = {https://ui.adsabs.harvard.edu/abs/1998SPIE.3352...34R}
}

@ARTICLE{cunnington/etal:2023a,
       author = {{Cunnington}, Steven and {Li}, Yichao and {Santos}, Mario G. and {Wang}, Jingying and {Carucci}, Isabella P. and {Irfan}, Melis O. and {Pourtsidou}, Alkistis and {Spinelli}, Marta and {Wolz}, Laura and {Soares}, Paula S. and {Blake}, Chris and {Bull}, Philip and {Engelbrecht}, Brandon and {Fonseca}, Jos{\'e} and {Grainge}, Keith and {Ma}, Yin-Zhe},
        title = "{H I intensity mapping with MeerKAT: power spectrum detection in cross-correlation with WiggleZ galaxies}",
      journal = {\mnras},
         year = 2023,
        month = feb,
       volume = {518},
       number = {4},
        pages = {6262-6272},
          doi = {10.1093/mnras/stac3060},
archivePrefix = {arXiv},
       eprint = {2206.01579},
 primaryClass = {astro-ph.CO},
       adsurl = {https://ui.adsabs.harvard.edu/abs/2023MNRAS.518.6262C}
}

@ARTICLE{kovetz/etal:2017,
       author = {{Kovetz}, Ely D. and {Viero}, Marco P. and {Lidz}, Adam and {Newburgh}, Laura and {Rahman}, Mubdi and {Switzer}, Eric and {Kamionkowski}, Marc and {Aguirre}, James and {Alvarez}, Marcelo and {Bock}, James and {Bond}, J. Richard and {Bower}, Goeffry and {Bradford}, C. Matt and {Breysse}, Patrick C. and {Bull}, Philip and {Chang}, Tzu-Ching and {Cheng}, Yun-Ting and {Chung}, Dongwoo and {Cleary}, Kieran and {Corray}, Asantha and {Crites}, Abigail and {Croft}, Rupert and {Dor{\'e}}, Olivier and {Eastwood}, Michael and {Ferrara}, Andrea and {Fonseca}, Jos{\'e} and {Jacobs}, Daniel and {Keating}, Garrett K. and {Lagache}, Guilaine and {Lakhlani}, Gunjan and {Liu}, Adrian and {Moodley}, Kavilan and {Murray}, Norm and {P{\'e}nin}, Aur{\'e}lie and {Popping}, Gerg{\"o} and {Pullen}, Anthony and {Reichers}, Dominik and {Saito}, Shun and {Saliwanchik}, Ben and {Santos}, Mario and {Somerville}, Rachel and {Stacey}, Gordon and {Stein}, George and {Villaescusa-Navarro}, Francesco and {Visbal}, Eli and {Weltman}, Amanda and {Wolz}, Laura and {Zemcov}, Micheal},
        title = "{Line-Intensity Mapping: 2017 Status Report}",
      journal = {arXiv e-prints},
         year = 2017,
        month = sep,
          eid = {arXiv:1709.09066},
        pages = {arXiv:1709.09066},
          doi = {10.48550/arXiv.1709.09066},
archivePrefix = {arXiv},
       eprint = {1709.09066},
 primaryClass = {astro-ph.CO},
       adsurl = {https://ui.adsabs.harvard.edu/abs/2017arXiv170909066K}
}

@ARTICLE{bernal/kovetz:2022,
       author = {{Bernal}, Jos{\'e} Luis and {Kovetz}, Ely D.},
        title = "{Line-intensity mapping: theory review with a focus on star-formation lines}",
      journal = {\aapr},
         year = 2022,
        month = dec,
       volume = {30},
       number = {1},
          eid = {5},
        pages = {5},
          doi = {10.1007/s00159-022-00143-0},
archivePrefix = {arXiv},
       eprint = {2206.15377},
 primaryClass = {astro-ph.CO},
       adsurl = {https://ui.adsabs.harvard.edu/abs/2022A&ARv..30....5B}
}

@ARTICLE{dunne/etal:2024,
       author = {{Dunne}, Delaney A. and {Cleary}, Kieran A. and {Breysse}, Patrick C. and {Chung}, Dongwoo T. and {Ihle}, H{\r{a}}vard T. and {Bond}, J. Richard and {Eriksen}, Hans Kristian and {Gundersen}, Joshua Ott and {Keating}, Laura C. and {Kim}, Junhan and {Lunde}, Jonas Gahr Sturtzel and {Murray}, Norman and {Padmanabhan}, Hamsa and {Philip}, Liju and {Stutzer}, Nils-Ole and {Tolgay}, Do{\u{g}}a and {Wehus}, Ingunn Katherine and {Church}, Sarah E. and {Gaier}, Todd and {Harris}, Andrew I. and {Hobbs}, Richard and {Lamb}, James W. and {Lawrence}, Charles R. and {Readhead}, Anthony C.~S. and {Woody}, David P.},
        title = "{COMAP Early Science. VIII. A Joint Stacking Analysis with eBOSS Quasars}",
      journal = {\apj},
         year = 2024,
        month = apr,
       volume = {965},
       number = {1},
          eid = {7},
        pages = {7},
          doi = {10.3847/1538-4357/ad2dfc},
archivePrefix = {arXiv},
       eprint = {2304.09832},
 primaryClass = {astro-ph.GA},
       adsurl = {https://ui.adsabs.harvard.edu/abs/2024ApJ...965....7D}
}

@ARTICLE{cleary/etal:2022,
       author = {{Cleary}, Kieran A. and {Borowska}, Jowita and {Breysse}, Patrick C. and {Catha}, Morgan and {Chung}, Dongwoo T. and {Church}, Sarah E. and {Dickinson}, Clive and {Eriksen}, Hans Kristian and {Foss}, Marie Kristine and {Gundersen}, Joshua Ott and {Harper}, Stuart E. and {Harris}, Andrew I. and {Hobbs}, Richard and {Ihle}, H{\r{a}}vard T. and {Kim}, Junhan and {Kocz}, Jonathon and {Lamb}, James W. and {Lunde}, Jonas G.~S. and {Padmanabhan}, Hamsa and {Pearson}, Timothy J. and {Philip}, Liju and {Powell}, Travis W. and {Rasmussen}, Maren and {Readhead}, Anthony C.~S. and {Rennie}, Thomas J. and {Silva}, Marta B. and {Stutzer}, Nils-Ole and {Uzgil}, Bade D. and {Watts}, Duncan J. and {Wehus}, Ingunn Kathrine and {Woody}, David P. and {Basoalto}, Lilian and {Bond}, J. Richard and {Dunne}, Delaney A. and {Gaier}, Todd and {Hensley}, Brandon and {Keating}, Laura C. and {Lawrence}, Charles R. and {Murray}, Norman and {Paladini}, Roberta and {Reeves}, Rodrigo and {Viero}, Marco P. and {Wechsler}, Risa H. and {Comap Collaboration}},
        title = "{COMAP Early Science. I. Overview}",
      journal = {\apj},
         year = 2022,
        month = jul,
       volume = {933},
       number = {2},
          eid = {182},
        pages = {182},
          doi = {10.3847/1538-4357/ac63cc},
archivePrefix = {arXiv},
       eprint = {2111.05927},
 primaryClass = {astro-ph.CO},
       adsurl = {https://ui.adsabs.harvard.edu/abs/2022ApJ...933..182C}
}

@ARTICLE{paul/etal:2023,
       author = {{Paul}, Sourabh and {Santos}, Mario G. and {Chen}, Zhaoting and {Wolz}, Laura},
        title = "{A first detection of neutral hydrogen intensity mapping on Mpc scales at $z\approx 0.32$ and $z\approx 0.44$}",
      journal = {arXiv e-prints},
         year = 2023,
        month = jan,
          eid = {arXiv:2301.11943},
        pages = {arXiv:2301.11943},
          doi = {10.48550/arXiv.2301.11943},
archivePrefix = {arXiv},
       eprint = {2301.11943},
 primaryClass = {astro-ph.CO},
       adsurl = {https://ui.adsabs.harvard.edu/abs/2023arXiv230111943P}
}

@ARTICLE{keenan/etal:2022,
       author = {{Keenan}, Ryan P. and {Keating}, Garrett K. and {Marrone}, Daniel P.},
        title = "{An Intensity Mapping Constraint on the CO-galaxy Cross-power Spectrum at Redshift  3}",
      journal = {\apj},
         year = 2022,
        month = mar,
       volume = {927},
       number = {2},
          eid = {161},
        pages = {161},
          doi = {10.3847/1538-4357/ac4888},
archivePrefix = {arXiv},
       eprint = {2110.02239},
 primaryClass = {astro-ph.GA},
       adsurl = {https://ui.adsabs.harvard.edu/abs/2022ApJ...927..161K}
}

@INPROCEEDINGS{santos/etal:2016,
       author = {{Santos}, M. and {Bull}, P. and {Camera}, S. and {Chen}, S. and {Fonseca}, J. and {Heywood}, I. and {Hilton}, M. and {Jarvis}, M. and {Jozsa}, G.~I.~G. and {Knowles}, K. and {Leeuw}, L. and {Maartens}, R. and {Malefahlo}, E. and {McAlpine}, K. and {Moodley}, K. and {Patel}, P. and {Pourtsidou}, A. and {Prescott}, M. and {Spekkens}, K. and {Taylor}, R. and {Witzemann}, A. and {Whittam}, I.~H.},
        title = "{A Large Sky Survey with MeerKAT}",
    booktitle = {MeerKAT Science: On the Pathway to the SKA},
         year = 2016,
        month = jan,
          eid = {32},
        pages = {32},
          doi = {10.22323/1.277.0032},
archivePrefix = {arXiv},
       eprint = {1709.06099},
 primaryClass = {astro-ph.CO},
       adsurl = {https://ui.adsabs.harvard.edu/abs/2016mks..confE..32S}
}

@ARTICLE{keating/etal:2016,
       author = {{Keating}, Garrett K. and {Marrone}, Daniel P. and {Bower}, Geoffrey C. and {Leitch}, Erik and {Carlstrom}, John E. and {DeBoer}, David R.},
        title = "{COPSS II: The Molecular Gas Content of Ten Million Cubic Megaparsecs at Redshift z {\ensuremath{\sim}} 3}",
      journal = {\apj},
         year = 2016,
        month = oct,
       volume = {830},
       number = {1},
          eid = {34},
        pages = {34},
          doi = {10.3847/0004-637X/830/1/34},
archivePrefix = {arXiv},
       eprint = {1605.03971},
 primaryClass = {astro-ph.GA},
       adsurl = {https://ui.adsabs.harvard.edu/abs/2016ApJ...830...34K}
}

@ARTICLE{keating/etal:2020,
       author = {{Keating}, Garrett K. and {Marrone}, Daniel P. and {Bower}, Geoffrey C. and {Keenan}, Ryan P.},
        title = "{An Intensity Mapping Detection of Aggregate CO Line Emission at 3 mm}",
      journal = {\apj},
         year = 2020,
        month = oct,
       volume = {901},
       number = {2},
          eid = {141},
        pages = {141},
          doi = {10.3847/1538-4357/abb08e},
archivePrefix = {arXiv},
       eprint = {2008.08087},
 primaryClass = {astro-ph.GA},
       adsurl = {https://ui.adsabs.harvard.edu/abs/2020ApJ...901..141K}
}

@ARTICLE{concerto/etal:2020,
       author = {{CONCERTO Collaboration} and {Ade}, P. and {Aravena}, M. and {Barria}, E. and {Beelen}, A. and {Benoit}, A. and {B{\'e}thermin}, M. and {Bounmy}, J. and {Bourrion}, O. and {Bres}, G. and {De Breuck}, C. and {Calvo}, M. and {Cao}, Y. and {Catalano}, A. and {D{\'e}sert}, F. -X. and {Dur{\'a}n}, C.~A. and {Fasano}, A. and {Fenouillet}, T. and {Garcia}, J. and {Garde}, G. and {Goupy}, J. and {Groppi}, C. and {Hoarau}, C. and {Lagache}, G. and {Lambert}, J. -C. and {Leggeri}, J. -P. and {Levy-Bertrand}, F. and {Mac{\'\i}as-P{\'e}rez}, J. and {Mani}, H. and {Marpaud}, J. and {Mauskopf}, P. and {Monfardini}, A. and {Pisano}, G. and {Ponthieu}, N. and {Prieur}, L. and {Roni}, S. and {Roudier}, S. and {Tourres}, D. and {Tucker}, C.},
        title = "{A wide field-of-view low-resolution spectrometer at APEX: Instrument design and scientific forecast}",
      journal = {\aap},
         year = 2020,
        month = oct,
       volume = {642},
          eid = {A60},
        pages = {A60},
          doi = {10.1051/0004-6361/202038456},
archivePrefix = {arXiv},
       eprint = {2007.14246},
 primaryClass = {astro-ph.IM},
       adsurl = {https://ui.adsabs.harvard.edu/abs/2020A&A...642A..60C}
}

@ARTICLE{ccat_prime/etal:2023,
       author = {{CCAT-Prime Collaboration} and {Aravena}, Manuel and {Austermann}, Jason E. and {Basu}, Kaustuv and {Battaglia}, Nicholas and {Beringue}, Benjamin and {Bertoldi}, Frank and {Bigiel}, Frank and {Bond}, J. Richard and {Breysse}, Patrick C. and {Broughton}, Colton and {Bustos}, Ricardo and {Chapman}, Scott C. and {Charmetant}, Maude and {Choi}, Steve K. and {Chung}, Dongwoo T. and {Clark}, Susan E. and {Cothard}, Nicholas F. and {Crites}, Abigail T. and {Dev}, Ankur and {Douglas}, Kaela and {Duell}, Cody J. and {D{\"u}nner}, Rolando and {Ebina}, Haruki and {Erler}, Jens and {Fich}, Michel and {Fissel}, Laura M. and {Foreman}, Simon and {Freundt}, R.~G. and {Gallardo}, Patricio A. and {Gao}, Jiansong and {Garc{\'\i}a}, Pablo and {Giovanelli}, Riccardo and {Golec}, Joseph E. and {Groppi}, Christopher E. and {Haynes}, Martha P. and {Henke}, Douglas and {Hensley}, Brandon and {Herter}, Terry and {Higgins}, Ronan and {Hlo{\v{z}}ek}, Ren{\'e}e and {Huber}, Anthony and {Huber}, Zachary and {Hubmayr}, Johannes and {Jackson}, Rebecca and {Johnstone}, Douglas and {Karoumpis}, Christos and {Keating}, Laura C. and {Komatsu}, Eiichiro and {Li}, Yaqiong and {Magnelli}, Benjamin and {Matthews}, Brenda C. and {Mauskopf}, Philip D. and {McMahon}, Jeffrey J. and {Meerburg}, P. Daniel and {Meyers}, Joel and {Muralidhara}, Vyoma and {Murray}, Norman W. and {Niemack}, Michael D. and {Nikola}, Thomas and {Okada}, Yoko and {Puddu}, Roberto and {Riechers}, Dominik A. and {Rosolowsky}, Erik and {Rossi}, Kayla and {Rotermund}, Kaja and {Roy}, Anirban and {Sadavoy}, Sarah I. and {Schaaf}, Reinhold and {Schilke}, Peter and {Scott}, Douglas and {Simon}, Robert and {Sinclair}, Adrian K. and {Sivakoff}, Gregory R. and {Stacey}, Gordon J. and {Stutz}, Amelia M. and {Stutzki}, Juergen and {Tahani}, Mehrnoosh and {Thanjavur}, Karun and {Timmermann}, Ralf A. and {Ullom}, Joel N. and {van Engelen}, Alexander and {Vavagiakis}, Eve M. and {Vissers}, Michael R. and {Wheeler}, Jordan D. and {White}, Simon D.~M. and {Zhu}, Yijie and {Zou}, Bugao},
        title = "{CCAT-prime Collaboration: Science Goals and Forecasts with Prime-Cam on the Fred Young Submillimeter Telescope}",
      journal = {\apjs},
         year = 2023,
        month = jan,
       volume = {264},
       number = {1},
          eid = {7},
        pages = {7},
          doi = {10.3847/1538-4365/ac9838},
archivePrefix = {arXiv},
       eprint = {2107.10364},
 primaryClass = {astro-ph.CO},
       adsurl = {https://ui.adsabs.harvard.edu/abs/2023ApJS..264....7C}
}

@ARTICLE{sun/etal:2021,
       author = {{Sun}, G. and {Chang}, T. -C. and {Uzgil}, B.~D. and {Bock}, J.~J. and {Bradford}, C.~M. and {Butler}, V. and {Caze-Cortes}, T. and {Cheng}, Y. -T. and {Cooray}, A. and {Crites}, A.~T. and {Hailey-Dunsheath}, S. and {Emerson}, N. and {Frez}, C. and {Hoscheit}, B.~L. and {Hunacek}, J. and {Keenan}, R.~P. and {Li}, C.~T. and {Madonia}, P. and {Marrone}, D.~P. and {Moncelsi}, L. and {Shiu}, C. and {Trumper}, I. and {Turner}, A. and {Weber}, A. and {Wei}, T.~S. and {Zemcov}, M.},
        title = "{Probing Cosmic Reionization and Molecular Gas Growth with TIME}",
      journal = {\apj},
         year = 2021,
        month = jul,
       volume = {915},
       number = {1},
          eid = {33},
        pages = {33},
          doi = {10.3847/1538-4357/abfe62},
archivePrefix = {arXiv},
       eprint = {2012.09160},
 primaryClass = {astro-ph.GA},
       adsurl = {https://ui.adsabs.harvard.edu/abs/2021ApJ...915...33S}
}

@ARTICLE{switzer/etal:2021,
       author = {{Switzer}, Eric R. and {Barrentine}, Emily M. and {Cataldo}, Giuseppe and {Essinger-Hileman}, Thomas and {Ade}, Peter A.~R. and {Anderson}, Christopher J. and {Barlis}, Alyssa and {Beeman}, Jeffrey and {Bellis}, Nicholas and {Bolatto}, Alberto D. and {Breysse}, Patrick C. and {Bulcha}, Berhanu T. and {Chevres-Fernanadez}, Lee-Roger and {Cho}, Chullhee and {Connors}, Jake A. and {Ehsan}, Negar and {Glenn}, Jason and {Golec}, Joseph and {Hays-Wehle}, James P. and {Hess}, Larry A. and {Jahromi}, Amir E. and {Jenkins}, Trevian and {Kimball}, Mark O. and {Kogut}, Alan J. and {Lowe}, Luke N. and {Mauskopf}, Philip and {McMahon}, Jeffrey and {Mirzaei}, Mona and {Moseley}, Harvey and {Mugge-Durum}, Jonas and {Noroozian}, Omid and {Oxholm}, Trevor M. and {Parekh}, Tatsat and {Pen}, Ue-Li and {Pullen}, Anthony R. and {Rahmani}, Maryam and {Ramirez}, Mathias M. and {Roselli}, Florian and {Shire}, Konrad and {Siebert}, Gage and {Sinclair}, Adrian K. and {Somerville}, Rachel S. and {Stephenson}, Ryan and {Stevenson}, Thomas R. and {Timbie}, Peter and {Termini}, Jared and {Trenkamp}, Justin and {Tucker}, Carole and {Visbal}, Elijah and {Volpert}, Carolyn G. and {Wollack}, Edward J. and {Yang}, Shengqi and {Yung}, L.~Y. Aaron},
        title = "{Experiment for cryogenic large-aperture intensity mapping: instrument design}",
      journal = {Journal of Astronomical Telescopes, Instruments, and Systems},
         year = 2021,
        month = oct,
       volume = {7},
          eid = {044004},
        pages = {044004},
          doi = {10.1117/1.JATIS.7.4.044004},
       adsurl = {https://ui.adsabs.harvard.edu/abs/2021JATIS...7d4004S}
}

@ARTICLE{vieira/etal:2020,
       author = {{Vieira}, Joaquin and {Aguirre}, James and {Bradford}, C. Matt and {Filippini}, Jeffrey and {Groppi}, Christopher and {Marrone}, Dan and {Bethermin}, Matthieu and {Chang}, Tzu-Ching and {Devlin}, Mark and {Dore}, Oliver and {Fu}, Jianyang Frank and {Hailey Dunsheath}, Steven and {Holder}, Gilbert and {Keating}, Garrett and {Keenan}, Ryan and {Kovetz}, Ely and {Lagache}, Guilaine and {Mauskopf}, Philip and {Narayanan}, Desika and {Popping}, Gergo and {Shirokoff}, Erik and {Somerville}, Rachel and {Trumper}, Isaac and {Uzgil}, Bade and {Zmuidzinas}, Jonas},
        title = "{The Terahertz Intensity Mapper (TIM): a Next-Generation Experiment for Galaxy Evolution Studies}",
      journal = {arXiv e-prints},
         year = 2020,
        month = sep,
          eid = {arXiv:2009.14340},
        pages = {arXiv:2009.14340},
          doi = {10.48550/arXiv.2009.14340},
archivePrefix = {arXiv},
       eprint = {2009.14340},
 primaryClass = {astro-ph.IM},
       adsurl = {https://ui.adsabs.harvard.edu/abs/2020arXiv200914340V}
}

@ARTICLE{dore/etal:2014,
       author = {{Dor{\'e}}, Olivier and {Bock}, Jamie and {Ashby}, Matthew and {Capak}, Peter and {Cooray}, Asantha and {de Putter}, Roland and {Eifler}, Tim and {Flagey}, Nicolas and {Gong}, Yan and {Habib}, Salman and {Heitmann}, Katrin and {Hirata}, Chris and {Jeong}, Woong-Seob and {Katti}, Raj and {Korngut}, Phil and {Krause}, Elisabeth and {Lee}, Dae-Hee and {Masters}, Daniel and {Mauskopf}, Phil and {Melnick}, Gary and {Mennesson}, Bertrand and {Nguyen}, Hien and {{\"O}berg}, Karin and {Pullen}, Anthony and {Raccanelli}, Alvise and {Smith}, Roger and {Song}, Yong-Seon and {Tolls}, Volker and {Unwin}, Steve and {Venumadhav}, Tejaswi and {Viero}, Marco and {Werner}, Mike and {Zemcov}, Mike},
        title = "{Cosmology with the SPHEREX All-Sky Spectral Survey}",
      journal = {arXiv e-prints},
         year = 2014,
        month = dec,
          eid = {arXiv:1412.4872},
        pages = {arXiv:1412.4872},
          doi = {10.48550/arXiv.1412.4872},
archivePrefix = {arXiv},
       eprint = {1412.4872},
 primaryClass = {astro-ph.CO},
       adsurl = {https://ui.adsabs.harvard.edu/abs/2014arXiv1412.4872D}
}

@ARTICLE{deboer/etal:2017,
       author = {{DeBoer}, David R. and {Parsons}, Aaron R. and {Aguirre}, James E. and {Alexander}, Paul and {Ali}, Zaki S. and {Beardsley}, Adam P. and {Bernardi}, Gianni and {Bowman}, Judd D. and {Bradley}, Richard F. and {Carilli}, Chris L. and {Cheng}, Carina and {de Lera Acedo}, Eloy and {Dillon}, Joshua S. and {Ewall-Wice}, Aaron and {Fadana}, Gcobisa and {Fagnoni}, Nicolas and {Fritz}, Randall and {Furlanetto}, Steve R. and {Glendenning}, Brian and {Greig}, Bradley and {Grobbelaar}, Jasper and {Hazelton}, Bryna J. and {Hewitt}, Jacqueline N. and {Hickish}, Jack and {Jacobs}, Daniel C. and {Julius}, Austin and {Kariseb}, MacCalvin and {Kohn}, Saul A. and {Lekalake}, Telalo and {Liu}, Adrian and {Loots}, Anita and {MacMahon}, David and {Malan}, Lourence and {Malgas}, Cresshim and {Maree}, Matthys and {Martinot}, Zachary and {Mathison}, Nathan and {Matsetela}, Eunice and {Mesinger}, Andrei and {Morales}, Miguel F. and {Neben}, Abraham R. and {Patra}, Nipanjana and {Pieterse}, Samantha and {Pober}, Jonathan C. and {Razavi-Ghods}, Nima and {Ringuette}, Jon and {Robnett}, James and {Rosie}, Kathryn and {Sell}, Raddwine and {Smith}, Craig and {Syce}, Angelo and {Tegmark}, Max and {Thyagarajan}, Nithyanandan and {Williams}, Peter K.~G. and {Zheng}, Haoxuan},
        title = "{Hydrogen Epoch of Reionization Array (HERA)}",
      journal = {\pasp},
         year = 2017,
        month = apr,
       volume = {129},
       number = {974},
        pages = {045001},
          doi = {10.1088/1538-3873/129/974/045001},
archivePrefix = {arXiv},
       eprint = {1606.07473},
 primaryClass = {astro-ph.IM},
       adsurl = {https://ui.adsabs.harvard.edu/abs/2017PASP..129d5001D}
}

@ARTICLE{planck/etal:2020,
       author = {{Planck Collaboration} and {Aghanim}, N. and {Akrami}, Y. and {Ashdown}, M. and {Aumont}, J. and {Baccigalupi}, C. and {Ballardini}, M. and {Banday}, A.~J. and {Barreiro}, R.~B. and {Bartolo}, N. and {Basak}, S. and {Battye}, R. and {Benabed}, K. and {Bernard}, J. -P. and {Bersanelli}, M. and {Bielewicz}, P. and {Bock}, J.~J. and {Bond}, J.~R. and {Borrill}, J. and {Bouchet}, F.~R. and {Boulanger}, F. and {Bucher}, M. and {Burigana}, C. and {Butler}, R.~C. and {Calabrese}, E. and {Cardoso}, J. -F. and {Carron}, J. and {Challinor}, A. and {Chiang}, H.~C. and {Chluba}, J. and {Colombo}, L.~P.~L. and {Combet}, C. and {Contreras}, D. and {Crill}, B.~P. and {Cuttaia}, F. and {de Bernardis}, P. and {de Zotti}, G. and {Delabrouille}, J. and {Delouis}, J. -M. and {Di Valentino}, E. and {Diego}, J.~M. and {Dor{\'e}}, O. and {Douspis}, M. and {Ducout}, A. and {Dupac}, X. and {Dusini}, S. and {Efstathiou}, G. and {Elsner}, F. and {En{\ss}lin}, T.~A. and {Eriksen}, H.~K. and {Fantaye}, Y. and {Farhang}, M. and {Fergusson}, J. and {Fernandez-Cobos}, R. and {Finelli}, F. and {Forastieri}, F. and {Frailis}, M. and {Fraisse}, A.~A. and {Franceschi}, E. and {Frolov}, A. and {Galeotta}, S. and {Galli}, S. and {Ganga}, K. and {G{\'e}nova-Santos}, R.~T. and {Gerbino}, M. and {Ghosh}, T. and {Gonz{\'a}lez-Nuevo}, J. and {G{\'o}rski}, K.~M. and {Gratton}, S. and {Gruppuso}, A. and {Gudmundsson}, J.~E. and {Hamann}, J. and {Handley}, W. and {Hansen}, F.~K. and {Herranz}, D. and {Hildebrandt}, S.~R. and {Hivon}, E. and {Huang}, Z. and {Jaffe}, A.~H. and {Jones}, W.~C. and {Karakci}, A. and {Keih{\"a}nen}, E. and {Keskitalo}, R. and {Kiiveri}, K. and {Kim}, J. and {Kisner}, T.~S. and {Knox}, L. and {Krachmalnicoff}, N. and {Kunz}, M. and {Kurki-Suonio}, H. and {Lagache}, G. and {Lamarre}, J. -M. and {Lasenby}, A. and {Lattanzi}, M. and {Lawrence}, C.~R. and {Le Jeune}, M. and {Lemos}, P. and {Lesgourgues}, J. and {Levrier}, F. and {Lewis}, A. and {Liguori}, M. and {Lilje}, P.~B. and {Lilley}, M. and {Lindholm}, V. and {L{\'o}pez-Caniego}, M. and {Lubin}, P.~M. and {Ma}, Y. -Z. and {Mac{\'\i}as-P{\'e}rez}, J.~F. and {Maggio}, G. and {Maino}, D. and {Mandolesi}, N. and {Mangilli}, A. and {Marcos-Caballero}, A. and {Maris}, M. and {Martin}, P.~G. and {Martinelli}, M. and {Mart{\'\i}nez-Gonz{\'a}lez}, E. and {Matarrese}, S. and {Mauri}, N. and {McEwen}, J.~D. and {Meinhold}, P.~R. and {Melchiorri}, A. and {Mennella}, A. and {Migliaccio}, M. and {Millea}, M. and {Mitra}, S. and {Miville-Desch{\^e}nes}, M. -A. and {Molinari}, D. and {Montier}, L. and {Morgante}, G. and {Moss}, A. and {Natoli}, P. and {N{\o}rgaard-Nielsen}, H.~U. and {Pagano}, L. and {Paoletti}, D. and {Partridge}, B. and {Patanchon}, G. and {Peiris}, H.~V. and {Perrotta}, F. and {Pettorino}, V. and {Piacentini}, F. and {Polastri}, L. and {Polenta}, G. and {Puget}, J. -L. and {Rachen}, J.~P. and {Reinecke}, M. and {Remazeilles}, M. and {Renzi}, A. and {Rocha}, G. and {Rosset}, C. and {Roudier}, G. and {Rubi{\~n}o-Mart{\'\i}n}, J.~A. and {Ruiz-Granados}, B. and {Salvati}, L. and {Sandri}, M. and {Savelainen}, M. and {Scott}, D. and {Shellard}, E.~P.~S. and {Sirignano}, C. and {Sirri}, G. and {Spencer}, L.~D. and {Sunyaev}, R. and {Suur-Uski}, A. -S. and {Tauber}, J.~A. and {Tavagnacco}, D. and {Tenti}, M. and {Toffolatti}, L. and {Tomasi}, M. and {Trombetti}, T. and {Valenziano}, L. and {Valiviita}, J. and {Van Tent}, B. and {Vibert}, L. and {Vielva}, P. and {Villa}, F. and {Vittorio}, N. and {Wandelt}, B.~D. and {Wehus}, I.~K. and {White}, M. and {White}, S.~D.~M. and {Zacchei}, A. and {Zonca}, A.},
        title = "{Planck 2018 results. VI. Cosmological parameters}",
      journal = {\aap},
         year = 2020,
        month = sep,
       volume = {641},
          eid = {A6},
        pages = {A6},
          doi = {10.1051/0004-6361/201833910},
archivePrefix = {arXiv},
       eprint = {1807.06209},
 primaryClass = {astro-ph.CO},
       adsurl = {https://ui.adsabs.harvard.edu/abs/2020A&A...641A...6P}
}

@ARTICLE{dunne/etal:2025,
       author = {{Dunne}, D.~A. and {Cleary}, K.~A. and {Breysse}, P.~C. and {Chung}, D.~T. and {Ihle}, H.~T. and {Lunde}, J.~G.~S. and {Padmanabhan}, H. and {Stutzer}, N. -O. and {Bond}, J.~R. and {Gundersen}, J.~O. and {Kim}, J. and {Readhead}, A.~C.~S.},
        title = "{Three-Dimensional Stacking as a Line Intensity Mapping Statistic}",
      journal = {arXiv e-prints},
         year = 2025,
        month = mar,
          eid = {arXiv:2503.21743},
        pages = {arXiv:2503.21743},
          doi = {10.48550/arXiv.2503.21743},
archivePrefix = {arXiv},
       eprint = {2503.21743},
 primaryClass = {astro-ph.CO},
       adsurl = {https://ui.adsabs.harvard.edu/abs/2025arXiv250321743D}
}

@ARTICLE{chen/etal:2025,
       author = {{Chen}, Zhaoting and {Cunnington}, Steven and {Pourtsidou}, Alkistis and {Wolz}, Laura and {Spinelli}, Marta and {Bernal}, Jos{\'e} Luis and {Barberi-Squarotti}, Matilde and {Camera}, Stefano and {Carucci}, Isabella P. and {Fonseca}, Jos{\'e} and {Grainge}, Keith and {Irfan}, Melis O. and {Santos}, Mario G. and {Wang}, Jingying and {Meerklass Collaboration}},
        title = "{Emission-line Stacking of 21 cm Intensity Maps with MeerKLASS: Inference Pipeline and Application to the L-band Deep-field Data}",
      journal = {\apjs},
         year = 2025,
        month = jul,
       volume = {279},
       number = {1},
          eid = {19},
        pages = {19},
          doi = {10.3847/1538-4365/add897},
archivePrefix = {arXiv},
       eprint = {2504.03908},
 primaryClass = {astro-ph.CO},
       adsurl = {https://ui.adsabs.harvard.edu/abs/2025ApJS..279...19C}
}

@ARTICLE{beers/flynn/gebhardt:1990,
       author = {{Beers}, Timothy C. and {Flynn}, Kevin and {Gebhardt}, Karl},
        title = "{Measures of Location and Scale for Velocities in Clusters of Galaxies---A Robust Approach}",
      journal = {\aj},
         year = 1990,
        month = jul,
       volume = {100},
        pages = {32},
          doi = {10.1086/115487},
       adsurl = {https://ui.adsabs.harvard.edu/abs/1990AJ....100...32B}
}

@ARTICLE{konno/etal:2016,
       author = {{Konno}, Akira and {Ouchi}, Masami and {Nakajima}, Kimihiko and {Duval}, Florent and {Kusakabe}, Haruka and {Ono}, Yoshiaki and {Shimasaku}, Kazuhiro},
        title = "{Bright and Faint Ends of Ly{\ensuremath{\alpha}} Luminosity Functions at z = 2 Determined by the Subaru Survey: Implications for AGNs, Magnification Bias, and ISM H I Evolution}",
      journal = {\apj},
         year = 2016,
        month = may,
       volume = {823},
       number = {1},
          eid = {20},
        pages = {20},
          doi = {10.3847/0004-637X/823/1/20},
archivePrefix = {arXiv},
       eprint = {1512.01854},
 primaryClass = {astro-ph.GA},
       adsurl = {https://ui.adsabs.harvard.edu/abs/2016ApJ...823...20K}
}

@ARTICLE{ouchi/etal:2008,
       author = {{Ouchi}, Masami and {Shimasaku}, Kazuhiro and {Akiyama}, Masayuki and {Simpson}, Chris and {Saito}, Tomoki and {Ueda}, Yoshihiro and {Furusawa}, Hisanori and {Sekiguchi}, Kazuhiro and {Yamada}, Toru and {Kodama}, Tadayuki and {Kashikawa}, Nobunari and {Okamura}, Sadanori and {Iye}, Masanori and {Takata}, Tadafumi and {Yoshida}, Michitoshi and {Yoshida}, Makiko},
        title = "{The Subaru/XMM-Newton Deep Survey (SXDS). IV. Evolution of Ly{\ensuremath{\alpha}} Emitters from z = 3.1 to 5.7 in the 1 deg$^{2}$ Field: Luminosity Functions and AGN}",
      journal = {\apjs},
         year = 2008,
        month = jun,
       volume = {176},
       number = {2},
        pages = {301-330},
          doi = {10.1086/527673},
archivePrefix = {arXiv},
       eprint = {0707.3161},
 primaryClass = {astro-ph},
       adsurl = {https://ui.adsabs.harvard.edu/abs/2008ApJS..176..301O}
}

@ARTICLE{schlegel/finkbeiner/davis:1998,
       author = {{Schlegel}, David J. and {Finkbeiner}, Douglas P. and {Davis}, Marc},
        title = "{Maps of Dust Infrared Emission for Use in Estimation of Reddening and Cosmic Microwave Background Radiation Foregrounds}",
      journal = {\apj},
         year = 1998,
        month = jun,
       volume = {500},
       number = {2},
        pages = {525-553},
          doi = {10.1086/305772},
archivePrefix = {arXiv},
       eprint = {astro-ph/9710327},
 primaryClass = {astro-ph},
       adsurl = {https://ui.adsabs.harvard.edu/abs/1998ApJ...500..525S}
}

@ARTICLE{schlafly/finkbeiner:2011,
       author = {{Schlafly}, Edward F. and {Finkbeiner}, Douglas P.},
        title = "{Measuring Reddening with Sloan Digital Sky Survey Stellar Spectra and Recalibrating SFD}",
      journal = {\apj},
         year = 2011,
        month = aug,
       volume = {737},
       number = {2},
          eid = {103},
        pages = {103},
          doi = {10.1088/0004-637X/737/2/103},
archivePrefix = {arXiv},
       eprint = {1012.4804},
 primaryClass = {astro-ph.GA},
       adsurl = {https://ui.adsabs.harvard.edu/abs/2011ApJ...737..103S}
}

@ARTICLE{green:2018,
       author = {{Green}, {Gregory M.}},
        title = "{dustmaps: A Python interface for maps of interstellar dust}",
      journal = {The Journal of Open Source Software},
         year = "2018",
        month = "Jun",
       volume = {3},
       number = {26},
        pages = {695},
          doi = {10.21105/joss.00695},
       adsurl = {https://ui.adsabs.harvard.edu/abs/2018JOSS....3..695G}
}

@ARTICLE{fitzpatrick:1999,
       author = {{Fitzpatrick}, Edward L.},
        title = "{Correcting for the Effects of Interstellar Extinction}",
      journal = {\pasp},
         year = 1999,
        month = jan,
       volume = {111},
       number = {755},
        pages = {63-75},
          doi = {10.1086/316293},
archivePrefix = {arXiv},
       eprint = {astro-ph/9809387},
 primaryClass = {astro-ph},
       adsurl = {https://ui.adsabs.harvard.edu/abs/1999PASP..111...63F}
}

@article{blas/lesgourges/tram:2011,
    author = "Blas, Diego and Lesgourgues, Julien and Tram, Thomas",
    title = "{The Cosmic Linear Anisotropy Solving System (CLASS) II: Approximation schemes}",
    eprint = "1104.2933",
    archivePrefix = "arXiv",
    primaryClass = "astro-ph.CO",
    reportNumber = "CERN-PH-TH-2011-082, LAPTH-010-11",
    doi = "10.1088/1475-7516/2011/07/034",
    journal = "JCAP",
    volume = "07",
    pages = "034",
    year = "2011"
}

@ARTICLE{smith/etal:2003,
       author = {{Smith}, R.~E. and {Peacock}, J.~A. and {Jenkins}, A. and {White}, S.~D.~M. and {Frenk}, C.~S. and {Pearce}, F.~R. and {Thomas}, P.~A. and {Efstathiou}, G. and {Couchman}, H.~M.~P.},
        title = "{Stable clustering, the halo model and non-linear cosmological power spectra}",
      journal = {\mnras},
         year = 2003,
        month = jun,
       volume = {341},
       number = {4},
        pages = {1311-1332},
          doi = {10.1046/j.1365-8711.2003.06503.x},
archivePrefix = {arXiv},
       eprint = {astro-ph/0207664},
 primaryClass = {astro-ph},
       adsurl = {https://ui.adsabs.harvard.edu/abs/2003MNRAS.341.1311S}
}

@ARTICLE{weiss/etal:2025,
       author = {{Weiss}, Laurel H. and {Gebhardt}, Karl and {Davis}, Dustin and {Mentuch Cooper}, Erin and {Lujan Niemeyer}, Maja and {Qezlou}, Mahdi and {Mirza Khanlari}, Mahan and {Ciardullo}, Robin and {Farrow}, Daniel and {Gawiser}, Eric and {Gazagnes}, Simon and {Gronwall}, Caryl and {Hill}, Gary J. and {Schneider}, Donald P.},
        title = "{Using Ly{\ensuremath{\alpha}} Absorption to Measure the Intensity and Variability of z {\ensuremath{\sim}} 2.4 Ultraviolet Background Light}",
      journal = {\apj},
         year = 2025,
        month = apr,
       volume = {983},
       number = {1},
          eid = {72},
        pages = {72},
          doi = {10.3847/1538-4357/adc0f9},
archivePrefix = {arXiv},
       eprint = {2504.13253},
 primaryClass = {astro-ph.GA},
       adsurl = {https://ui.adsabs.harvard.edu/abs/2025ApJ...983...72W}
}

@ARTICLE{khanlari/etal:2025,
       author = {{Khanlari}, Mahan Mirza and {Gebhardt}, Karl and {Weiss}, Laurel H. and {Davis}, Dustin and {Cooper}, Erin Mentuch and {Qezlou}, Mahdi and {Niemeyer}, Maja Lujan and {Ciardullo}, Robin and {Schneider}, Donald P. and {Mukae}, Shiro and {Liu}, Chenxu and {Farrow}, Daniel and {Hill}, Gary J. and {Zeimann}, Gregory R. and {Kollatschny}, Wolfram},
        title = "{The HETDEX Survey: Probing Neutral Hydrogen in the Circumgalactic Medium of {\ensuremath{\sim}}88,000 Ly{\ensuremath{\alpha}} Emitters}",
      journal = {\apj},
         year = 2025,
        month = aug,
       volume = {989},
       number = {2},
          eid = {169},
        pages = {169},
          doi = {10.3847/1538-4357/adf10e},
archivePrefix = {arXiv},
       eprint = {2507.15942},
 primaryClass = {astro-ph.GA},
       adsurl = {https://ui.adsabs.harvard.edu/abs/2025ApJ...989..169K}
}

@ARTICLE{byrohl/nelson:2023,
       author = {{Byrohl}, Chris and {Nelson}, Dylan},
        title = "{The cosmic web in Lyman-alpha emission}",
      journal = {\mnras},
         year = 2023,
        month = aug,
       volume = {523},
       number = {4},
        pages = {5248-5273},
          doi = {10.1093/mnras/stad1779},
archivePrefix = {arXiv},
       eprint = {2212.08666},
 primaryClass = {astro-ph.GA},
       adsurl = {https://ui.adsabs.harvard.edu/abs/2023MNRAS.523.5248B}
}

@ARTICLE{umeda/etal:2025,
       author = {{Umeda}, Hiroya and {Ouchi}, Masami and {Kikuta}, Satoshi and {Harikane}, Yuichi and {Ono}, Yoshiaki and {Shibuya}, Takatoshi and {Inoue}, Akio K. and {Shimasaku}, Kazuhiro and {Liang}, Yongming and {Matsumoto}, Akinori and {Saito}, Shun and {Kusakabe}, Haruka and {Kageura}, Yuta and {Nakane}, Minami},
        title = "{SILVERRUSH. XIV. Ly{\ensuremath{\alpha}} Luminosity Functions and Angular Correlation Functions from 20,000 Ly{\ensuremath{\alpha}} Emitters at z {\ensuremath{\sim}} 2.2{\textendash}7.3 from up to 24 deg$^{2}$ HSC-SSP and CHORUS Surveys: Linking the Postreionization Epoch to the Heart of Reionization}",
      journal = {\apjs},
         year = 2025,
        month = apr,
       volume = {277},
       number = {2},
          eid = {37},
        pages = {37},
          doi = {10.3847/1538-4365/adb1c0},
archivePrefix = {arXiv},
       eprint = {2411.15495},
 primaryClass = {astro-ph.GA},
       adsurl = {https://ui.adsabs.harvard.edu/abs/2025ApJS..277...37U}
}

@ARTICLE{dijkstra:2019,
       author = {{Dijkstra}, Mark},
        title = "{Physics of Ly{\ensuremath{\alpha}} Radiative Transfer}",
      journal = {Saas-Fee Advanced Course},
         year = 2019,
        month = jan,
       volume = {46},
        pages = {1},
          doi = {10.1007/978-3-662-59623-4_1},
       adsurl = {https://ui.adsabs.harvard.edu/abs/2019SAAS...46....1D}
}

@ARTICLE{zheng/etal:2011,
       author = {{Zheng}, Zheng and {Cen}, Renyue and {Trac}, Hy and {Miralda-Escud{\'e}}, Jordi},
        title = "{Radiative Transfer Modeling of Ly{\ensuremath{\alpha}} Emitters. II. New Effects on Galaxy Clustering}",
      journal = {\apj},
         year = 2011,
        month = jan,
       volume = {726},
       number = {1},
          eid = {38},
        pages = {38},
          doi = {10.1088/0004-637X/726/1/38},
archivePrefix = {arXiv},
       eprint = {1003.4990},
 primaryClass = {astro-ph.CO},
       adsurl = {https://ui.adsabs.harvard.edu/abs/2011ApJ...726...38Z}
}

@ARTICLE{behrens/niemeyer:2013,
       author = {{Behrens}, C. and {Niemeyer}, J.},
        title = "{Effects of Lyman-alpha scattering in the IGM on clustering statistics of Lyman-alpha emitters}",
      journal = {\aap},
         year = 2013,
        month = aug,
       volume = {556},
          eid = {A5},
        pages = {A5},
          doi = {10.1051/0004-6361/201321172},
archivePrefix = {arXiv},
       eprint = {1306.0920},
 primaryClass = {astro-ph.CO},
       adsurl = {https://ui.adsabs.harvard.edu/abs/2013A&A...556A...5B}
}

@ARTICLE{behrens/etal:2018,
       author = {{Behrens}, C. and {Byrohl}, C. and {Saito}, S. and {Niemeyer}, J.~C.},
        title = "{The impact of Lyman-{\ensuremath{\alpha}} radiative transfer on large-scale clustering in the Illustris simulation}",
      journal = {\aap},
         year = 2018,
        month = jun,
       volume = {614},
          eid = {A31},
        pages = {A31},
          doi = {10.1051/0004-6361/201731783},
archivePrefix = {arXiv},
       eprint = {1710.06171},
 primaryClass = {astro-ph.GA},
       adsurl = {https://ui.adsabs.harvard.edu/abs/2018A&A...614A..31B}
}

@ARTICLE{rowan-robinson/etal:2016,
       author = {{Rowan-Robinson}, Michael and {Oliver}, Seb and {Wang}, Lingyu and {Farrah}, Duncan and {Clements}, David L. and {Gruppioni}, Carlotta and {Marchetti}, Lucia and {Rigopoulou}, Dimitra and {Vaccari}, Mattia},
        title = "{The star formation rate density from z = 1 to 6}",
      journal = {\mnras},
         year = 2016,
        month = sep,
       volume = {461},
       number = {1},
        pages = {1100-1111},
          doi = {10.1093/mnras/stw1169},
archivePrefix = {arXiv},
       eprint = {1605.03937},
 primaryClass = {astro-ph.GA},
       adsurl = {https://ui.adsabs.harvard.edu/abs/2016MNRAS.461.1100R}
}

@ARTICLE{byrohl/etal:2021,
       author = {{Byrohl}, Chris and {Nelson}, Dylan and {Behrens}, Christoph and {Kostyuk}, Ivan and {Glatzle}, Martin and {Pillepich}, Annalisa and {Hernquist}, Lars and {Marinacci}, Federico and {Vogelsberger}, Mark},
        title = "{The physical origins and dominant emission mechanisms of Lyman alpha haloes: results from the TNG50 simulation in comparison to MUSE observations}",
      journal = {\mnras},
         year = 2021,
        month = oct,
       volume = {506},
       number = {4},
        pages = {5129-5152},
          doi = {10.1093/mnras/stab1958},
archivePrefix = {arXiv},
       eprint = {2009.07283},
 primaryClass = {astro-ph.GA},
       adsurl = {https://ui.adsabs.harvard.edu/abs/2021MNRAS.506.5129B}
}

@ARTICLE{agrawal/etal:2017,
       author = {{Agrawal}, Aniket and {Makiya}, Ryu and {Chiang}, Chi-Ting and {Jeong}, Donghui and {Saito}, Shun and {Komatsu}, Eiichiro},
        title = "{Generating log-normal mock catalog of galaxies in redshift space}",
      journal = {\jcap},
         year = 2017,
        month = oct,
       volume = {2017},
       number = {10},
          eid = {003},
        pages = {003},
          doi = {10.1088/1475-7516/2017/10/003},
archivePrefix = {arXiv},
       eprint = {1706.09195},
 primaryClass = {astro-ph.CO},
       adsurl = {https://ui.adsabs.harvard.edu/abs/2017JCAP...10..003A}
}

@Article{         harris2020array,
 title         = {Array programming with {NumPy}},
 author        = {Charles R. Harris and K. Jarrod Millman and St{\'{e}}fan J.
                 van der Walt and Ralf Gommers and Pauli Virtanen and David
                 Cournapeau and Eric Wieser and Julian Taylor and Sebastian
                 Berg and Nathaniel J. Smith and Robert Kern and Matti Picus
                 and Stephan Hoyer and Marten H. van Kerkwijk and Matthew
                 Brett and Allan Haldane and Jaime Fern{\'{a}}ndez del
                 R{\'{i}}o and Mark Wiebe and Pearu Peterson and Pierre
                 G{\'{e}}rard-Marchant and Kevin Sheppard and Tyler Reddy and
                 Warren Weckesser and Hameer Abbasi and Christoph Gohlke and
                 Travis E. Oliphant},
 year          = {2020},
 month         = sep,
 journal       = {Nature},
 volume        = {585},
 number        = {7825},
 pages         = {357--362},
 doi           = {10.1038/s41586-020-2649-2},
 publisher     = {Springer Science and Business Media {LLC}},
 url           = {https://doi.org/10.1038/s41586-020-2649-2}
}

@ARTICLE{2020SciPy-NMeth,
  author  = {Virtanen, Pauli and Gommers, Ralf and Oliphant, Travis E. and
            Haberland, Matt and Reddy, Tyler and Cournapeau, David and
            Burovski, Evgeni and Peterson, Pearu and Weckesser, Warren and
            Bright, Jonathan and {van der Walt}, St{\'e}fan J. and
            Brett, Matthew and Wilson, Joshua and Millman, K. Jarrod and
            Mayorov, Nikolay and Nelson, Andrew R. J. and Jones, Eric and
            Kern, Robert and Larson, Eric and Carey, C J and
            Polat, {\.I}lhan and Feng, Yu and Moore, Eric W. and
            {VanderPlas}, Jake and Laxalde, Denis and Perktold, Josef and
            Cimrman, Robert and Henriksen, Ian and Quintero, E. A. and
            Harris, Charles R. and Archibald, Anne M. and
            Ribeiro, Ant{\^o}nio H. and Pedregosa, Fabian and
            {van Mulbregt}, Paul and {SciPy 1.0 Contributors}},
  title   = {{{SciPy} 1.0: Fundamental Algorithms for Scientific
            Computing in Python}},
  journal = {Nature Methods},
  year    = {2020},
  volume  = {17},
  pages   = {261--272},
  adsurl  = {https://rdcu.be/b08Wh},
  doi     = {10.1038/s41592-019-0686-2},
}

@Article{matplotlib,
  Author    = {Hunter, J. D.},
  Title     = {Matplotlib: A 2D graphics environment},
  Journal   = {Computing in Science \& Engineering},
  Volume    = {9},
  Number    = {3},
  Pages     = {90--95},
  publisher = {IEEE COMPUTER SOC},
  doi       = {10.1109/MCSE.2007.55},
  year      = 2007
}

@ARTICLE{kaiser:1987,
       author = {{Kaiser}, Nick},
        title = "{Clustering in real space and in redshift space}",
      journal = {\mnras},
         year = 1987,
        month = jul,
       volume = {227},
        pages = {1-21},
          doi = {10.1093/mnras/227.1.1},
       adsurl = {https://ui.adsabs.harvard.edu/abs/1987MNRAS.227....1K}
}

@ARTICLE{tornotti/etal:2025,
       author = {{Tornotti}, Davide and {Fumagalli}, Michele and {Fossati}, Matteo and {Arrigoni Battaia}, Fabrizio and {Benitez-Llambay}, Alejandro and {Dayal}, Pratika and {Dutta}, Rajeshwari and {Peroux}, Celine and {Rafelski}, Marc and {Revalski}, Mitchell},
        title = "{The MUSE Ultra Deep Field: A 5 Mpc Stretch of the z {\ensuremath{\approx}} 4 Cosmic Web Revealed in Emission}",
      journal = {\apjl},
         year = 2025,
        month = feb,
       volume = {980},
       number = {2},
          eid = {L43},
        pages = {L43},
          doi = {10.3847/2041-8213/adb0ba},
archivePrefix = {arXiv},
       eprint = {2412.06895},
 primaryClass = {astro-ph.GA},
       adsurl = {https://ui.adsabs.harvard.edu/abs/2025ApJ...980L..43T}
}

@ARTICLE{chavezortiz/etal:2023,
       author = {{Ch{\'a}vez Ortiz}, {\'O}scar A. and {Finkelstein}, Steven L. and {Davis}, Dustin and {Leung}, Gene and {Mentuch Cooper}, Erin and {Bagley}, Micaela and {Larson}, Rebecca and {Casey}, Caitlin M. and {McCarron}, Adam P. and {Gebhardt}, Karl and {Guo}, Yuchen and {Liu}, Chenxu and {Laseter}, Isaac and {Rhodes}, Jason and {Bender}, Ralf and {Fabricius}, Max and {S{\'a}nchez}, Ariel G. and {Scarlata}, Claudia and {Capak}, Peter and {Zalesky}, Lukas and {Sanders}, David and {Szapudi}, Istvan and {Baxter}, Eric and {McPartland}, Conor and {Weaver}, John R. and {Toft}, Sune and {Mobasher}, Bahram and {Suzuki}, Nao and {Chartab}, Nima},
        title = "{Introducing the Texas Euclid Survey for Ly{\ensuremath{\alpha}} (TESLA) Survey: Initial Study Correlating Galaxy Properties to Ly{\ensuremath{\alpha}} Emission}",
      journal = {\apj},
         year = 2023,
        month = aug,
       volume = {952},
       number = {2},
          eid = {110},
        pages = {110},
          doi = {10.3847/1538-4357/acc403},
archivePrefix = {arXiv},
       eprint = {2304.03258},
 primaryClass = {astro-ph.GA},
       adsurl = {https://ui.adsabs.harvard.edu/abs/2023ApJ...952..110C}
}

@ARTICLE{ouchi/etal:2020,
       author = {{Ouchi}, Masami and {Ono}, Yoshiaki and {Shibuya}, Takatoshi},
        title = "{Observations of the Lyman-{\ensuremath{\alpha}} Universe}",
      journal = {\araa},
         year = 2020,
        month = aug,
       volume = {58},
        pages = {617-659},
          doi = {10.1146/annurev-astro-032620-021859},
archivePrefix = {arXiv},
       eprint = {2012.07960},
 primaryClass = {astro-ph.GA},
       adsurl = {https://ui.adsabs.harvard.edu/abs/2020ARA&A..58..617O}
}

@article{karkare/etal:2022,
   title={SPT-SLIM: A Line Intensity Mapping Pathfinder for the South Pole Telescope},
   volume={209},
   ISSN={1573-7357},
   url={http://dx.doi.org/10.1007/s10909-022-02702-2},
   DOI={10.1007/s10909-022-02702-2},
   number={5–6},
   journal={Journal of Low Temperature Physics},
   publisher={Springer Science and Business Media LLC},
   author={Karkare, K. S. and Anderson, A. J. and Barry, P. S. and Benson, B. A. and Carlstrom, J. E. and Cecil, T. and Chang, C. L. and Dobbs, M. A. and Hollister, M. and Keating, G. K. and Marrone, D. P. and McMahon, J. and Montgomery, J. and Pan, Z. and Robson, G. and Rouble, M. and Shirokoff, E. and Smecher, G.},
   year={2022},
   month=mar, pages={758–765} }

@ARTICLE{gawiser/etal:2007,
       author = {{Gawiser}, Eric and {Francke}, Harold and {Lai}, Kamson and {Schawinski}, Kevin and {Gronwall}, Caryl and {Ciardullo}, Robin and {Quadri}, Ryan and {Orsi}, Alvaro and {Barrientos}, L. Felipe and {Blanc}, Guillermo A. and {Fazio}, Giovanni and {Feldmeier}, John J. and {Huang}, Jia-sheng and {Infante}, Leopoldo and {Lira}, Paulina and {Padilla}, Nelson and {Taylor}, Edward N. and {Treister}, Ezequiel and {Urry}, C. Megan and {van Dokkum}, Pieter G. and {Virani}, Shanil N.},
        title = "{Ly{\ensuremath{\alpha}}-Emitting Galaxies at z = 3.1: L* Progenitors Experiencing Rapid Star Formation}",
      journal = {\apj},
         year = 2007,
        month = dec,
       volume = {671},
       number = {1},
        pages = {278-284},
          doi = {10.1086/522955},
archivePrefix = {arXiv},
       eprint = {0710.2697},
 primaryClass = {astro-ph},
       adsurl = {https://ui.adsabs.harvard.edu/abs/2007ApJ...671..278G}
}

@ARTICLE{guaita/etal:2010,
       author = {{Guaita}, Lucia and {Gawiser}, Eric and {Padilla}, Nelson and {Francke}, Harold and {Bond}, Nicholas A. and {Gronwall}, Caryl and {Ciardullo}, Robin and {Feldmeier}, John J. and {Sinawa}, Shawn and {Blanc}, Guillermo A. and {Virani}, Shanil},
        title = "{Ly{\ensuremath{\alpha}}-emitting Galaxies at z = 2.1 in ECDF-S: Building Blocks of Typical Present-day Galaxies?}",
      journal = {\apj},
         year = 2010,
        month = may,
       volume = {714},
       number = {1},
        pages = {255-269},
          doi = {10.1088/0004-637X/714/1/255},
archivePrefix = {arXiv},
       eprint = {0910.2244},
 primaryClass = {astro-ph.CO},
       adsurl = {https://ui.adsabs.harvard.edu/abs/2010ApJ...714..255G}
}

@ARTICLE{kikuchihara/etal:2022,
       author = {{Kikuchihara}, Shotaro and {Harikane}, Yuichi and {Ouchi}, Masami and {Ono}, Yoshiaki and {Shibuya}, Takatoshi and {Itoh}, Ryohei and {Kakuma}, Ryota and {Inoue}, Akio K. and {Kusakabe}, Haruka and {Shimasaku}, Kazuhiro and {Momose}, Rieko and {Sugahara}, Yuma and {Kikuta}, Satoshi and {Saito}, Shun and {Kashikawa}, Nobunari and {Zhang}, Haibin and {Lee}, Chien-Hsiu},
        title = "{SILVERRUSH. XII. Intensity Mapping for Ly{\ensuremath{\alpha}} Emission Extending over 100-1000 Comoving Kpc around z   2-7 LAEs with Subaru HSC-SSP and CHORUS Data}",
      journal = {\apj},
         year = 2022,
        month = jun,
       volume = {931},
       number = {2},
          eid = {97},
        pages = {97},
          doi = {10.3847/1538-4357/ac69de},
archivePrefix = {arXiv},
       eprint = {2108.09288},
 primaryClass = {astro-ph.GA},
       adsurl = {https://ui.adsabs.harvard.edu/abs/2022ApJ...931...97K}
}

@ARTICLE{kakuma/etal:2021,
       author = {{Kakuma}, Ryota and {Ouchi}, Masami and {Harikane}, Yuichi and {Ono}, Yoshiaki and {Inoue}, Akio K. and {Komiyama}, Yutaka and {Kusakabe}, Haruka and {Lee}, Chien-Hsiu and {Matsuda}, Yuichi and {Matsuoka}, Yoshiki and {Mawatari}, Ken and {Momose}, Rieko and {Shibuya}, Takatoshi and {Taniguchi}, Yoshiaki},
        title = "{SILVERRUSH. IX. Ly{\ensuremath{\alpha}} Intensity Mapping with Star-forming Galaxies at z = 5.7 and 6.6: A Possible Detection of Extended Ly{\ensuremath{\alpha}} Emission at {\ensuremath{\gtrsim}}100 Comoving Kiloparsecs around and beyond the Virial-radius Scale of Galaxy Dark Matter Halos}",
      journal = {\apj},
         year = 2021,
        month = jul,
       volume = {916},
       number = {1},
          eid = {22},
        pages = {22},
          doi = {10.3847/1538-4357/ac0725},
archivePrefix = {arXiv},
       eprint = {1906.00173},
 primaryClass = {astro-ph.GA},
       adsurl = {https://ui.adsabs.harvard.edu/abs/2021ApJ...916...22K}
}

@ARTICLE{wisotzki/etal:2018,
       author = {{Wisotzki}, L. and {Bacon}, R. and {Brinchmann}, J. and {Cantalupo}, S. and {Richter}, P. and {Schaye}, J. and {Schmidt}, K.~B. and {Urrutia}, T. and {Weilbacher}, P.~M. and {Akhlaghi}, M. and {Bouch{\'e}}, N. and {Contini}, T. and {Guiderdoni}, B. and {Herenz}, E.~C. and {Inami}, H. and {Kerutt}, J. and {Leclercq}, F. and {Marino}, R.~A. and {Maseda}, M. and {Monreal-Ibero}, A. and {Nanayakkara}, T. and {Richard}, J. and {Saust}, R. and {Steinmetz}, M. and {Wendt}, M.},
        title = "{Nearly all the sky is covered by Lyman-{\ensuremath{\alpha}} emission around high-redshift galaxies}",
      journal = {\nat},
         year = 2018,
        month = oct,
       volume = {562},
       number = {7726},
        pages = {229-232},
          doi = {10.1038/s41586-018-0564-6},
archivePrefix = {arXiv},
       eprint = {1810.00843},
 primaryClass = {astro-ph.GA},
       adsurl = {https://ui.adsabs.harvard.edu/abs/2018Natur.562..229W}
}

@ARTICLE{kusakabe/etal:2022,
       author = {{Kusakabe}, Haruka and {Verhamme}, Anne and {Blaizot}, J{\'e}r{\'e}my and {Garel}, Thibault and {Wisotzki}, Lutz and {Leclercq}, Floriane and {Bacon}, Roland and {Schaye}, Joop and {Gallego}, Sofia G. and {Kerutt}, Josephine and {Matthee}, Jorryt and {Maseda}, Michael and {Nanayakkara}, Themiya and {Pell{\'o}}, Roser and {Richard}, Johan and {Tresse}, Laurence and {Urrutia}, Tanya and {Vitte}, Elo{\"\i}se},
        title = "{The MUSE eXtremely Deep Field: Individual detections of Ly{\ensuremath{\alpha}} haloes around rest-frame UV-selected galaxies at z $\simeq$ 2.9-4.4}",
      journal = {\aap},
         year = 2022,
        month = apr,
       volume = {660},
          eid = {A44},
        pages = {A44},
          doi = {10.1051/0004-6361/202142302},
archivePrefix = {arXiv},
       eprint = {2201.07257},
 primaryClass = {astro-ph.GA},
       adsurl = {https://ui.adsabs.harvard.edu/abs/2022A&A...660A..44K}
}

@ARTICLE{steidel/etal:2011,
       author = {{Steidel}, Charles C. and {Bogosavljevi{\'c}}, Milan and {Shapley}, Alice E. and {Kollmeier}, Juna A. and {Reddy}, Naveen A. and {Erb}, Dawn K. and {Pettini}, Max},
        title = "{Diffuse Ly{\ensuremath{\alpha}} Emitting Halos: A Generic Property of High-redshift Star-forming Galaxies}",
      journal = {\apj},
         year = 2011,
        month = aug,
       volume = {736},
       number = {2},
          eid = {160},
        pages = {160},
          doi = {10.1088/0004-637X/736/2/160},
archivePrefix = {arXiv},
       eprint = {1101.2204},
 primaryClass = {astro-ph.CO},
       adsurl = {https://ui.adsabs.harvard.edu/abs/2011ApJ...736..160S}
}

@ARTICLE{trainor/etal:2025,
       author = {{Trainor}, Ryan F. and {Lamb}, Noah R. and {Steidel}, Charles C. and {Chen}, Yuguang and {Erb}, Dawn K. and {Trenholm}, Elizabeth and {McClain}, Rebecca L. and {Kovach}, Io},
        title = "{The Lyman-alpha Halos of Galaxies at z=2-3 in the Keck Baryonic Structure Survey}",
      journal = {arXiv e-prints},
         year = 2025,
        month = may,
          eid = {arXiv:2505.15881},
        pages = {arXiv:2505.15881},
          doi = {10.48550/arXiv.2505.15881},
archivePrefix = {arXiv},
       eprint = {2505.15881},
 primaryClass = {astro-ph.GA},
       adsurl = {https://ui.adsabs.harvard.edu/abs/2025arXiv250515881T}
}

@ARTICLE{kikuta/etal:2023,
       author = {{Kikuta}, Satoshi and {Matsuda}, Yuichi and {Inoue}, Shigeki and {Steidel}, Charles C. and {Cen}, Renyue and {Zheng}, Zheng and {Yajima}, Hidenobu and {Momose}, Rieko and {Imanishi}, Masatoshi and {Komiyama}, Yutaka},
        title = "{UV and Ly{\ensuremath{\alpha}} Halos of Ly{\ensuremath{\alpha}} Emitters across Environments at z = 2.84}",
      journal = {\apj},
         year = 2023,
        month = apr,
       volume = {947},
       number = {2},
          eid = {75},
        pages = {75},
          doi = {10.3847/1538-4357/acbf30},
archivePrefix = {arXiv},
       eprint = {2302.12848},
 primaryClass = {astro-ph.GA},
       adsurl = {https://ui.adsabs.harvard.edu/abs/2023ApJ...947...75K}
}

@ARTICLE{house/etal:2024,
       author = {{House}, Lindsay R. and {Gebhardt}, Karl and {Finkelstein}, Keely and {Mentuch Cooper}, Erin and {Davis}, Dustin and {Farrow}, Daniel J. and {Schneider}, Donald P.},
        title = "{Participatory Science and Machine Learning Applied to Millions of Sources in the Hobby{\textendash}Eberly Telescope Dark Energy Experiment}",
      journal = {\apj},
         year = 2024,
        month = nov,
       volume = {975},
       number = {2},
          eid = {172},
        pages = {172},
          doi = {10.3847/1538-4357/ad782c},
archivePrefix = {arXiv},
       eprint = {2409.08359},
 primaryClass = {astro-ph.IM},
       adsurl = {https://ui.adsabs.harvard.edu/abs/2024ApJ...975..172H}
}

@ARTICLE{house/etal:2023,
       author = {{House}, Lindsay R. and {Gebhardt}, Karl and {Finkelstein}, Keely and {Cooper}, Erin Mentuch and {Davis}, Dustin and {Ciardullo}, Robin and {Farrow}, Daniel J. and {Finkelstein}, Steven L. and {Gronwall}, Caryl and {Jeong}, Donghui and {Johnson}, L. Clifton and {Liu}, Chenxu and {Thomas}, Benjamin P. and {Zeimann}, Gregory},
        title = "{Using Dark Energy Explorers and Machine Learning to Enhance the Hobby-Eberly Telescope Dark Energy Experiment}",
      journal = {\apj},
         year = 2023,
        month = jun,
       volume = {950},
       number = {2},
          eid = {82},
        pages = {82},
          doi = {10.3847/1538-4357/accdd0},
archivePrefix = {arXiv},
       eprint = {2304.07348},
 primaryClass = {astro-ph.IM},
       adsurl = {https://ui.adsabs.harvard.edu/abs/2023ApJ...950...82H}
}

@ARTICLE{dong/etal:2023,
       author = {{Dong}, Chenze and {Lee}, Khee-Gan and {Ata}, Metin and {Horowitz}, Benjamin and {Momose}, Rieko},
        title = "{Observational Evidence for Large-scale Gas Heating in a Galaxy Protocluster at z = 2.30}",
      journal = {\apjl},
         year = 2023,
        month = mar,
       volume = {945},
       number = {2},
          eid = {L28},
        pages = {L28},
          doi = {10.3847/2041-8213/acba89},
archivePrefix = {arXiv},
       eprint = {2303.07619},
 primaryClass = {astro-ph.GA},
       adsurl = {https://ui.adsabs.harvard.edu/abs/2023ApJ...945L..28D}
}

@ARTICLE{miller/etal:2021,
       author = {{Miller}, Joel S.~A. and {Bolton}, James S. and {Hatch}, Nina A.},
        title = "{Searching for the shadows of giants - II. The effect of local ionization on the Ly {\ensuremath{\alpha}} absorption signatures of protoclusters at redshift z   2.4}",
      journal = {\mnras},
         year = 2021,
        month = oct,
       volume = {506},
       number = {4},
        pages = {6001-6013},
          doi = {10.1093/mnras/stab2083},
archivePrefix = {arXiv},
       eprint = {2107.07307},
 primaryClass = {astro-ph.CO},
       adsurl = {https://ui.adsabs.harvard.edu/abs/2021MNRAS.506.6001M}
}

@ARTICLE{byrohl/etal:2019,
       author = {{Byrohl}, Chris and {Saito}, Shun and {Behrens}, Christoph},
        title = "{Radiative transfer distortions of Lyman {\ensuremath{\alpha}} emitters: a new Fingers-of-God damping in the clustering in redshift space}",
      journal = {\mnras},
         year = 2019,
        month = nov,
       volume = {489},
       number = {3},
        pages = {3472-3491},
          doi = {10.1093/mnras/stz2260},
archivePrefix = {arXiv},
       eprint = {1906.02173},
 primaryClass = {astro-ph.CO},
       adsurl = {https://ui.adsabs.harvard.edu/abs/2019MNRAS.489.3472B}
}

@ARTICLE{gurung-lopez/etal:2021,
       author = {{Gurung-L{\'o}pez}, Siddhartha and {Saito}, Shun and {Baugh}, Carlton M. and {Bonoli}, Silvia and {Lacey}, Cedric G. and {Orsi}, {\'A}lvaro A.},
        title = "{Determining the systemic redshift of Lyman {\ensuremath{\alpha}} emitters with neural networks and improving the measured large-scale clustering}",
      journal = {\mnras},
         year = 2021,
        month = jan,
       volume = {500},
       number = {1},
        pages = {603-626},
          doi = {10.1093/mnras/staa3269},
archivePrefix = {arXiv},
       eprint = {2005.12931},
 primaryClass = {astro-ph.GA},
       adsurl = {https://ui.adsabs.harvard.edu/abs/2021MNRAS.500..603G}
}

@ARTICLE{tinker/etal:2010,
       author = {{Tinker}, Jeremy L. and {Robertson}, Brant E. and {Kravtsov}, Andrey V. and {Klypin}, Anatoly and {Warren}, Michael S. and {Yepes}, Gustavo and {Gottl{\"o}ber}, Stefan},
        title = "{The Large-scale Bias of Dark Matter Halos: Numerical Calibration and Model Tests}",
      journal = {\apj},
         year = 2010,
        month = dec,
       volume = {724},
       number = {2},
        pages = {878-886},
          doi = {10.1088/0004-637X/724/2/878},
archivePrefix = {arXiv},
       eprint = {1001.3162},
 primaryClass = {astro-ph.CO},
       adsurl = {https://ui.adsabs.harvard.edu/abs/2010ApJ...724..878T}
}

@ARTICLE{raiter/etal:2010,
       author = {{Raiter}, A. and {Schaerer}, D. and {Fosbury}, R.~A.~E.},
        title = "{Predicted UV properties of very metal-poor starburst galaxies}",
      journal = {\aap},
         year = 2010,
        month = nov,
       volume = {523},
          eid = {A64},
        pages = {A64},
          doi = {10.1051/0004-6361/201015236},
archivePrefix = {arXiv},
       eprint = {1008.2114},
 primaryClass = {astro-ph.CO},
       adsurl = {https://ui.adsabs.harvard.edu/abs/2010A&A...523A..64R}
}

@ARTICLE{wyithe/dijkstra:2011,
       author = {{Wyithe}, J. Stuart B. and {Dijkstra}, Mark},
        title = "{Non-gravitational contributions to the clustering of Ly{\ensuremath{\alpha}} selected galaxies: implications for cosmological surveys}",
      journal = {\mnras},
         year = 2011,
        month = aug,
       volume = {415},
       number = {4},
        pages = {3929-3950},
          doi = {10.1111/j.1365-2966.2011.19007.x},
archivePrefix = {arXiv},
       eprint = {1104.0712},
 primaryClass = {astro-ph.CO},
       adsurl = {https://ui.adsabs.harvard.edu/abs/2011MNRAS.415.3929W},
}

@ARTICLE{schechter:1976,
       author = {{Schechter}, P.},
        title = "{An analytic expression for the luminosity function for galaxies.}",
      journal = {\apj},
         year = 1976,
        month = jan,
       volume = {203},
        pages = {297-306},
          doi = {10.1086/154079},
       adsurl = {https://ui.adsabs.harvard.edu/abs/1976ApJ...203..297S}
}

@ARTICLE{herrera/etal:2025,
       author = {{Herrera}, Danisbel and {Gawiser}, Eric and {Benda}, Barbara and {Firestone}, Nicole M. and {Ramakrishnan}, Vandana and {Moon}, Byeongha and {Lee}, Kyoung-Soo and {Park}, Changbom and {Valdes}, Francisco and {Yang}, Yujin and {Artale}, Mar{\'\i}a Celeste and {Ciardullo}, Robin and {Gronwall}, Caryl and {Guaita}, Lucia and {Hwang}, Ho Seong and {Kennedy}, Jacob and {Kumar}, Ankit and {Zabludoff}, Ann},
        title = "{ODIN: Clustering Analysis of 14,000 Ly{\ensuremath{\alpha}}-emitting Galaxies at z = 2.4, 3.1, and 4.5}",
      journal = {\apjl},
         year = 2025,
        month = aug,
       volume = {988},
       number = {2},
          eid = {L57},
        pages = {L57},
          doi = {10.3847/2041-8213/adec82},
archivePrefix = {arXiv},
       eprint = {2503.17824},
 primaryClass = {astro-ph.GA},
       adsurl = {https://ui.adsabs.harvard.edu/abs/2025ApJ...988L..57H}
}

@ARTICLE{renard/etal:2024,
       author = {{Renard}, Pablo and {Spinoso}, Daniele and {Montero-Camacho}, Paulo and {Sun}, Zechang and {Zou}, Hu and {Cai}, Zheng},
        title = "{Probing the cosmic web in Ly{\ensuremath{\alpha}} emission over large scales: an intensity mapping forecast for DECaLS/BASS and DESI}",
      journal = {\mnras},
         year = 2024,
        month = nov,
       volume = {535},
       number = {1},
        pages = {826-852},
          doi = {10.1093/mnras/stae2358},
archivePrefix = {arXiv},
       eprint = {2406.18775},
 primaryClass = {astro-ph.CO},
       adsurl = {https://ui.adsabs.harvard.edu/abs/2024MNRAS.535..826R}
}

@INPROCEEDINGS{renard/etal:2025,
       author = {{Renard}, P. and {Spinoso}, D. and {Sun}, Zechang and {Zou}, Hu and {Montero-Camacho}, P. and {Cai}, Zheng},
        title = "{Can We Detect the Cosmic Web in Ly{\ensuremath{\alpha}} Emission? Ly{\ensuremath{\alpha}} Intensity Mapping Forecast for DECaLS/BASS-DESI}",
    booktitle = {Highlights of Spanish Astrophysics XII},
         year = 2025,
       editor = {{Manteiga}, M. and {Gonz{\'a}lez-Galindo}, F. and {Labiano-Ortega}, A. and {Mart{\'\i}nez-Gonz{\'a}lez}, M.~J. and {Rea}, N. and {Romero-G{\'o}mez}, M. and {Ulla-Miguel}, A. and {Yepes}, G. and {Rodr{\'\i}guez-L{\'o}pez}, C. and {G{\'o}mez-Garc{\'\i}a}, A. and et al.},
        month = may,
        pages = {4},
       adsurl = {https://ui.adsabs.harvard.edu/abs/2025hsa..conf....4R}
}

@ARTICLE{fonseca/etal:2017,
       author = {{Fonseca}, Jos{\'e} and {Silva}, Marta B. and {Santos}, M{\'a}rio G. and {Cooray}, Asantha},
        title = "{Cosmology with intensity mapping techniques using atomic and molecular lines}",
      journal = {\mnras},
         year = 2017,
        month = jan,
       volume = {464},
       number = {2},
        pages = {1948-1965},
          doi = {10.1093/mnras/stw2470},
archivePrefix = {arXiv},
       eprint = {1607.05288},
 primaryClass = {astro-ph.CO},
       adsurl = {https://ui.adsabs.harvard.edu/abs/2017MNRAS.464.1948F}
}
\bibliographystyle{aasjournal}

\end{document}